\documentclass[twocolumn,superscriptaddress,nopacs,preprintnumbers,amsmath,amssymb,nofootinbib]{revtex4}

\usepackage{graphicx,verbatim}

\def\P{\mathcal{P}}
\def\Pi{\mathcal{P}_\infty}
\def\tr{{\rm tr}\,}
\def\o{\omega}
\def\pa{\partial}
\def\bo{\bar{\omega}}

\def\z{\vec{z}_1}
\def\zz{\vec{z}_2}
\def\vd{\vec{\delta}}

\def\s{4\pi \o s/\beta}
\def\r{4\pi \bo r/\beta}

\def\tc{\tilde{\chi}}

\def\T{{\sf T}}

\def\sig{\sigma}

\def\Eqn#1{Eq.~(\ref{#1})}

\def\fig#1{Fig.~\ref{#1}}

\begin{document}

\title{The Vortex Structure of SU(2) Calorons
} 

\author{Falk Bruckmann}
\affiliation{Institut f\"ur Theoretische Physik, Universit\"at
  Regensburg, D-93040 Regensburg, Germany} 
\author{Ernst-Michael Ilgenfritz}
\affiliation{Institut f\"ur Physik, Humboldt-Universit\"at, D-12489 Berlin, Germany}
\affiliation{Institut f\"ur Theoretische Physik, Universit\"at
  Heidelberg, D-69120 Heidelberg, Germany}
\author{Boris Martemyanov}
\affiliation{Institute for Theoretical and Experimental Physics, B.\ Cheremushkinskaya 25, 117259 Moscow, Russia}
\author{Bo Zhang}
\affiliation{Institut f\"ur Theoretische Physik, Universit\"at
  Regensburg, D-93040 Regensburg, Germany} 

\begin{abstract}
  We reveal the center vortex content of $SU(2)$ calorons and ensembles of them. We use Laplacian Center Gauge as well as Maximal Center Gauges to show that the vortex in a single caloron consists of two parts. The first one connects the constituent dyons of the caloron (which are monopoles in Laplacian Abelian Gauge) and extends in time. The second part is predominantly spatial, encloses one of the dyons and can be related to the twist in the caloron gauge field. This part depends strongly on the caloron holonomy and degenerates to a plane when the holonomy is maximally nontrivial, i.e. when the asymptotic Polyakov loop is traceless. Correspondingly, we find the spatial vortices in caloron ensembles to percolate in this case. This finding fits perfectly in the confinement scenario of vortices and shows that calorons are suitable to facilitate the vortex confinement mechanism.
\vspace{-7.5cm}
\begin{flushright}

HU-EP-09/61

\end{flushright}
\vspace{7.5cm}
\end{abstract}

\maketitle

\section{Introduction}
\label{sect:Introduction}
To answer the question of what drives confinement and other nonperturbative 
effects in QCD, basically three sorts of topological excitations
have been
intensively examined over the years: instantons, magnetic monopoles and 
center vortices.  Instantons as solutions of the 
equations of motion are special: they are the relevant objects in a semiclassical approach. 
While the generation of a chiral condensate is very natural via the (quasi) zero 
modes\footnote{This mechanism is based on the index theorem and thus will work for any object with topological charge.}, confinement remained unexplained in this model. 

At finite temperature, where the classical solutions are called calorons \cite{Harrington:1978ve,Kraan:1998pm,Lee:1998bb}, 
there has been quite some progress recently, due to two effects. First of 
all, the asymptotic Polyakov loop plays a key role determining the properties of a 
new type of caloron solutions (for more details on calorons see 
Sect.~\ref{sect:Calorons} and the reviews \cite{Bruckmann:2003yq}). Under the 
conjecture that 
the asymptotic Polyakov loop is related to the average Polyakov loop, 
the order parameter of confinement, calorons are sensitive to the 
phase of QCD under consideration.

Secondly, the new calorons with nontrivial holonomy consist of $N$ 
dyons/magnetic monopoles for the gauge group $SU(N)$. In this way, 
contact is seemingly made to the 
Dual Superconductor scenario. 
We stress that the dyon constituents of calorons appear in an unambiguous way as classical objects.

This is in contrast to Abelian monopoles and also center vortices, the other
sorts of objects used to explain confinement.
In the sense, that they are widely accepted, 
they are not of semiclassical nature. 
They represent gauge defects of codimension 3 and 2, respectively,  
which remain after the respective gauge fixing and projection.
Their interrelation and the fact that they are a pre\-re\-qui\-si\-te for the
occurrence of topological charge in general has been quantitatively 
studied in the past~\cite{Boyko:2006ic}.

Monopoles are usually obtained by applying the Maximal Abelian Gauge (MAG).
Center vortices in lattice QCD can be defined through a center projection (therefore called P-vortices) after the lattice
gauge field has been transformed into the Maximal Center Gauge (directly 
by Direct Maximal Center Gauge [DMCG] or indirectly 
by Indirect Maximal Center Gauge [IMCG], 
with the MAG as preconditioner) or into the Laplacian Center Gauge (LCG).
We refer to Sect.~\ref{sect:CenterVortices} for the technicalities.

We would also like to mention another important development during recent years. 
Fermionic methods have become available to study topological
structures without the necessity of smoothing. 
Singular, codimension 1 sheets of sign-coherent topological charge have been 
found and proposed to be characteristic for genuine quantum configurations 
\cite{Horvath:2003yj,Ilgenfritz:2007xu} and potentially important for 
the confinement property. 
The relation to the other low-dimensional singular topological excitations is still not completely understood. 
In this scenario, (anti)selfdual objects like calorons typically appear as topological lumps after smearing \cite{Ilgenfritz:2004zz},
i.e.\ at a resolution length bigger than the lattice spacing \cite{Ilgenfritz:2007xu}.

The physical mechanisms assigned to calorons would be based on their quantum 
weight \cite{Diakonov:2004jn}, their moduli space metric \cite{Kraan:1998kp}, 
the particular features of their fermionic zero modes \cite{GarciaPerez:1999ux} 
and the specific suppression of dyons 
according to the action they acquire in different 
phases \cite{Bornyakov:2008bg}.
The last observation suggests an overall description of confinement 
\cite{Diakonov:2007nv} and deconfinement in terms of calorons' dyon 
constituents as independent degrees of freedom. 
The proposed generalized (approximative) moduli space metric, however, 
presents some
difficulties \cite{Bruckmann:2009nw} which are not yet overcome.

Since calorons unify instanton and monopoles, it is natural to ask for the 
relation to  
center vortices. In four dimensions the latter are two-dimensional worldsheets 
with Wilson loops taking values in the center of the gauge group if they 
are linked with the vortex sheet. Vortices that randomly penetrate a given 
Wilson loop 
very naturally give rise to an area law. Since vortices are 
closed surfaces, the necessary randomness can be facilitated only by large vortices. 
This is further translated into the percolation of vortices, meaning that 
the size of the (largest) vortex clusters becomes comparable to the extension 
of the space itself. This 
percolation has been observed in lattice 
simulations of the confined phase \cite{Langfeld:1997jx}, while in the deconfined phase the vortices align in the timelike direction 
and the percolation mechanism remains working only for spatial Wilson loops \cite{Langfeld:1998cz,Engelhardt:1999fd}.
This parallels percolation properties of monopoles. Moreover, 
it conforms with the observation at high temperatures 
that the spatial Wilson loops keep a string tension in contrast to the 
correlators of Polyakov loops.

In this paper we will merge the caloron and vortex picture focussing on two 
aspects: {\it (i)} to demonstrate 
how the vortex content of individual 
calorons depends on the parameters of the caloron solution -- in particular 
the holonomy -- and {\it (ii)} to obtain the vortices in corresponding caloron 
ensembles and analyze 
their percolation properties. For simplicity we will 
restrict ourselves to gauge group $SU(2)$. We will mainly use LCG, 
which has found a correlation of vortices to instantons cores  
in \cite{Alexandrou:1999vx,Alexandrou:1999iy}. 
We recall that LCG has been abandoned
for finding vortices in $SU(2)$ Monte Carlo configurations because
the vortex density did not possess a good continuum limit \cite{Langfeld:2001nz}.
This observation does not invalidate the application of LCG to smooth 
(semiclassical) field configurations.
We also 
compare with results obtained by DMCG and IMCG. 
In order to enable the application of these gauge-fixing techniques we discretize 
calorons on a lattice, which is known to reproduce continuum results very well.

In the combination of Laplacian Abelian Gauge (LAG) and LCG, magnetic monopole 
worldlines are known to reside on the vortex sheets, and we will confirm this 
for the calorons' constituent dyons (see \cite{Alexandrou:1999iy} for some 
first findings). 

In addition, we find another -- mainly spatial -- part of the vortex surfaces.
It is strongly related to the twist of the dyons within the caloron. 
We explain this by analytic arguments that yield good approximations for the locations of these spatial vortices. 

For a single caloron, both parts of the vortex system together generate two 
intersection points 
needed to constitute the topological charge in the vortex 
language.\footnote{Of course, for the caloron as a classical object the 
topological charge density is continuously distributed.} 

In caloron ensembles we find the spatial vortex surfaces to percolate only at 
low temperatures (where the holonomy is maximally non-trivial), while the 
space-time vortex surfaces are rather independent of the phase (i.e. the
holonomy), both in agreement with physical expectations and with observations
in caloron gas simulations that 
have evaluated the $Q\bar{Q}$ free energy on one 
hand and the space like Wilson loops on the other \cite{Gerhold:2006sk}. 

From the results for 
individual calorons it is clear that the Polyakov loop, which we treat as 
an input parameter for the caloron solution, is responsible for the percolation
and hence the 
string tension in the confined phase. This lends 
support for the hypothesis that the holonomy is important as the ``correct 
background'' for the classical objects featuring in a semiclassical 
understanding of finite temperature QCD.

The paper is organised as follows. In the next two Sections~\ref{sect:Calorons}
and~\ref{sect:CenterVortices} we review the properties of calorons and 
vortices, including technicalities of how to discretize the former and how to 
detect the latter. In Section~\ref{sect:IndividualCalorons} we describe 
the vortex content of single calorons. In Section~\ref{sect:VorticesCaloronEnsembles}
we demonstrate how the vortex content of caloron gases changes with the holonomy 
parameter.  We conclude with a summary and a brief outlook.
Part of our results have been published in \cite{Zhang:2009et}.

\section{Calorons}
\label{sect:Calorons}

\subsection{Generalities}
\label{subsect:Generalities}

Calorons are instantons, i.e. selfdual\footnote{The results for antiselfdual 
calorons with negative topological charge are completely analogous.} 
Yang-Mills fields and therefore solutions of the equations of motion, at 
finite temperature. In other words, their base space is $R^3\times S^1$ 
where the circle $S^1$ has circumference $\beta=1/k_B T$ as usual.

As it turns out from the explicit solutions \cite{Harrington:1978ve,Kraan:1998pm,Lee:1998bb}, 
calorons consist of localised lumps of topological charge density, which --
due to selfduality -- are lumps of action density, too. For the gauge group 
$SU(N)$ one can have up to $N$ lumps per unit topological charge. 
When well separated, these lumps are static\footnote{
The gauge field, generically, can and will be time dependent, 
see Sect.~\ref{subsect:Twist}.}.
Moreover, they possess (quantised) magnetic charge equal to their electric 
charge and hence are called \emph{dyons}. 
Consequently, the moduli of calorons are the spatial locations of the dyons, which can take any value, plus phases \cite{Kraan:1998kp}.

Another important (superselection) parameter of the new solutions by 
Kraan/van Baal and Lee/Lu \cite{Kraan:1998pm,Lee:1998bb} is the holonomy, 
the limit of the (untraced) Polyakov loop at spatial infinity,
\begin{equation}
 \Pi=\lim_{|\vec{x}|\to\infty}\P\exp\left(i\int_0^\beta\!\!\! A_0 d x_0\right) \, .
\end{equation}
Due to the magnetic neutrality of the dyons within a caloron, 
this limit is independent of the direction the limit is taken.
(In our convention the gauge fields are hermitean, we basically follow the notation of \cite{Kraan:1998pm} but multiply their antihermitean gauge fields by $i$ and reinstate $\beta$.)

In $SU(2)$ we diagonalise $\Pi$,
\begin{equation}
 \Pi=\exp\left(2\pi i\o \sig_3\right)
\label{eq:eqn_holonomy_diagonal}
\end{equation}
with $\sig_i$ the Pauli matrices.
Note that $\o=0$ or $1/2$ amount to trivial holonomies $\Pi=\pm 1_2$, 
whereas the case $\o=1/4$, i.e. $\tr\Pi=0$ is referred to as maximal 
nontrivial holonomy.

\begin{figure}
\includegraphics[width=0.9\linewidth]{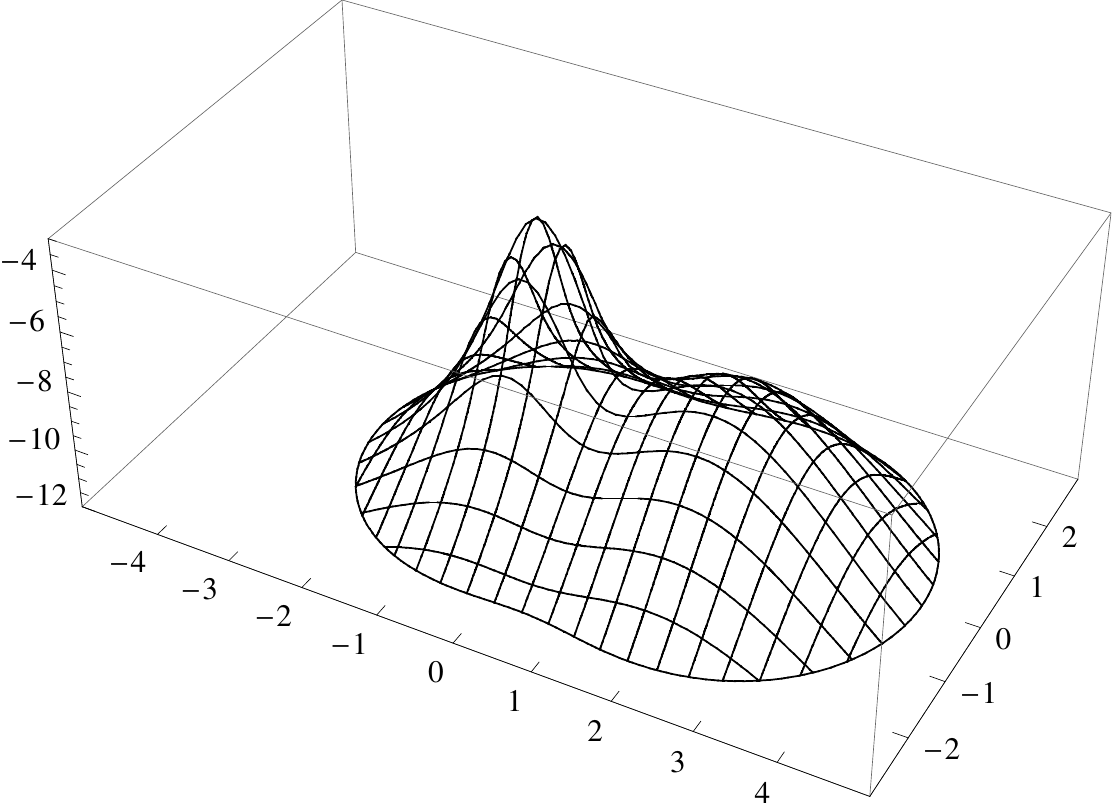}
\includegraphics[width=0.9\linewidth]{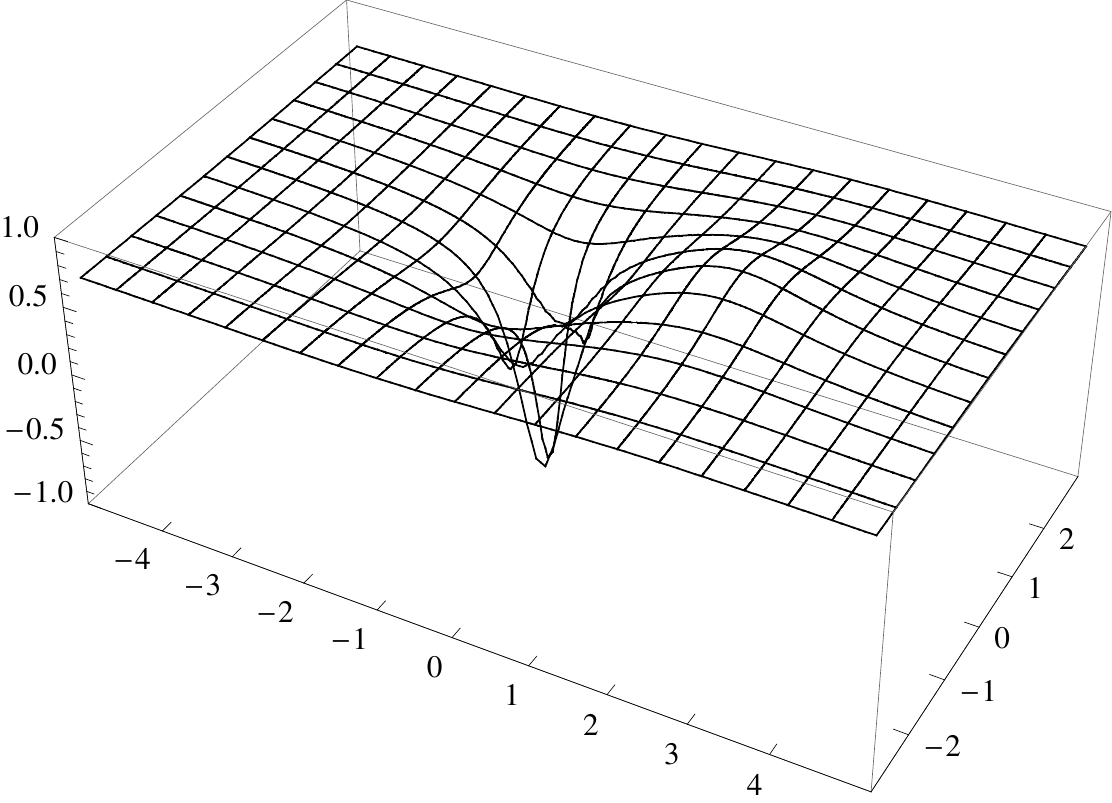}
\caption{Action density (top, shown in logarithmic scale and cut below $e^{-12}$) 
and Polyakov loop (bottom) in the $(x_1,x_3)$-plane (measured in units of $\beta$) at $x_0=x_2=0$ for a 
caloron with intermediate holonomy $\o=0.12$ and size $\rho=0.9\beta$ as 
discretized on a $8\times 48^2\times 80$ lattice. 
The dyon locations are $\z=(0,0,-0.61)$ 
and $\zz=(0,0,1.93)$.} 
\label{fig:Figure1}
\end{figure}

The constituent dyons have fractional topological charges (``masses'') 
governed by the holonomy, namely $2\omega$ and $2\bo\equiv 1-2\omega$, 
cf.\ Fig.~\ref{fig:Figure1} upper panel, such that -- from the point of 
view of the
topological density -- the constituent dyons are identical in the case 
of maximal nontrivial holonomy $\o=1/4$.

To be more concrete, the gauge field of a unit charge caloron in the periodic 
gauge\footnote{This gauge is in contrast to the nonperiodic ``algebraic gauge'' 
where $A_0$ asymptotically vanishes and the holonomy is carried by the 
transition function.} is given by 
\begin{equation}
\begin{split}
A_\mu^3&=-\frac{1}{2}\,\bar{\eta}^3_{\mu\nu}\pa_\nu\log\phi-\frac{2\pi\o}{\beta}\delta_{\mu,0}\,\\
A_\mu^1-iA_\mu^2&=-\frac{1}{2}\,\phi\,(\bar{\eta}^1_{\mu\nu}-i\bar{\eta}^2_{\mu\nu})(\pa_\nu+\frac{4\pi i \o}{\beta}\delta_{\nu,0})\,\tc \, ,
\label{eq:eqn_A_caloron}
\end{split}
\end{equation}
where $\bar{\eta}$ is the 't Hooft tensor (we use the convention in \cite{Kraan:1998pm}) and $\phi$ and $\chi$ are 
($x_0$-periodic) combinations of trigonometric and hyperbolic functions of 
$x_0$ and $\vec{x}$, respectively, see appendix \ref{app_functions} and 
\cite{Kraan:1998pm}. 
They are given in terms of the distances 
\begin{equation}
 r=|\vec{x}-\z|\,,\quad s=|\vec{x}-\zz|
\end{equation}
from the following constituent dyon locations
\begin{equation}
 \z=(0,0,-2\pi\o\rho^2/\beta),\quad \zz=(0,0,2\pi\bo\rho^2/\beta) \, ,
\end{equation}
which we have put on the $x_3$-axis with the center of mass at the origin  
(which can always be achieved by space rotations and translations) and at a 
distance of $d\equiv\pi\rho^2/\beta$ 
to each other.

In case of large $\rho$, the action consists of approximately static lumps
(of radius $\beta/4\pi\o$ and $\beta/4\pi\bo$ in spatial directions) near 
$\z$ and
$\zz$. 
In the small $\rho$ limit the action profile approaches a single 4d instanton-like lump at the
origin. In Ref.~\cite{Kraan:1998pm} one can find more plots of the action 
density of $SU(2)$ calorons with different sizes and holonomies.

In the far-field limit, away from both dyons the function $\tc$ 
behaves like
\begin{eqnarray}
 \tc&=&\frac{4 d}{(r+s+d)^2} \left\{ r
e^{-\r} e^{-2\pi i x_0} + s e^{-\s} \right\}\nonumber\\
&& \times [ 1+\mathcal{O}(e^{-
{\rm min}(\r,\s)}) ] \, ,
\label{eq:eqn_ff_chi}
\end{eqnarray}
and hence the off-diagonal part of $A_\mu$ decays exponentially, 
while the Abelian part from 
\begin{equation}
 \phi=\frac{r+s + d}{r+s - d} +\mathcal{O}(e^{-{\rm min}(\r,\s)})
\label{eq:eqn_ff_phi}
\end{equation}
becomes a dipole field \cite{Kraan:1998pm}.  

The Polyakov loop in the bulk plays a role similar to an exponentiated Higgs field in the 
gauge group: it is  $+1_2$ and $-1_2$ in the vicinity of  
$\z$ and $\zz$
\footnote{On the line connecting 
the dyons the Polyakov loop can actually be computed exactly 
\cite{GarciaPerez:1999hs}.}, 
respectively, cf.\ Fig.~\ref{fig:Figure1} lower panel. The existence of such points is of topological 
origin \cite{Reinhardt:1997rm}. Thus the Polyakov 
loop is a more 
suitable pointer to the constituent dyon locations, which agrees with the 
maxima of topological density for the limiting case of well-separated dyons, 
but is valid even 
in case the two topological lumps merged into one for small $\rho$.

\subsection{The twist}
\label{subsect:Twist}

A less-known feature of the caloron we want to describe 
next is the Taubes twist. 
It basically means that the gauge field\footnote{The twist can be formulated 
in a gauge-invariant way by field strength correlators between points 
connected by Schwinger lines \cite{Ilgenfritz:2002qs}.} of one of the dyons 
is rotated by a time dependent gauge transformation (rotated in the direction 
of the holonomy, here the third direction in color space) w.r.t.\ the gauge 
field of the other dyon when they are combined into a caloron.
This is the way 
the dyons generate the unit topological charge~\cite{Kraan:1998pm}. 

The simplest way to reveal the twist is to consider the limit of 
well-separated dyons, i.e. when their distance $d$ is much larger than 
their radius $\beta/4\pi\o$ and $\beta/4\pi\bo$.
Let us consider points near the first dyon, 
$\vec{x}=\z+\vd$, where the distance $\delta\equiv|\vd|$ 
is small compared to the separation $d$, but not necessarily compared to the dyon size.
Then the relevant distances are obviously
\begin{equation}
 r=\delta\, ,\quad s=|(0,0,-d)+\vd|=d-\delta_3+\mathcal{O}(\delta^2/d) \, .
\end{equation}
In the appendix we derive the form of the functions $\phi$ and $\tc$ in 
this limit, 
\begin{eqnarray}
 \phi(\vec{x} = \z+\vd)&\simeq&\frac{2d}{\delta\coth(4\pi\bo \delta/\beta)-\delta_3}
\label{eq:eqn_fields_first_phi} \, ,\\
 \tc(\vec{x}=\z+\vd)&\simeq&e^{-2\pi i x_0/\beta}\frac{1}{2d}\frac{\delta}{\sinh(4\pi\bo \delta/\beta)}\, .
\label{eq:eqn_fields_first_tc}
\end{eqnarray}
The large factors of $2d$ cancel in the expressions $\partial_\mu\log\phi$ and $\phi\,\partial_\mu\tc$ relevant for $A_\mu$, \Eqn{eq:eqn_A_caloron}.

In the vicinity of the other dyon, $\vec{x}=\zz+\vd$, with 
\begin{equation}
s=\delta\,,\quad r=d+\delta_3+\mathcal{O}(\delta^2/d)\,,
\end{equation} 
we get very similar expressions 
with $\bo$ replaced by $\o$ and $\delta_3$ by $-\delta_3$, but the time-dependent phase factor is absent,
\begin{eqnarray}
 \phi(\vec{x}=\zz+\vd)&\simeq&\frac{2d}{\delta\coth(4\pi\o\delta/\beta)+\delta_3}
\label{eq:eqn_fields_second_phi}\, , \\
 \tc(\vec{x}=\zz+\vd)&\simeq&\frac{1}{2d}\frac{\delta}{\sinh(4\pi\o \delta/\beta)}
\, .
\label{eq:eqn_fields_second_tc}
\end{eqnarray}
This staticity of course also holds for 
$A_\mu$ of this dyon and all quantities computed from it.

Plugging in those functions into the gauge field of Eq.~(\ref{eq:eqn_A_caloron}) 
one can find that the corresponding gauge field components are connected via a 
PT transformation, and the exchange of $\o$ and $\bo$
\begin{eqnarray}
 (A_\mu^1-iA_\mu^2)(x_0,\zz+\vd;\o)&=&-(A_\mu^1-iA_\mu^2)(-x_0,\z-\vd;\bo)\nonumber\\
&&e^{-2\pi i x_0/\beta} 
\label{eq:eqn_relation_A_12}\\
 A_\mu^3(x_0,\zz+\vd;\o)&=&-A_\mu^3(-x_0,\z-\vd;\bo) \nonumber\\
&&-\frac{\pi}{\beta}\delta_{\mu,0} \, ,
\label{eq:eqn_relation_A_3} 
\end{eqnarray}
and a gauge transformation, namely
\begin{eqnarray}
 A_\mu(x_0,\zz+\vd;\o)&=&-\,^\T\!A_\mu(-x_0,\z-\vd;\bo)
\end{eqnarray}
with the time-dependent twist gauge transformation
\begin{eqnarray}
 \T(x_0)=\exp(-\pi i \frac{x_0}{\beta} \sig_3)\,.
\label{eq:eqn_twist_T}
\end{eqnarray}
This gauge transformation is nonperiodic, $\T(\beta)=-1_2$ (but acts in the adjoint representation).

The Polyakov loop values inside the dyons are obtained from 
$\tc(\vec{x}=\vec{z}_{1,2}+\delta)=\mathcal{O}(\delta^2)$ and
\begin{eqnarray}
 \phi(\vec{x}=\z+\delta)&=&\frac{2d}{\beta/4\pi\bo-\delta_3+\mathcal{O}(\delta^2)} \, ,\\
 \phi(\vec{x}=\zz+\delta)&=&\frac{2d}{\beta/4\pi\o+\delta_3+\mathcal{O}(\delta^2)}\, ,
\end{eqnarray}
which results in 
\begin{eqnarray}
 A_0(\z)&=&-\frac{\pi}{\beta}\,\sigma_3 \qquad \P(\z)=-1_2 \, ,
\label{eq:eqn_polloop_first}\\
 A_0(\zz)&=& 0 \qquad\:\:\:\:\:\:\:\:\:\: \P(\zz)=+1_2 \, .
\label{eq:eqn_polloop_second}
\end{eqnarray}

Actually, the gauge field around $\zz$ is that of a static magnetic monopole 
with the Higgs field $\Phi$ identified  with $A_0$ through dimensional 
reduction. Indeed, it vanishes at the core according to 
(\ref{eq:eqn_polloop_second}) and approaches the ``vacuum expectation value'' 
$|\Phi|=2\pi\o/\beta$ away from the core. Accordingly, $D_i\Phi$ is 
identified with $D_i A_0=F_{i0}=E_i$, and the Bogomolnyi equation with the 
selfduality equation.

The gauge field around $\z$ is that of a twisted monopole,
i.e.\ a monopole gauge rotated with $\T$. 
The corresponding Higgs field is obtained from that of a static monopole by the same $\T$, transforming in the adjoint representation. Therefore, the Higgs field $\Phi$ of the twisted monopole agrees with the gauge field $A_0$ apart from the inhomogeneous term in Eq.\ (\ref{eq:eqn_relation_A_3}).
$\Phi$ vanishes at the core, too, and approaches the 
vacuum expectation value $2\pi\bo/\beta$. 

The electric and magnetic charges, as measured in the $\Phi$ direction 
through the 't Hooft field strength tensor, are equal and the same for both 
dyons. This is consistent with the fact that selfdual configurations fulfilling 
the BPS bound must have positive magnetic charge. 

These fields are in some unusual gauge: around the dyon cores the Higgs field 
has the hedgehog form $\Phi^a\sim (\vec{x}-\vec{z}_{1,2})_a$ 
which is called the radial gauge.
Far away from the dyons the Higgs field $\Phi$ becomes diagonal up to exponentially small corrections. 
Indeed, if one neglects the exponentially small $\tc$'s of (\ref{eq:eqn_fields_first_tc}) and (\ref{eq:eqn_fields_second_tc})
and replaces 
the hyperbolic cotangent by $1$ in the denominator of 
(\ref{eq:eqn_fields_first_phi}) and 
(\ref{eq:eqn_fields_second_phi}), this would be the so-called unitary 
gauge with diagonal Higgs field and a Dirac string singularity 
(along the line connecting the dyons). Far away from the caloron's dyons, 
however, the ``hedgehog'' 
$\Phi$ is not ``combed'' compeletely and 
there is no need for a singularity\footnote{In contrast, the gauge 
field $A_4$ written down in Sect. IIA of \cite{Diakonov:2004jn} is diagonal 
and $A_\varphi$ has a singularity at the $x_3$-axis.}. In other words the covering of the color space happens in an exponentially small but finite solid angle. 

More precisely, the Higgs
field $\Phi$ 
approaches $-2\pi\o\sigma_3/\beta$ and $+2\pi\bo\sigma_3/\beta$ 
away from the static and twisting dyon, respectively, for almost all directions. 
These values differ by $\pi\sigma_3/\beta$, and hence the corresponding $A_0$'s can be glued together (apart from a gauge singularity at the origin).
Moreover, in $A_0$ the leading far field corrections to the asymptotic value, 
namely monopole terms, are of opposite sign w.r.t.\ the fixed color direction $\sigma_3$ and therefore do not induce a net 
winding number in the asymptotic Polyakov loop. Hence the  holonomy is independent 
of the direction.

We remind the reader that this subsection has been dealing with the limit 
of well-separated dyons, i.e. all formulae are correct up to exponential 
corrections in $\beta/d$ and algebraic ones in $\delta/d$. 

\subsection{Discretization} 
\label{subsect:Discretization}

In order perform the necessary gauge transformations or diagonalizations of the Laplace operator in numerical form we translate 
the caloron solutions -- and later caloron gas configurations --
into lattice configurations.
For a space-time grid (with a temporal 
extent $N_0=8$ 
and spatial sizes of $N_i=48,\ldots,80$, see specifications later) 
we compute the links $U_\mu(x)$
as path-ordered exponentials of the gauge field $A_\mu(x)$ 
(for single-caloron solutions given by Eqn.~(\ref{eq:eqn_A_caloron})). 
Practically, the integral
\begin{equation}
U_\mu(x)={\cal P} \exp \left(- i \int_x^{x+a\hat{\mu}} A_\mu(y) dy_\mu \right)
\label{eq:link_expression}
\end{equation} 
is decomposed into at least $N=20$ subintervals, for which the exponential
(\ref{eq:link_expression}) is obtained by exponentiation of 
$i A_\mu(\tilde{y}) a/N$ with 
$A_\mu(\tilde{y})$ evaluated in the midpoint of the subinterval.
These exponential expressions are then multiplied in the required order
(from $x$ left to $x+a\hat{\mu}$ right).
A necessary condition for the validity of this approximation is that 
$a/N \ll \rho$ with $\rho$ characterizing the caloron size or a typical 
caloron size in the multicaloron configurations.

Still this might be not sufficient to ensure that the potential $A_\mu(y)$
is reasonably constant within the subinterval of all links and give a converged result.
In particular, the gauge field (\ref{eq:eqn_A_caloron}) is singular at the origin and has big gradients near the line connecting the dyons, as visualised in Fig.~2 of \cite{Gerhold:2006bh}.
Hence we dynamically adjust the number of subintervals $N$ for every link, 
ensuring that further increasing $N$ 
would leave unchanged all entries of the resulting link matrix $U_\mu(x)$.

The lattice field constructed this way is not strictly 
periodic in the three spatial directions, but this is not important
for the lattices at hand with $N_i \gg N_0$. 
The action is already very close to $8\pi^2$, 
the maximal deviation occurs for large calorons ($\omega \gtrsim 0.9\beta$)
and is about 15 \%.

Lateron, we will make heavy use of the lowest
Laplacian eigenmodes in the LCG. 
When computing these modes in the caloron backgrounds we enforce spatial periodicity by hand.
In Maximal Center gauges we also consider the caloron gauge field as spatially periodic. 



\subsection{Caloron ensembles}
\label{subsect:CaloronEnsembles}

The caloron gas configurations considered later 
in this paper have been created along the lines of Ref.~\cite{Gerhold:2006sk}.
The four-dimensional center of mass locations of the calorons are sampled randomly 
as well as the spatial orientation of the ``dipole axis'' 
connecting the two dyons and the angle of a global $U(1)$ 
rotation around the axis $\sigma_3$ in color space. 
The caloron size is sampled from a suitable size distribution 
$D(\rho,T)$. 
  
The superposition is performed in the so--called algebraic gauge with the same 
holonomy parameter $\omega$ taken for all calorons and 
anticalorons\footnote{Superposing (anti)calorons with \emph{different holonomies} would 
create jumps of $A_0$ in the transition regions.}.
Finally, the additive superposition is gauge-rotated into the periodic gauge. 
Then the field $A_\mu(x)$   
is periodic in Euclidean time and possesses the required asymptotic
holonomy. 
We have applied cooling to the superpositions in order to ensure spatial 
periodicity of the gauge field.

In Sect.~\ref{sect:VorticesCaloronEnsembles} we will compare sequences of random caloron gas configurations 
which differ in nothing else than the global holonomy parameter $\omega$.

\section{Center vortices}
\label{sect:CenterVortices}

To detect center vortices, we will mainly use the Laplacian Center Gauge (LCG)
procedure, which can be viewed as a generalization of Direct Maximal Center Gauge 
(DMCG) with the advantage to avoid 
the Gribov problem of the latter \cite{deForcrand:2000pg}. We will compare our 
results to vortices from the maximal center gauges DMCG and 
IMCG in Sect.\ \ref{sect:direct_gauges}.

In LCG one has to compute the two lowest 
eigenvectors of minus the gauge covariant 
Laplacian operator in the adjoint 
representation\footnote{We use $\phi$ with a subindex for the eigenmodes of 
the Laplacian, not to be confused with the auxiliary function $\phi$ involved 
in the caloron gauge field, Eq. (\ref{eq:eqn_A_caloron}).},
\begin{eqnarray}
 -\Delta [U^A]\phi_{0,1} & = & \lambda_{0,1}\phi_{0,1}
\label{eq:eqn_eigen}\\
 \Delta^{ab}_{xy} [U^A] & = & \frac{1}{a^2}\sum_\mu 
 \Big(   U_\mu^A(x)^{ab} \delta_{x+a\hat\mu,y} 
\Big. \nonumber \\ & & \Big. 
+ U_\mu^A(x-\hat{\mu})^{ba} \delta_{x-a\hat{\mu},y}  - 2 \delta^{ab} \delta_{xy} \Big)\:\: \\ 
& & a,b=1,2,3  \, , \nonumber
\end{eqnarray}
which we do by virtue of the ARPACK package~\cite{arpack}.

For the vortex detection, the lowest mode $\phi_0$ 
is rotated to the third color direction, i.e.\ diagonalised,
\begin{equation}
^V\!\phi_0=|\phi_0|\,\sig_3\,.
\label{eq:eqn_gauge_abelian}
\end{equation}
The remaining Abelian freedom of rotations around the third axis, 
$V \to vV$ with $v=\exp(i\alpha\sig_3)$ is fixed (up to center 
elements) by demanding for $\phi_1$ a particular form with vanishing 
second component and positive first component, respectively,
\begin{equation}
 (^{vV}\!\phi_1)^{a=2}=0\,,\qquad (^{vV}\!\phi_1)^{a=1}>0 \, .
\label{eq:eqn_gauge_vortex}
\end{equation}

Defects of this gauge fixing procedure appear when $\phi_0$ and $\phi_1$ 
are collinear, because then the Abelian freedom parametrised by $v$ remains 
unfixed. In \cite{deForcrand:2000pg} it was shown, that the points $x$, 
where $\phi_0(x)$ and $\phi_1(x)$ are collinear, define the generically 
two-dimensional vortex surface, as the Wilson loops in perpendicular 
planes take center elements. This includes points $x$, where $\phi_0$ 
vanishes, $\phi_0(x)=0$, which define monopole worldlines in the 
Laplacian Abelian Gauge (LAG) \cite{vanderSijs:1996gn}.

\begin{figure}
\includegraphics[width=0.8\linewidth]{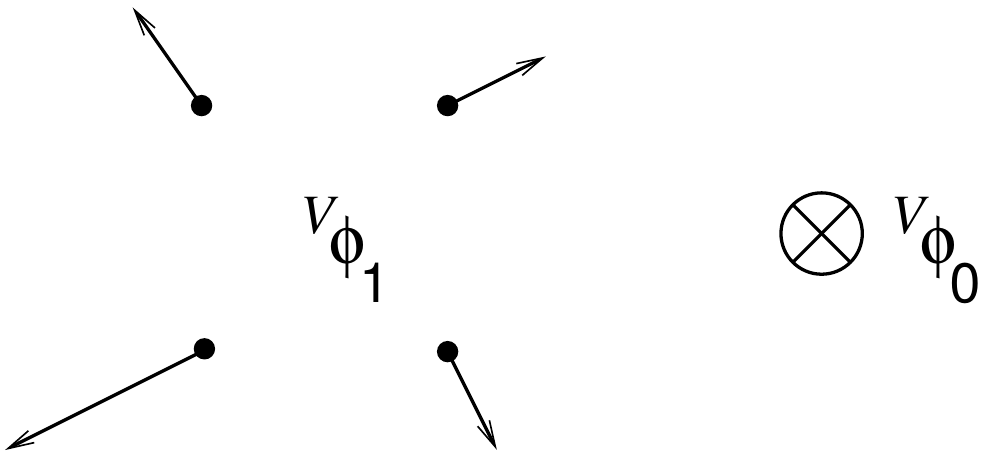}
\caption{The topological argument to detect vortices on a given plaquette: 
The transverse components of the first excited mode $\phi_1$ to the 
direction of the lowest mode $\phi_0$ (after both have been gauge transformed 
by $V$) are plotted for the four sites of a plaquette. The configuration 
shown here has a nonvanishing winding number, which implies that
the two eigenvectors are collinear in color space somewhere inside the plaquette.
}
\label{fig:Figure2}
\end{figure}

We detect the center vortices in LCG with the help of a topological 
argument: after having diagonalised $\phi_0$ by virtue of $V$, 
Eq.~(\ref{eq:eqn_gauge_abelian}), the question whether $\phi_0$ and $\phi_1$
are collinear amounts to $^V\!\phi_1$ being diagonal too, i.e. having zero
first and second component. We therefore inspect 
each plaquette, take all four corners and consider the projections of 
$^V\!\phi_1$ taken in these points onto the $(\sig_1,\sig_2)$-plane, 
see Fig. \ref{fig:Figure2}.

By assuming continuity\footnote{The continuity assumption underlies all 
attempts to measure topological objects on lattices. For semiclassical 
objects it is certainly justified.} of the field $^V\!\phi_1$ (more 
precisely, its $(\sig_1,\sig_2)$-projection) between the lattice sites 
of this plaquette,
we can easily assign a winding number to it. By normalization of the 
two-dimensional arrows this is actually a discretization of a mapping 
from a circle in 
coordinate space to a circle in color space. In the continuum this 
could give rise to any integer winding number, while with four 
discretization points the winding number can only take values 
$\{-1,0,1\}$. This winding number can easily be computed by adding 
the angles between the two-dimensional vectors on neighbouring sites.

A nontrivial winding number 
around the plaquette implies that the $(\sig_1,\sig_2)$-components 
of $^V\!\phi_1$ have a 
zero point inside the plaquette, which in turn means 
that the two eigenvectors are there collinear in color space. In this case
we can declare the midpoint of that plaquette 
belonging to the vortex surface.
The vortex surface is a two-dimensional closed surface formed by the plaquettes
of the dual lattice. 
The plaquettes of the dual lattice are orthogonal to and shifted by $a/2$ in all 
directions relative to the plaquettes of the original lattice.

At face value the above procedure is plagued by points where the lowest
eigenvector $\phi_0$ is close to the negative $\sig_3$-direction. 
Such situations are inevitable when $\phi_0$ has a hedgehog behavior around one of its zeroes, i.e.\ for monopoles in the LAG.
Then the diagonalising gauge transformation $V$ changes
drastically in space.
The corresponding transformed first excited mode  $^V\!\phi_1$ may give artificial
winding numbers and thus unphysical vortices if we insist on the 
continuity assumption in this case.

Actually, to detect vortices, the lowest eigenvector can be fixed 
to {\it any} color direction \cite{deForcrand:2000pg}, i.e.\ to different 
directions on different plaquettes.
Using this we rotate $\phi_0$
plaquette by plaquette to the direction of
the average $\bar{\phi}_0$ over the four corners of the plaquette. This
gauge rotation is in most cases a small rotation.
Afterwards we inspect $\phi_1$'s color components 
perpendicular to the average direction (this can be done by inspecting $^V\!\phi_1$ in the $(\sig_1,\sig_2)$-plane 
after diagonalising the four-site averaged lowest eigenvector, the resulting
gauge transformation now changes only mildly throughout the four sites of the 
plaquette).

Note that the winding number changes sign under $\phi_0\to -\phi_0$, 
but not under $\phi_1 \to -\phi_1$ 
(both changes of sign do not change the fact that these fields are eigenmodes 
of the Laplacian). 
Hence the global signs of $\phi_0$, $\phi_1$ and also 
the signs of the winding numbers are ambiguous. 

\section{Vortices in individual calorons}
\label{sect:IndividualCalorons}

\begin{figure}
 \includegraphics[height=0.9\linewidth]{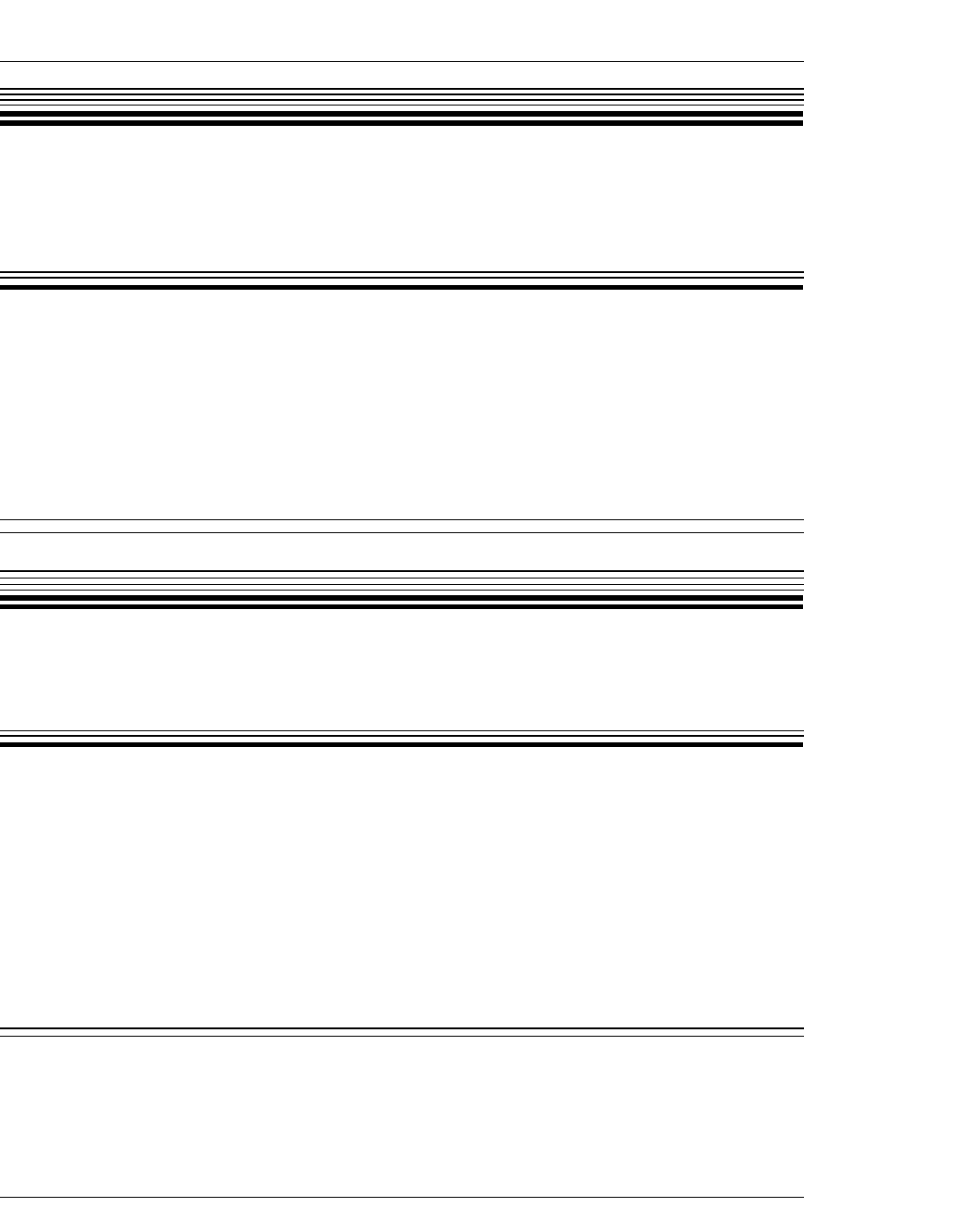}
 \includegraphics[height=0.9\linewidth]{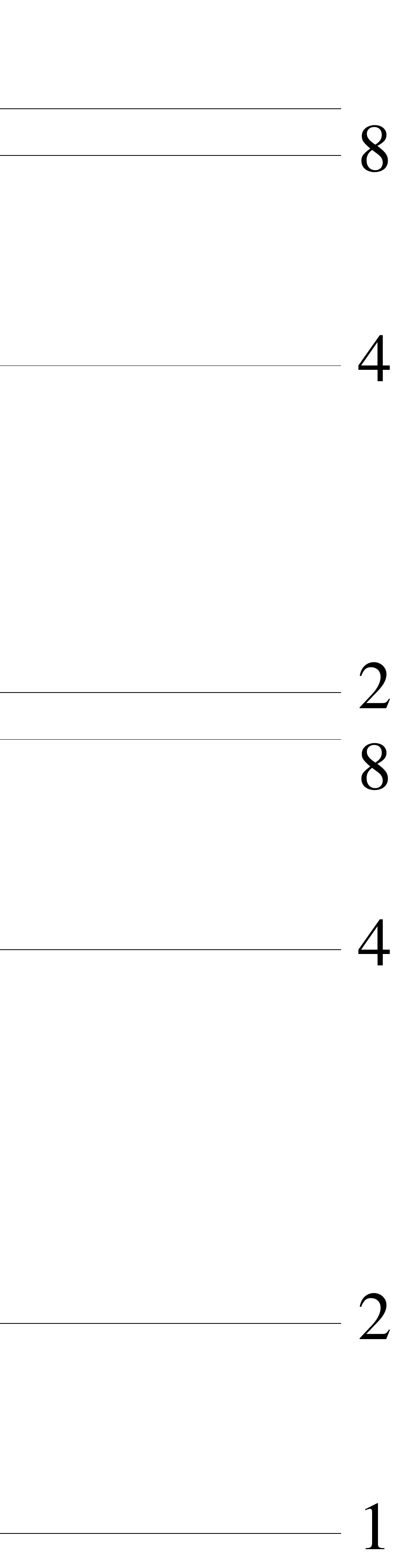}\\
\vspace{1cm}
\includegraphics[height=0.9\linewidth]{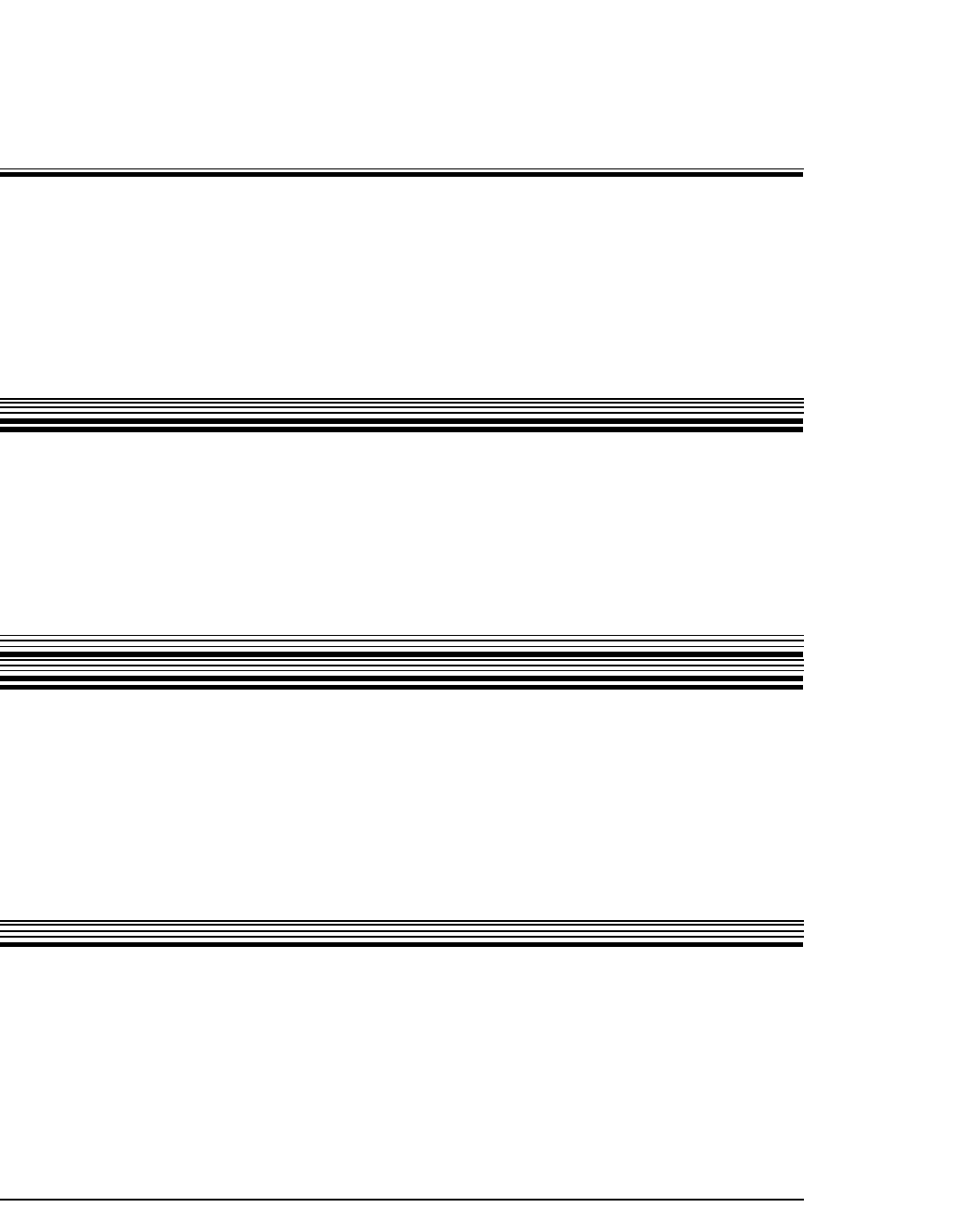}
 \includegraphics[height=0.9\linewidth]{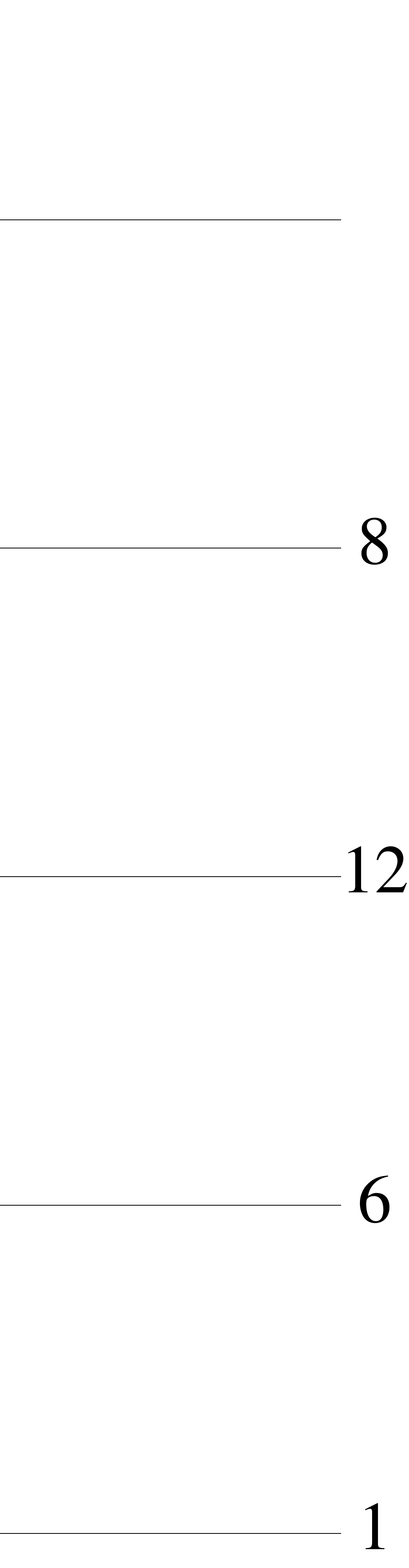}
\caption{The lowest 30 eigenvalues of the adjoint Laplacian operator for a caloron with  $\omega=0.12$ and $\rho=0.7\beta$ discretized on $8\times 48^2\times 80$ (top left) and $8\times 64^2\times 64$ (bottom left). Two-fold degeneracies are plotted as bold lines (and some eigenvalues have been slightly shifted to be distinguishable at this resolution).
For comparison we plotted in the right panels the free spectra on the same lattices marking their degeneracies by numbers. The lowest singlets on the rhs.\ 
always belong to the eigenvalue $\lambda=0$.  }
\label{fig:Figure3}
\end{figure}

The following results are obtained for single calorons discretized 
on space-time lattices with $N_0=8$ (meaning that our lattice spacing 
is $a=\beta/8$) and $N_1=N_2=48,\,N_3=80$ 
or $N_1=N_2=N_3=64$ points.

For the LCG vortices we have to take an ambiguity into account, 
namely the dependence of the adjoint Laplacian spectrum on the lattice 
discretization, in particular the ratio $N_3/N_{1,2}$. 
From experience we can summarize 
that the lowest adjoint eigenmode $\phi_0$ is rather independent 
of that 
``aspect ratio''.  
The first excited mode $\phi_1$ depends on it in the 
following way, cf. Fig.~\ref{fig:Figure3}: for large $N_3/N_{1,2}$ the 
first excited mode $\phi_1$ is a singlet, whereas for intermediate and 
small $N_3/N_{1,2}$ it is a doublet.

This ambiguity reflects the fact that we are forcing states of a continuous 
spectrum into a finite volume, which -- like waves in a potential well -- 
are then sensitive to the periodic boundary conditions\footnote{
A similar effect has been observed in Fig.~1 of \cite{Bruckmann:2000ay},
where the adjoint modes in the background of an instanton over the four-sphere
have been shown to depend on the radius of the sphere.}. 
Localised bound 
states, on the contary, 
should not depend much on the discretization.

Indeed, the absolute values and degeneracies of the eigenvalues can be 
understood by mimicking the caloron with constant links,
\begin{equation}
 U_0=\exp(2\pi i \o\sigma_3/N_0)\,,\quad U_i=1_2\,,
\end{equation}
that reproduce the holonomy 
(and have zero action).
For Laplacian modes in the fundamental representation this 
approximation was shown to be useful 
in \cite{Bruckmann:2005hy}.

In this free-field configuration the eigenmodes are waves proportional 
to $\prod_\mu\exp(2\pi i n_\mu x_\mu/N_\mu a)$ with integer $n_\mu$.
At nontrivial holonomies and on our lattices with $N_0\ll N_{1,2}\leq N_3$ one can easily convince oneself, that the lowest part of the spectrum is formed by modes in the third color direction, $\phi\sim\sigma_3$, which do not depend on $x_0$, $n_0=0$. The eigenvalues are then given by trigonometric functions of $2\pi n_i/N_i$, which for large $N_i$ can be well approximated by
\begin{equation}
 \lambda\simeq \frac{1}{a^2}\sum_i\left(2\pi\frac{n_i}{N_i}\right)^2\quad(\mbox{lowest } \lambda) \, .
\end{equation}
In other words, a wave in the $i$th direction contributes $n_i^2$ ``quanta'' 
of $(2\pi/N_i)^2$ to the eigenvalue.
The lowest eigenvalue in this approximation is always zero. This fits our numerical findings quite well, see Fig.\ \ref{fig:Figure3}.

In the asymmetric case, $N_3=80,\, N_{1,2}=48$ obviously the ``cheapest excitation''
is a wave along the $x_3$-axis (connecting the dyons), i.e.\ $n_3=\pm 1$. 
This gives a doublet, which in the presence of the caloron
is split into two lines, see Fig.\ \ref{fig:Figure3} top, the first excited mode 
is thus a singlet (the next modes are those with nontrivial $n_1=\pm 1$ or 
$n_2=\pm 1$ forming an approximate quartet and so on). 

In the symmetric case, $N_i=64$, on the other hand, excitations along all $x_i$ give equal energy contribution. For the excited modes this gives a sextet, which is again split by the caloron, see Fig.\ \ref{fig:Figure3} bottom. It turns out that 
the first excited mode remains two-fold degenerate. 
The eigenmodes are close to combinations of waves with nontrivial $n_1=\pm1 $ 
and with nontrivial $n_2=\pm 1$, reflecting the calorons' 
axial symmetry around the $x_3$-axis.

This finally explains the different spectra and different shape of the eigenmodes on the different lattices.

\subsection{The lowest eigenvector and the LAG monopoles}
\label{subsect:LAG_Monopoles}

As it turns out, away from the dyons the lowest mode $\phi_0$ becomes 
diagonal\footnote{The third direction in color space is distinguished 
by our (gauge) choice of the holonomy, Eqn. (\ref{eq:eqn_holonomy_diagonal}).} 
and constant, for normalisability reasons it is then approximately 
$(0,0,1/\sqrt{{\rm Vol}})^T$ with ${\rm Vol}=N_0N_1N_2N_3$. 

\begin{figure}[b]
 \includegraphics[width=0.9\linewidth]{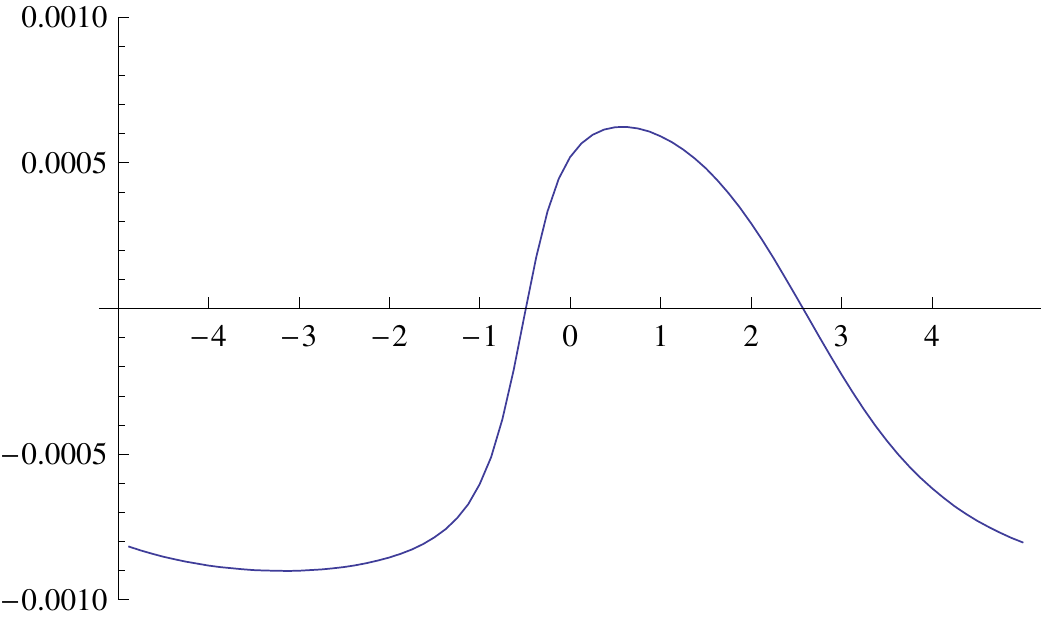}\\
 \qquad
 \includegraphics[width=0.85\linewidth]{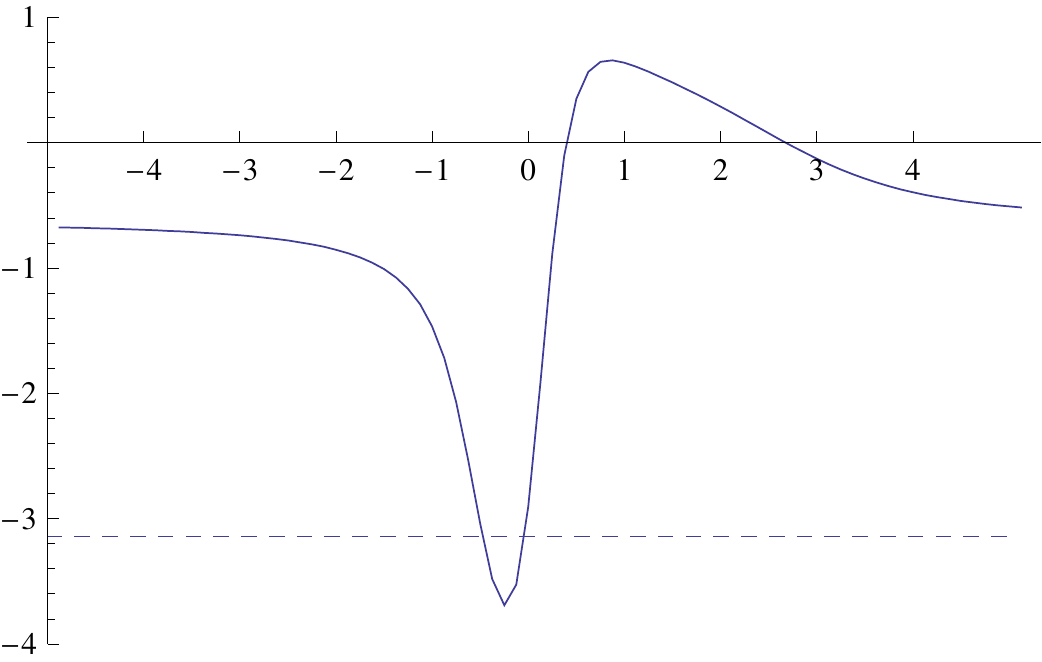}
\caption{Top: The third component of the lowest mode, $\phi_0^{a=3}$, 
along the $x_3$-axis (in units of $\beta$ at $x_0=\beta/2$) for a caloron 
with intermediate 
holonomy $\omega=0.1$ and size $\rho=1.0$ 
discretized on a $8\times 48^2\times 80$ lattice. 
The dyons have $x_3$-locations $-0.63\beta$ and $2.51\beta$. 
Note that for that lattice $1/\sqrt{{\rm Vol}}=0.00082$, a value
that is indeed taken on by the lowest mode far away from the dyons.
The other components $\phi_0^{a=1,2}$ are found to be of 
order $10^{-8}$ [not shown].
Bottom: the gauge field $A_0^{a=3}$ (in units of inverse $\beta$), which is 
related to the Higgs field $\Phi$ used to
explain the behaviour of the lowest mode around the dyons (see text). 
Note that $A_0^3$ takes
the value 
$-\pi/\beta$ near the twisting dyon.}
\label{fig:Figure4}
\end{figure}

Near each dyon core we find a zero of the third component of the lowest mode,
$\phi_0^{a=3}$, see Fig.~\ref{fig:Figure4} top panel. Together with the first 
and second component being very small on the whole $x_3$-axis, we expect 
zeroes in the modulus $|\phi_0|$ at the constituent dyons, which means 
that \emph{the dyons are LAG-monopoles}, cf.\ Fig.~3 in \cite{Alexandrou:1999iy} and Fig.~10 
in \cite{Bruckmann:2005hy}.

\begin{figure}[!b]
\includegraphics[width=\linewidth]{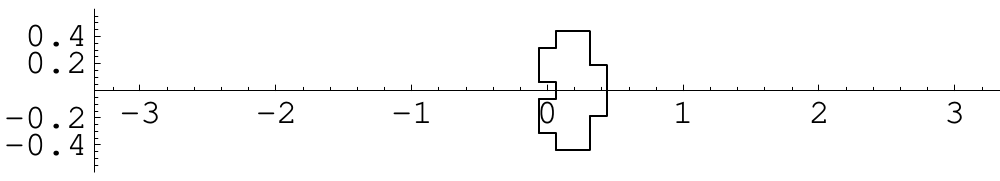}\\
\includegraphics[width=\linewidth]{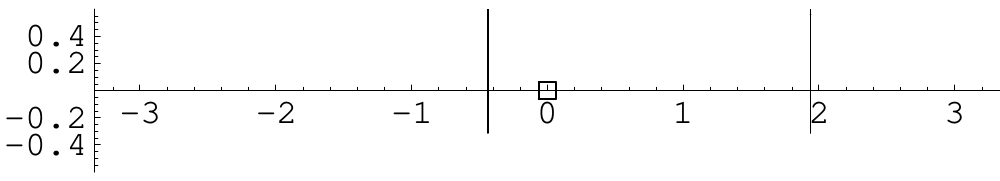}
\caption{Zeroes of the lowest adjoint mode, i.e.\ monopoles in Laplacian Abelian Gauge, in 
the $(x_0,x_3)$-plane (both in units of $\beta$, $x_3$ horizontally, at $x_1=x_2=0$) for
 calorons of holonomy $\o=0.1$ and sizes $\rho=0.5\beta$ (upper panel, $\z=(0,0,-0.16)$, $\zz=(0,0,0.63)$) and $\rho=0.9\beta$ (lower panel, $\z=(0,0,-0.51)$, $\zz=(0,0,2.04)$). At the origin a closed monopole wordline of minimal size occurs, which we ascribe to the gauge singularity in the caloron gauge field.}
\label{fig:Figure5}
\end{figure}

\begin{figure}
 \includegraphics[width=0.8\linewidth]{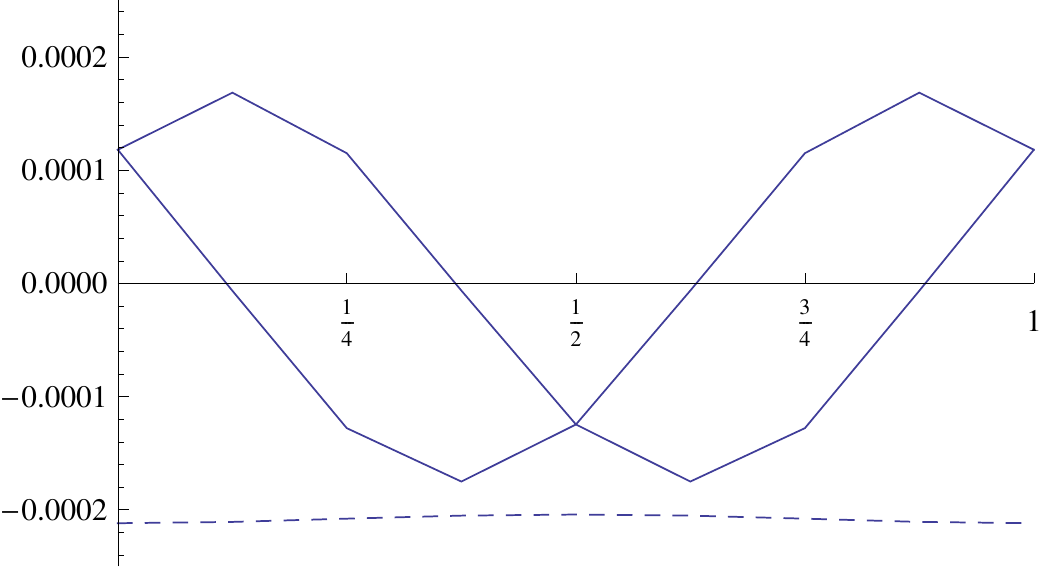}\\
 \includegraphics[width=0.8\linewidth]{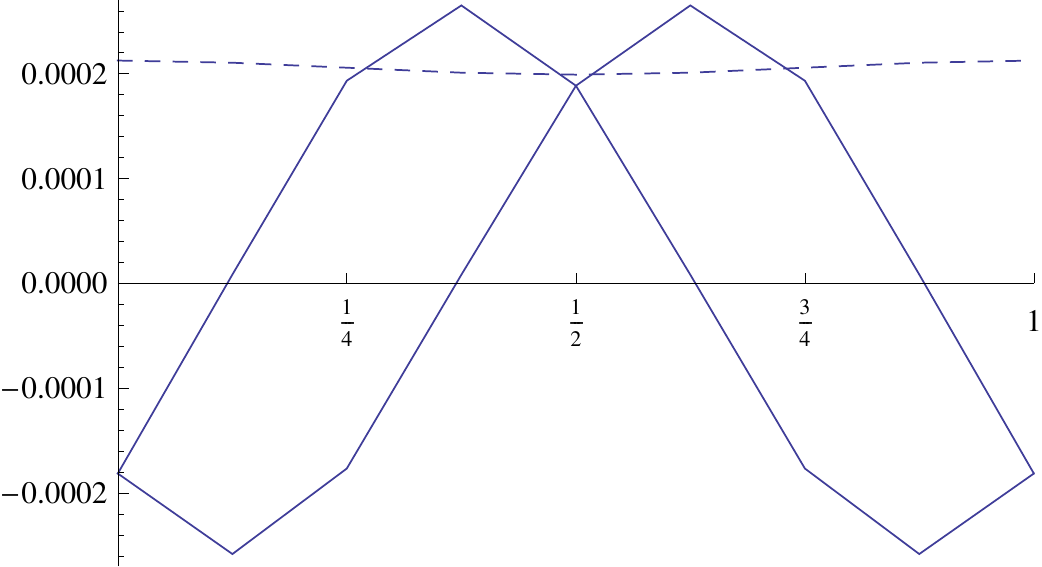}
\caption{The twist of the caloron gauge field reflected in the behaviour of 
the adjoint Laplacian modes. Shown are the 
individual 
color components of 
the lowest mode (top) and the first excited mode (bottom) as a function of 
time $x_0$ in the vicinity of the twisting dyon at $\z$.
The dashed curve depicts the (almost constant) third color component. 
The corresponding plots in the vicinity of the static dyon would simply show
static lines.}
\label{fig:Figure6}
\end{figure}

Such zeroes can be unambiguously detected by a winding number on lattice 
cubes similar to that of Sect.~\ref{sect:CenterVortices}. 
As a result we find almost static LAG-monopole worldlines for large calorons 
at the locations of their dyons, while monopole loops around 
the caloron center of mass are seen for small calorons (with $\rho\lesssim 0.5$, where the action density is 
strongly time-dependent as well), see Fig. \ref{fig:Figure5}.
Note that these locations are part of the LCG vortex surface by definition. 
Similar monopole worldlines have been obtained in the 
MAG \cite{Brower:1998ep,Ilgenfritz:2004zz}.
Adjoint fermionic zero modes, on the other hand, detect the constituent dyons by maxima \cite{GarciaPerez:2009mg}.

The lowest mode also reflects the twist of the caloron: the first and second 
component of $\phi_0$ near 
the dyon core are either static or rotate once with time $x_0$ evolving 
from $0$ to $\beta$. 
Fig.~\ref{fig:Figure6} shows this for the lowest mode as well as for the
first excited mode. Our results are essentially equal to Fig. 9 
of \cite{Bruckmann:2005hy}, just with a resolution of $N_0=8$ 
(instead of $N_0=4$) more clearly revealing the
sine- and cosine-like behaviours.\\

In order to understand the behaviour found for the lowest adjoint mode
$\phi_0$, we propose to compare it to the 
Higgs field $\Phi$ discussed in Sect.~\ref{subsect:Twist}.
For the static dyon one has from time-independence $D_0 \Phi=0$ and from the equation of motion 
$D_i(D_i \Phi)=D_i F_{i0}=0$. Therefore $\Phi$ of a single static dyon is a zero 
mode of the adjoint Laplacian $-\Delta=-D_\mu^2$. For the twisting dyon 
the same equations apply due to the transformation properties of $\Phi$ (under $\T$) and the latter is again a zero mode of the Laplacian.
These zero modes approach a constant (the vev) asymptotically, so they are
normalizable like a plane wave.

Around each dyon core, the lowest adjoint mode $\phi_0$ behaves similar 
to 
$\Phi$ of that dyon: it vanishes at the dyon core, becomes constant and dominated 
by the third component away from the dyons, it reveals the Taubes twist (around the twisting dyon) 
and is in the same gauge as $\Phi$. Since the latter 
are zero modes of $-\Delta$ in the background of isolated dyons,
a combination of them is a natural candidate to be the 
lowest mode of that (non-negative) operator in the caloron background.

The lowest adjoint mode $\phi_0$ for calorons with well-separated dyons 
is therefore best described in the following way, 
cf.\ Fig.~\ref{fig:Figure4}: Around the static dyon at $\zz$ one has 
$\phi_0 \sim \Phi=A_0$, where the proportionality constant of course disappears 
from the eigenvalue equation (\ref{eq:eqn_eigen}), but is approximately given 
by the normalization: 
$|\phi_0| \to (0,0,1/\sqrt{{\rm Vol}})^T$. 
Around the twisting dyon at $\z$, one has to compensate for the inhomogeneous 
term $\phi_0 \sim \Phi=A_0 + \sig_3 \, (\pi/\beta)$ 
(cf.\ eqn. (\ref{eq:eqn_relation_A_3})). The proportionality constant there 
turns out to be negative, such that the lowest mode is able to interpolate 
between these shapes with a rather mild variation throughout the remaining 
space, see Fig.~\ref{fig:Figure4} upper panel.

\subsection{
Dyon charge induced vortex}
\label{subsect:SpaceTimeVortices}

\begin{figure}
\includegraphics[width=0.9\linewidth]{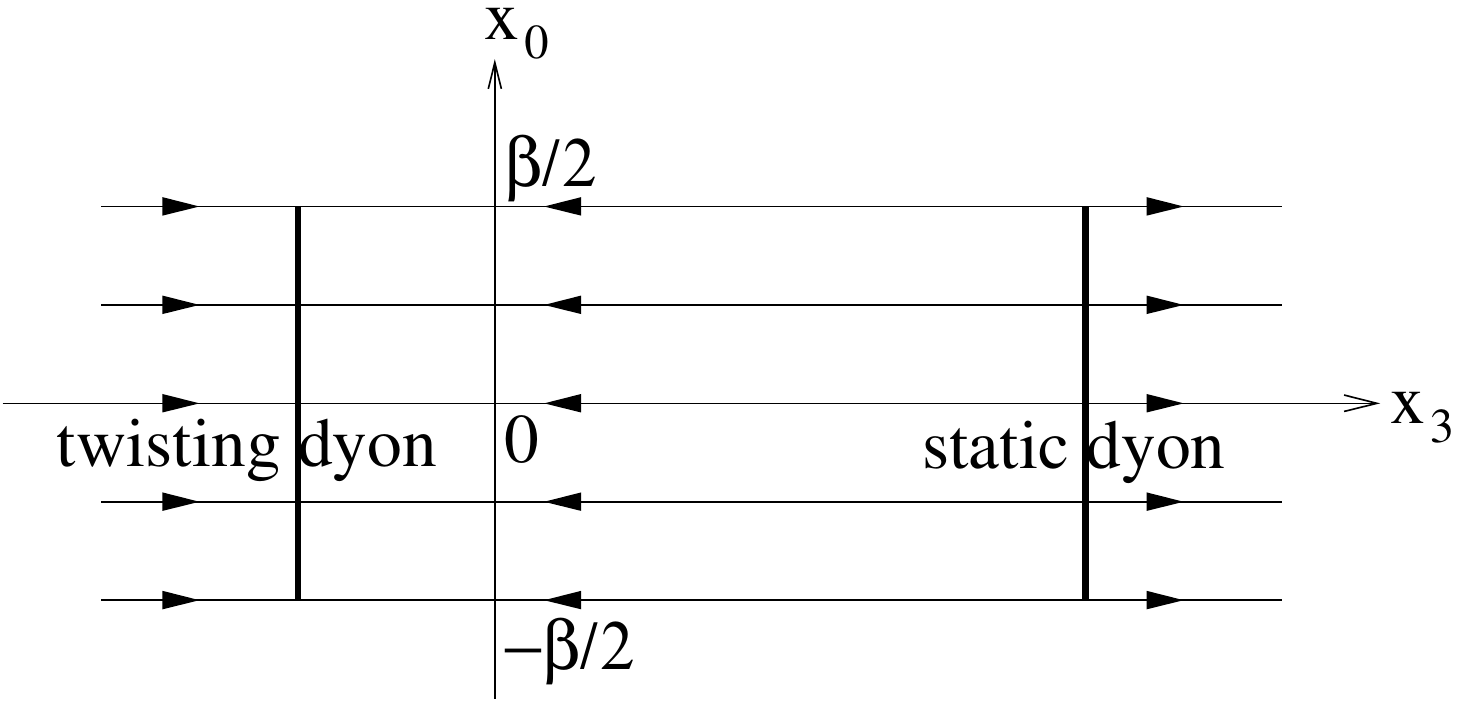}\\
\vspace{0.5cm}
\includegraphics[width=0.85\linewidth]{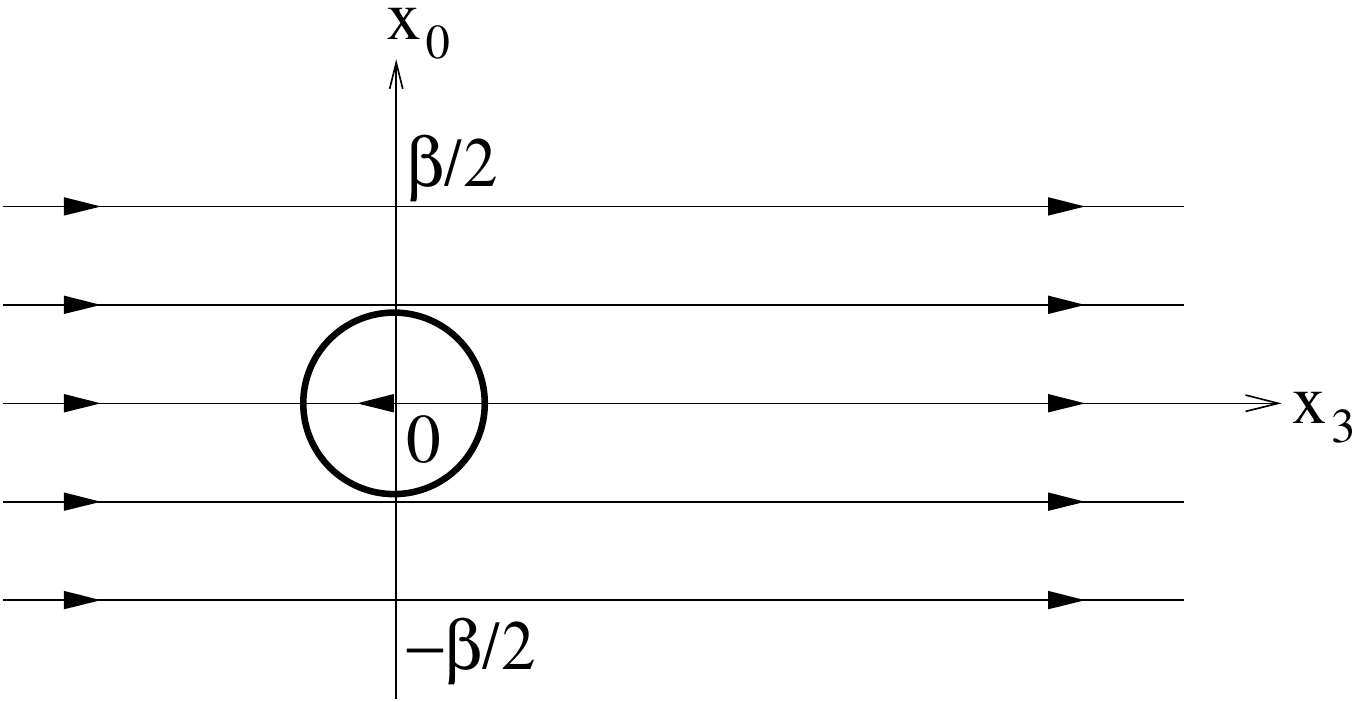}
\caption{The dyon charge induced part of the vortex 
in case the first excited mode is a singlet:
for a large caloron (top) and for a small caloron (bottom), 
shown schematically in the plane $x_1=x_2=0$.}
\label{fig:Figure7}
\end{figure}

In the following we present and discuss one 
part of the calorons' vortex that is caused by 
the magnetic charge of constituent dyons. 
Our findings are summarized schematically 
in Figs.~\ref{fig:Figure7} and \ref{fig:Figure9}.  

\begin{figure}[b]
\includegraphics[width=0.8\linewidth]{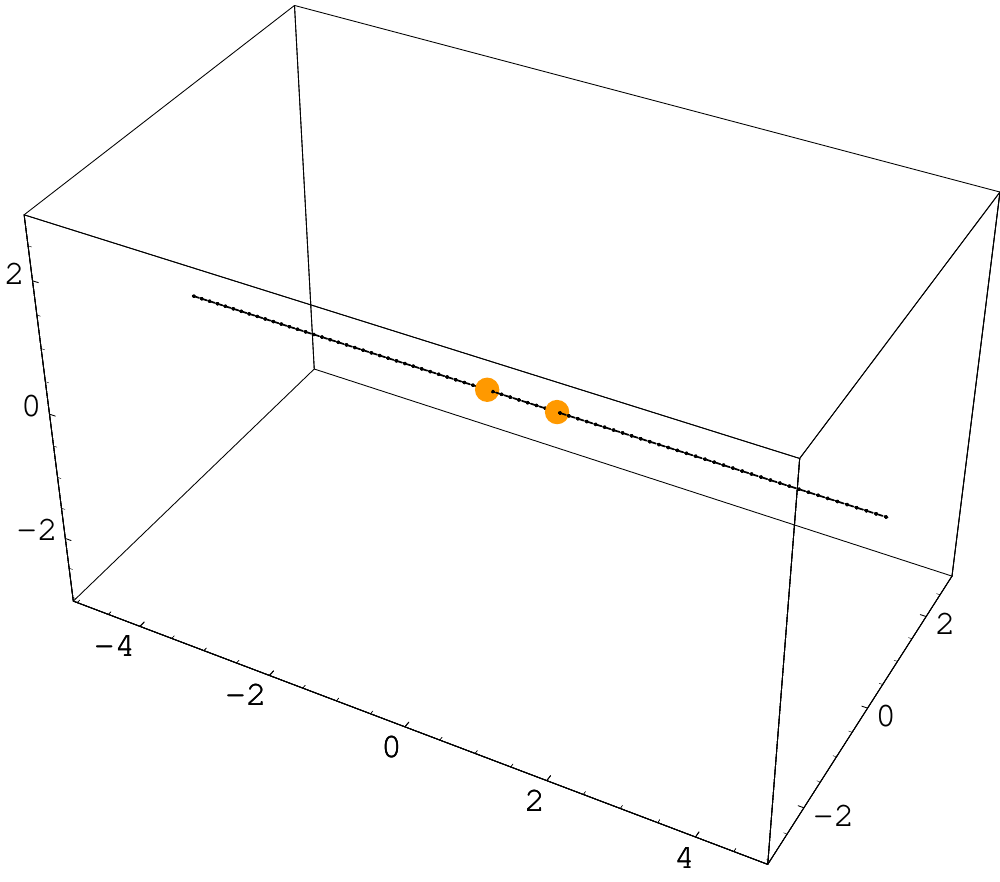}
\caption{The dyon charge induced part
of the vortex from the singlet first excited mode
as measured in a caloron with holonomy $\o=0.25$ and $\rho=0.6\beta$ in a
time slice. The outcome is identical to the $x_3$-axis and the same for all time slices.
The dots denote points on the vortex (lines along $x_0$) where the flux changes, 
i.e.\ the LAG-monopoles.}
\label{fig:Figure8}
\end{figure}

\begin{figure}[t]
\includegraphics[width=\linewidth]{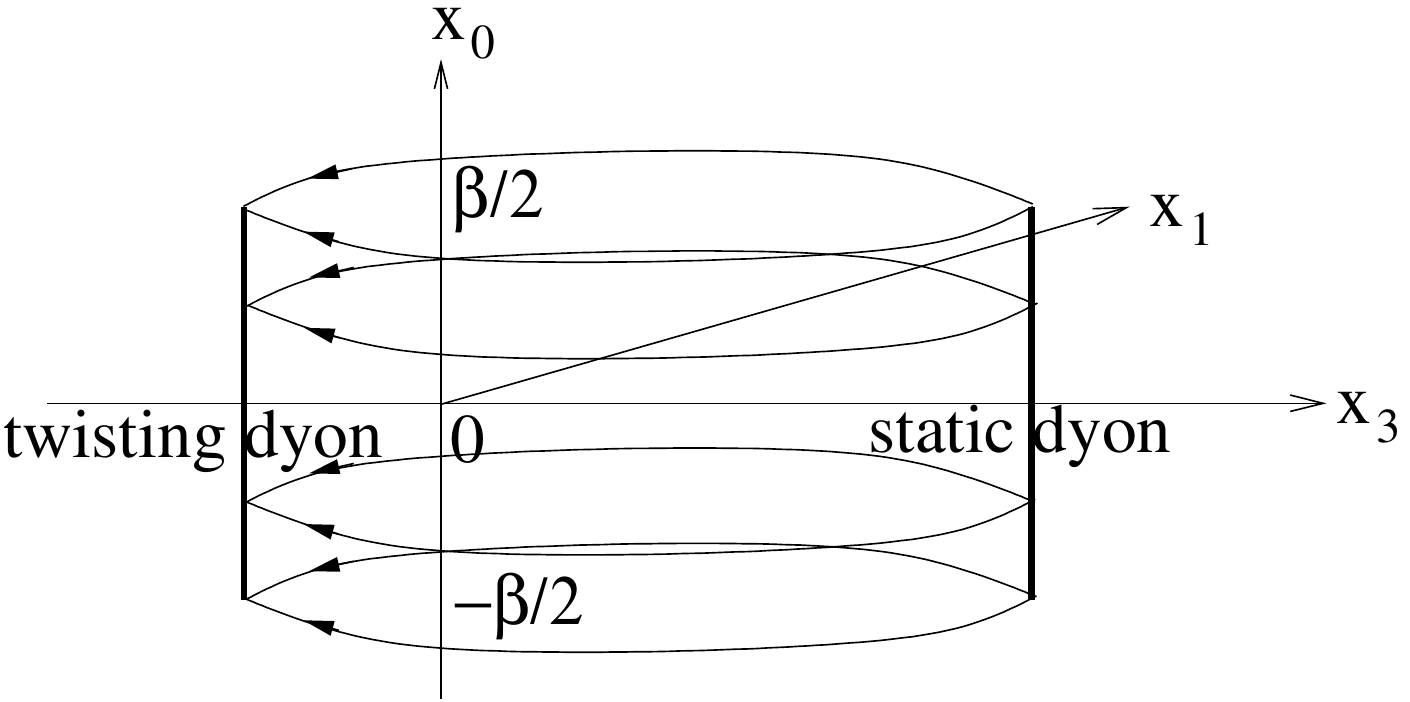}\\
\vspace{0.5cm}
\includegraphics[width=\linewidth]{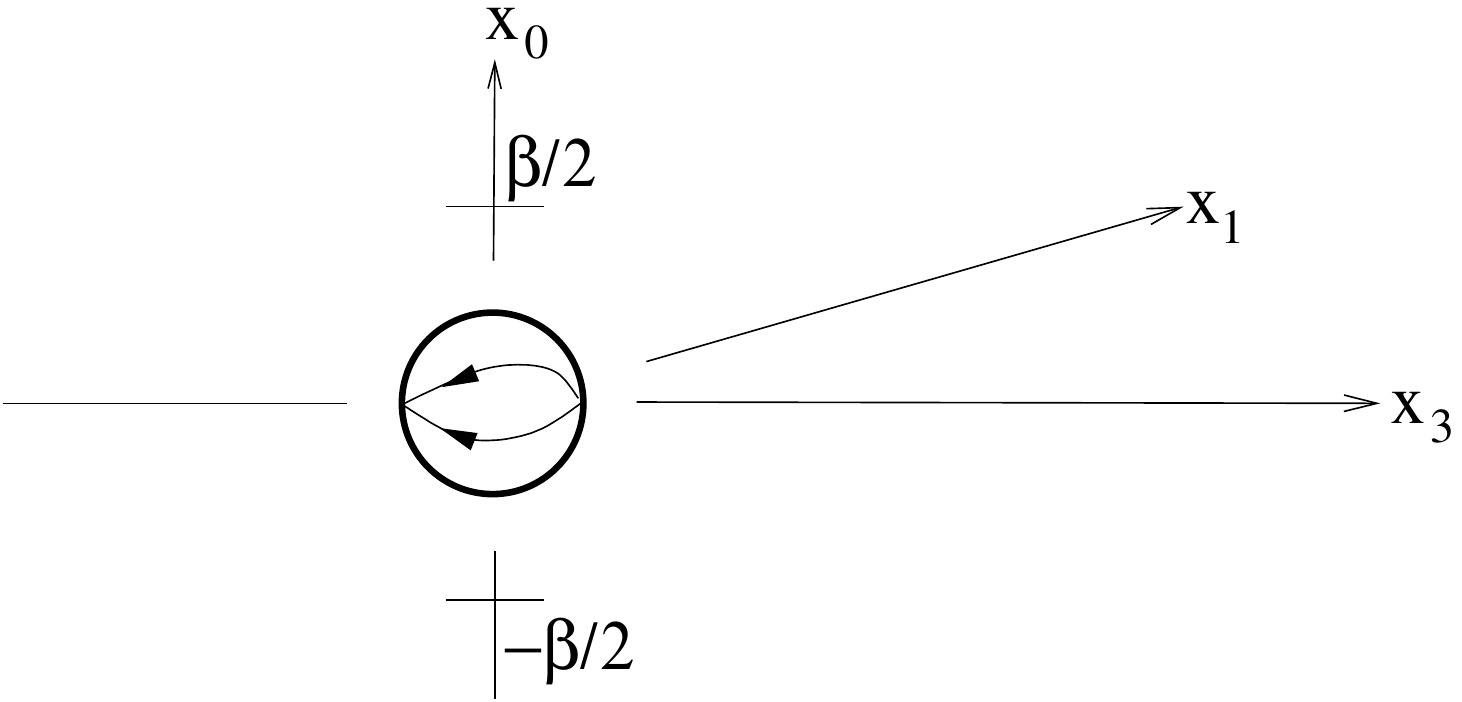}
\caption{The dyon charge induced part 
of the vortex from the doublet first excited mode 
for a large caloron (top) and for a small (bottom) caloron schematically at $x_2=0$.
}
\label{fig:Figure9}
\end{figure}

The ambiguity of the first excited mode $\phi_1$ of the adjoint Laplacian 
influences this part of the vortex most
such that we have to discuss the singlet and doublet cases separately. 
We find that for the 
singlet $\phi_1$, e.g.\ for $N_3/N_{1,2}=80/48$, the vortex consists of the 
whole $(x_0,x_3)$-plane at $x_1=x_2=0$ only, see Figs.~\ref{fig:Figure7} 
and \ref{fig:Figure8}. 
Hence this part of the vortex is space-time like. 
It includes the LAG-monopole worldlines, 
which are either two open (straight) lines or form one closed loop in 
that plane. In other words, the space-time vortex connects the dyons 
once through the center of mass of the caloron
and once through the periodic spatial boundary of the lattice.

The magnetic flux (measured through the winding number as described in
Sect.~\ref{sect:CenterVortices}) at every time slice points into the 
$\pm x_3$-direction. Its sign changes at the dyons as indicated by 
arrows\footnote{We have fixed the ambiguity in the winding number 
described in Sect.~\ref{sect:CenterVortices} by fixing the asymptotic
behaviour of the lowest mode.} in Fig.~\ref{fig:Figure7}.
The flux is always pointing towards the twisting dyon.

Independently of the flux one can investigate the alignment 
between the lowest and first excited mode. 
It changes from parallel\footnote{
In itself, calling $\phi_0$ and $\phi_1$ parallel is 
ambiguous as that changes when one of these eigenfunctions is multiplied 
by -1. The \emph{transition} from  parallel to antiparallel or vice versa,
however, is an unambiguous statement.}
to antiparallel near the static dyon, because the lowest mode $\phi_0$ 
vanishes (i.e.\ the dyon is a LAG-monopole) \cite{deForcrand:2000pg}.
In addition we find 
two other important facts  
not mentioned in \cite{deForcrand:2000pg}:
the alignment does not change at the twisting dyon since both modes $\phi_0$ 
and $\phi_1$ vanish there and it changes at some other locations outside 
of the calorons' dyons because $\phi_1$ has another zero there [not shown].\\

\begin{figure}[h]
\includegraphics[width=0.8\linewidth]{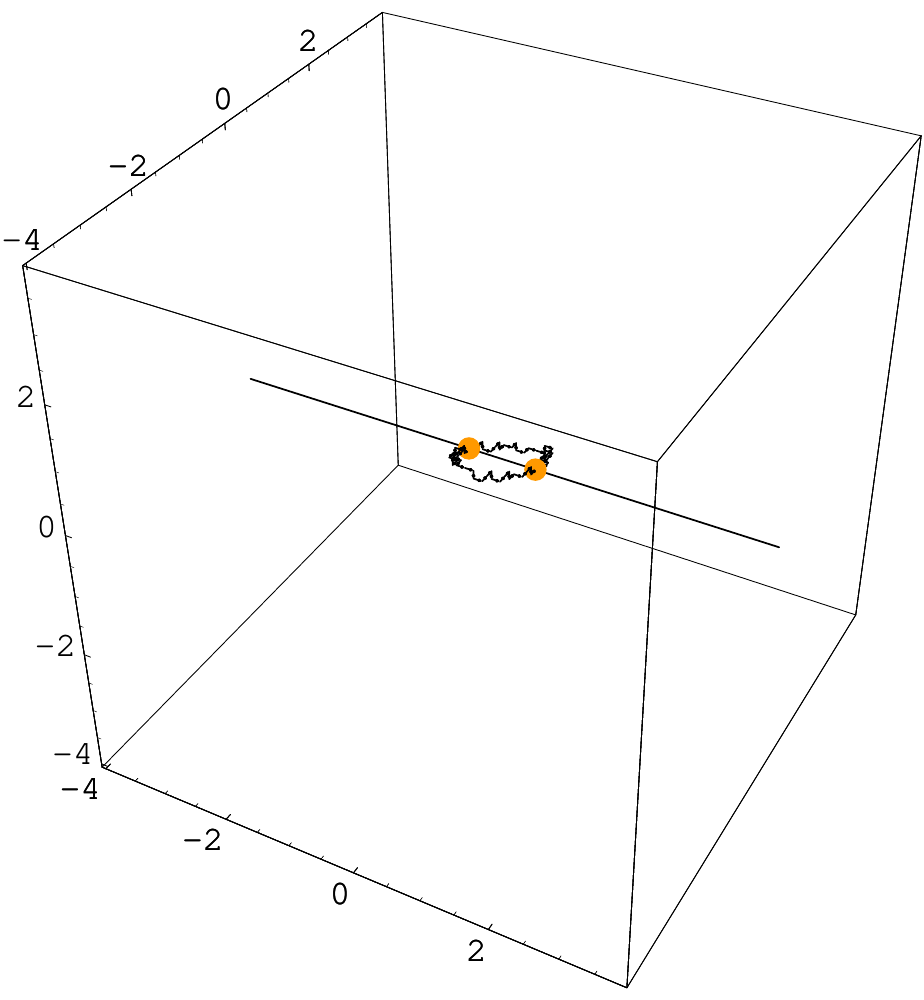}
\caption{The dyon charge induced part of the vortex from the doublet first 
excited state as measured 
in a caloron with holonomy $\o=0.25$
and $\rho=0.6\beta$ (same as in Fig.\ \protect\ref{fig:Figure8})
at a fixed time slice.
Like in Fig.~\protect\ref{fig:Figure8} the dots denote points on the vortex
where the flux changes.
The $x_3$-axis has been added to guide the eye, it is not part of the vortex surface here.
}
\label{fig:Figure10}
\end{figure}

For the doublet excited mode, i.e.\ at smaller $N_3/N_{1,2}=64/64$, 
the dyon charge induced vortex is slightly different: again it
connects the dyons, but now (for a fixed time) via two 
lines in the ``interior'' of the caloron, 
passing near the center of mass, see Figs.~\ref{fig:Figure9} and \ref{fig:Figure10}. 
These lines exist for all times for which 
the monopole worldline exists, that is \emph{for all times} if the caloron is 
large and \emph{for some subinterval of $x_0$} if the caloron is small 
(and the monopole worldline is a closed loop existing during the subinterval).

These two vortex surfaces spread away from the $x_3$-axis which connects 
the dyons. The axial symmetry around this axis is seemingly broken. However, 
using other linear combinations of the doublet in the role of the first 
excited mode 
(keeping the lowest one) in the procedure of center projection,
the vortex surface is rotated around the $x_3$-axis. The situation 
is very similar to the ``breaking'' of spherical symmetry in the hydrogen atom 
by choosing a state of particular quantum number $m$ out of a multiplet with 
fixed angular momentum $l$.

The magnetic flux flips at the dyons, just like in the case with singlet $\phi_1$.


Notice that these vortices are predominantly space-time like,
but have parts that are purely spatial, in particular for small calorons,
namely at 
minimal and maximal $x_0$ of the dyon charge induced vortex surface
(and at other locations in addition,
when the smooth continuum surface is approximated by plaquettes).

\begin{figure*}
\includegraphics[width=0.32\linewidth]{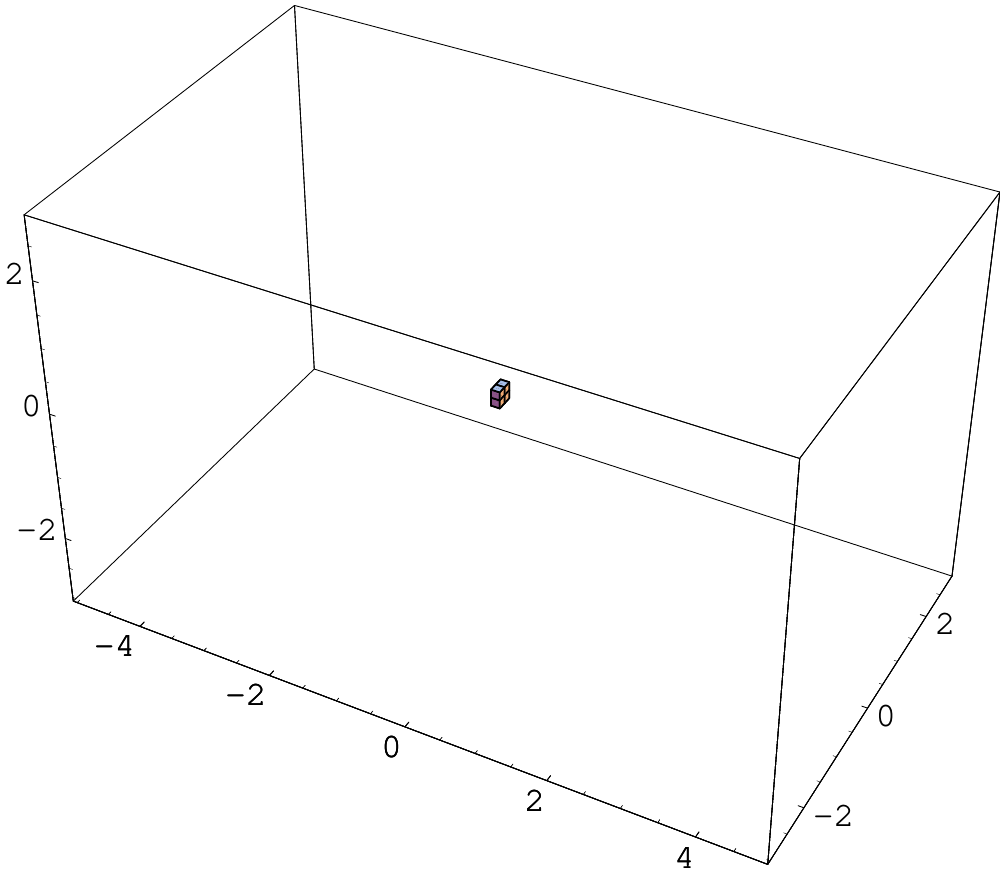}
\includegraphics[width=0.32\linewidth]{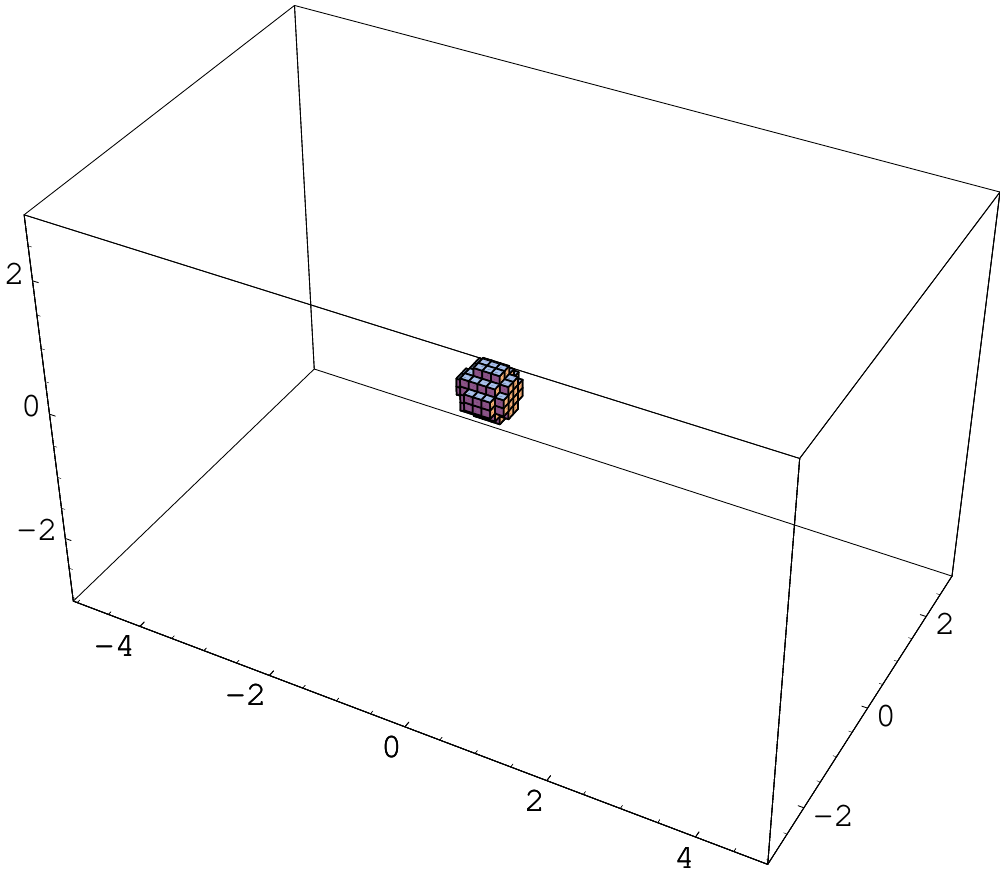}
\includegraphics[width=0.32\linewidth]{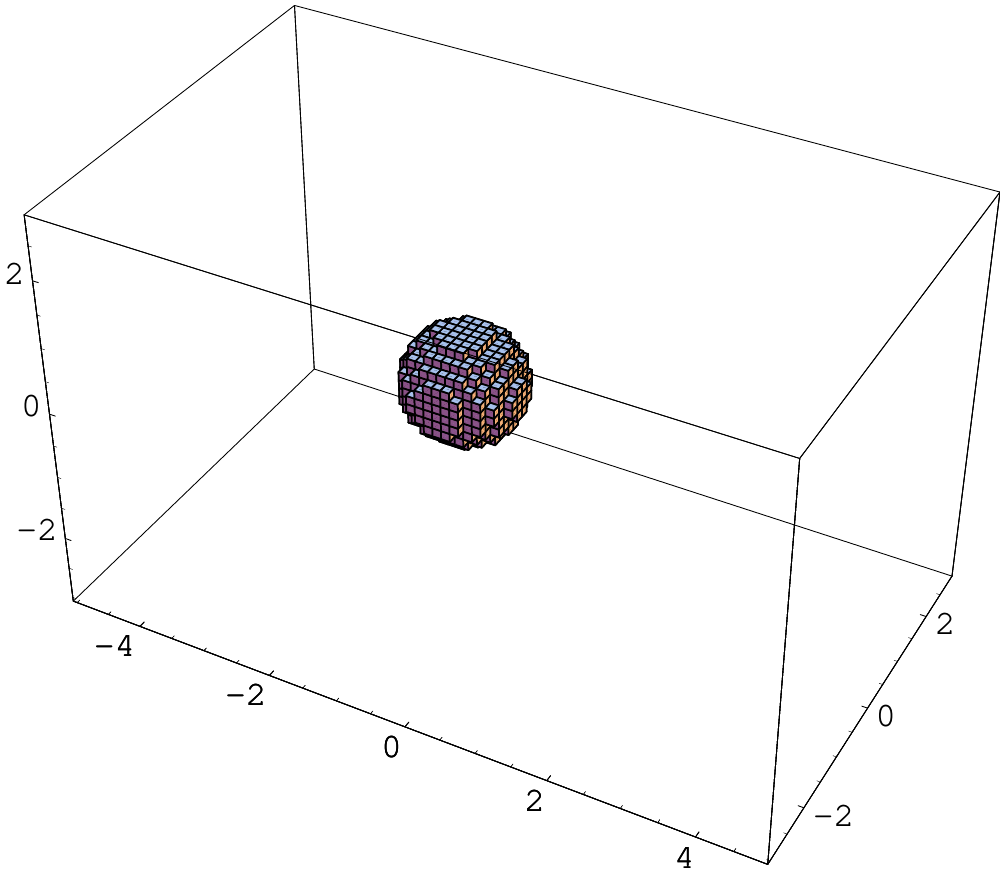}\\
\includegraphics[width=0.32\linewidth]{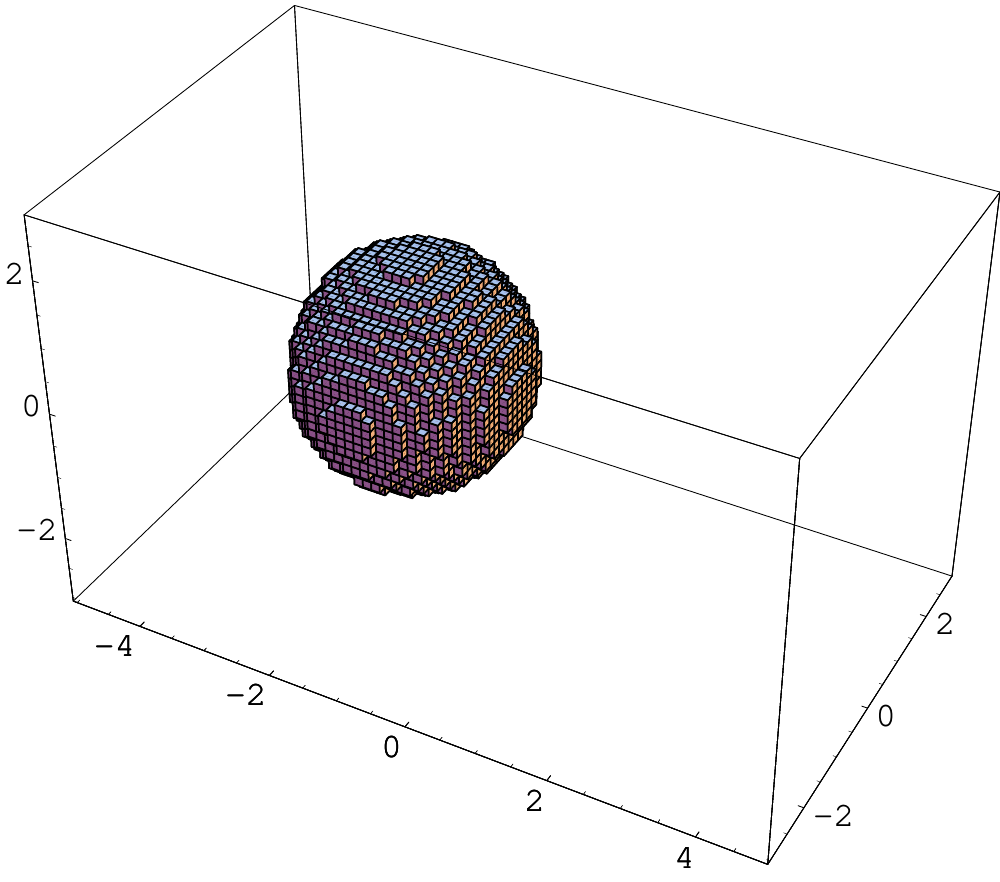}
\includegraphics[width=0.32\linewidth]{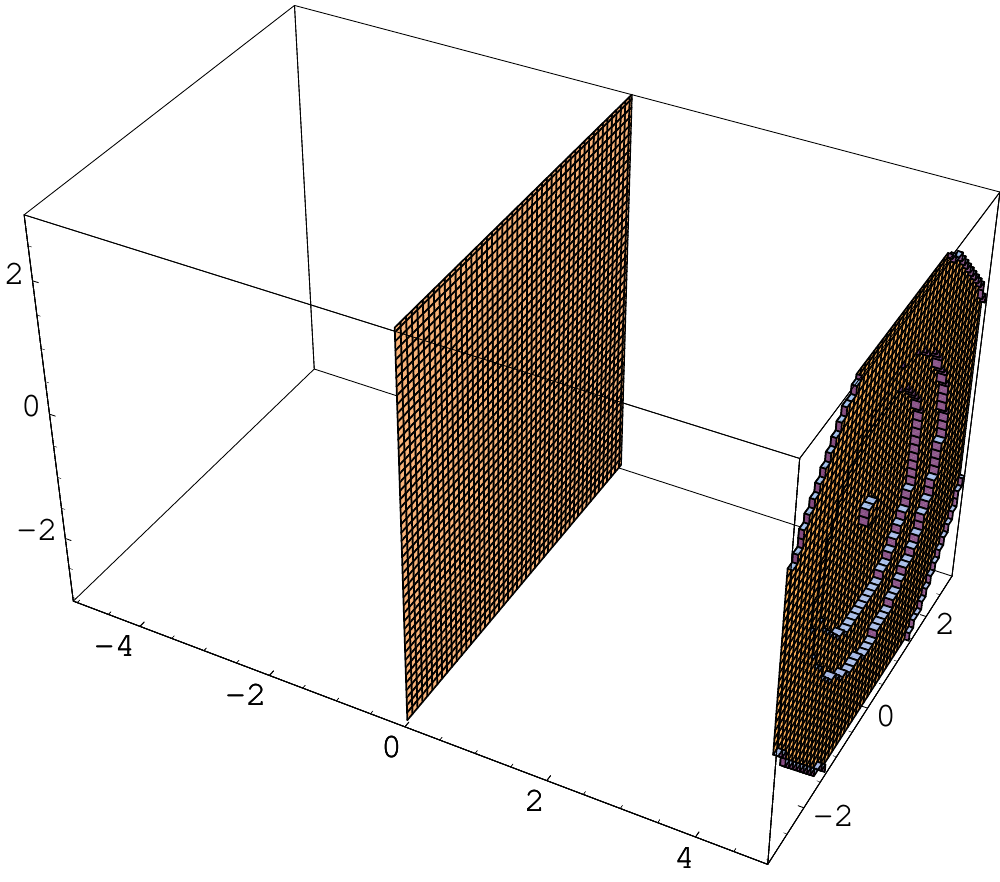}
\includegraphics[width=0.32\linewidth]{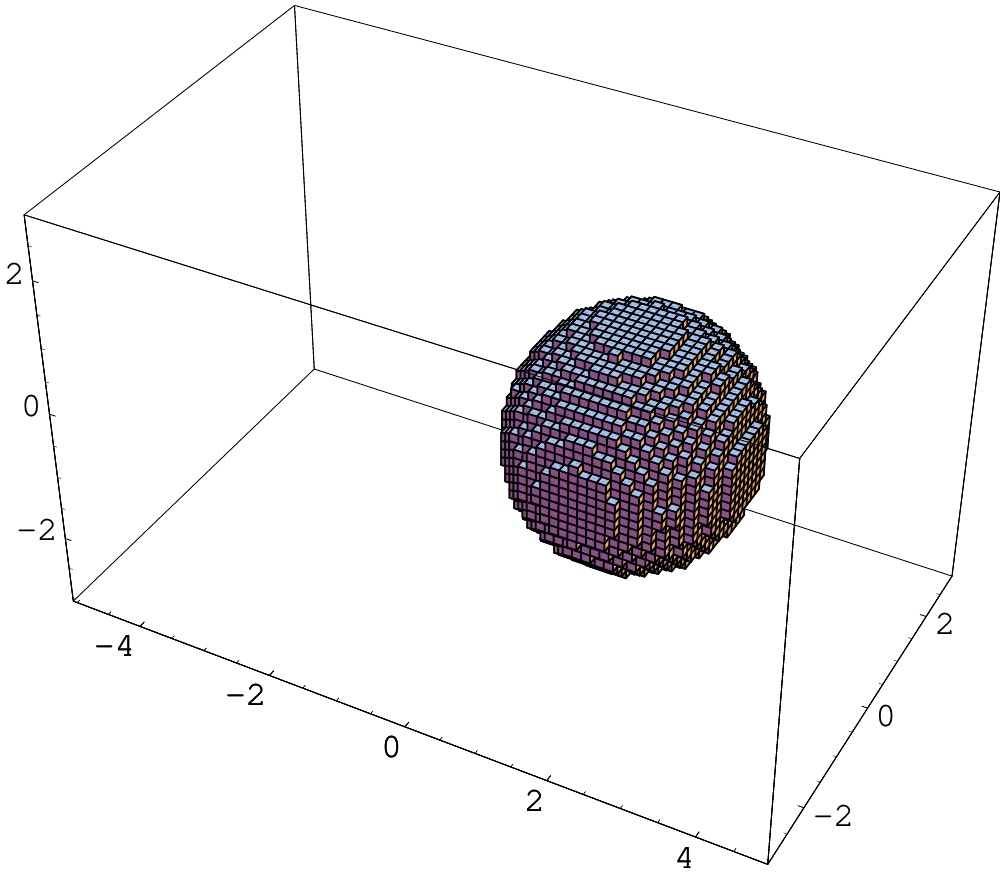}\\
\vspace{0.4cm}
\includegraphics[width=0.32\linewidth]{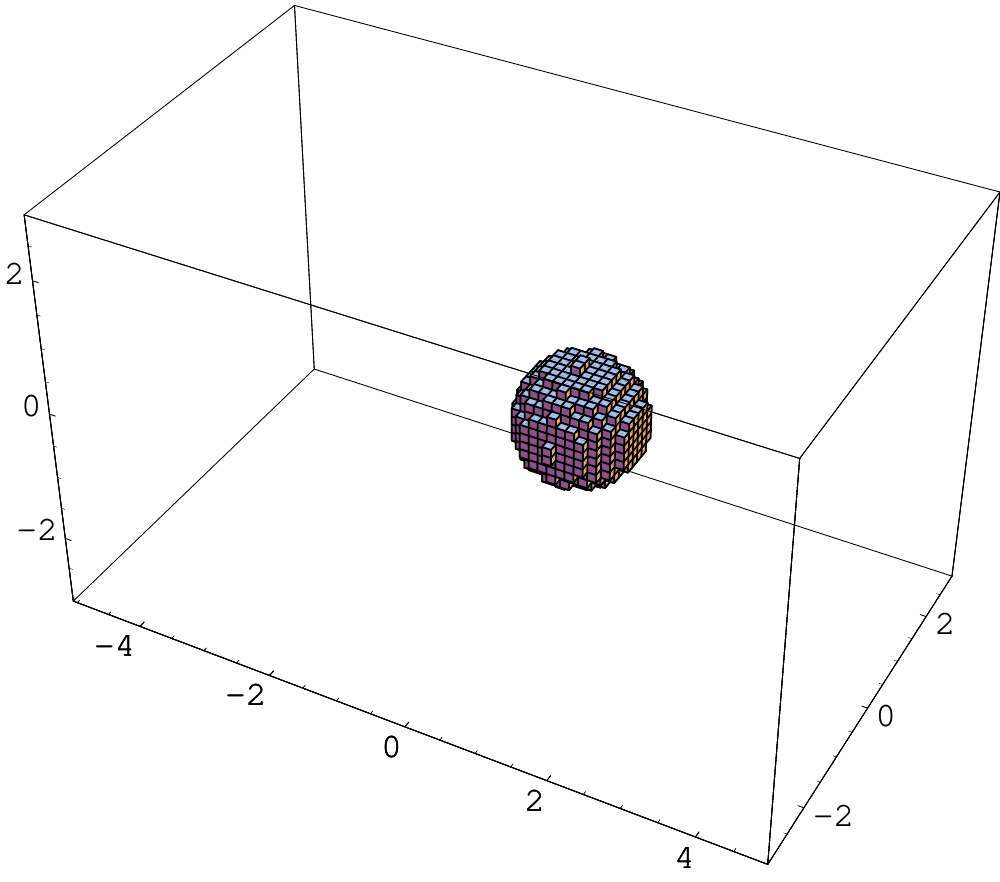}\hfill
\includegraphics[width=0.45\linewidth]{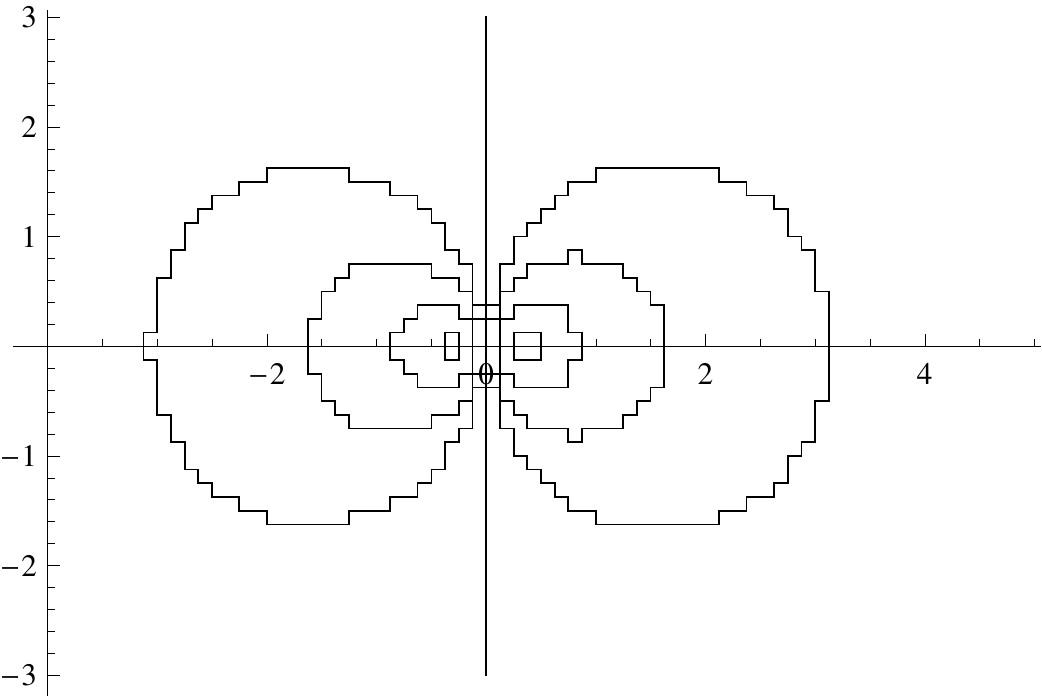}
\caption{Twist-induced part of the vortex (``bubble'') from singlet first excited modes
for calorons of size $\rho=0.6\beta$
and holonomies from left to right: $\o=0.1,0.12,0.16$ (upper row) $\o=0.2, 0.25,0.3$
(middle row) and $\o=0.34$ (lower row, left  panel). The plot in the lower right panel
summarises the results for $\o=0.1,\,0.12,\,0.16,\,0.2,\,0.25,\,0.3,\,0.34,
$ at $x_1=0$,
i.e. the bubbles are cut to circles. The plane near the boundary in the $\o=0.25$ picture
is an artifact caused by periodic boundary conditions.}
\label{fig:Figure11}
\end{figure*}

\begin{figure*}
\includegraphics[width=0.24\linewidth,]{LCG_CDLk8484880_Om012_Rho060_SpaceSpaceVS_TimeSlice3.pdf}
\includegraphics[width=0.24\linewidth,]{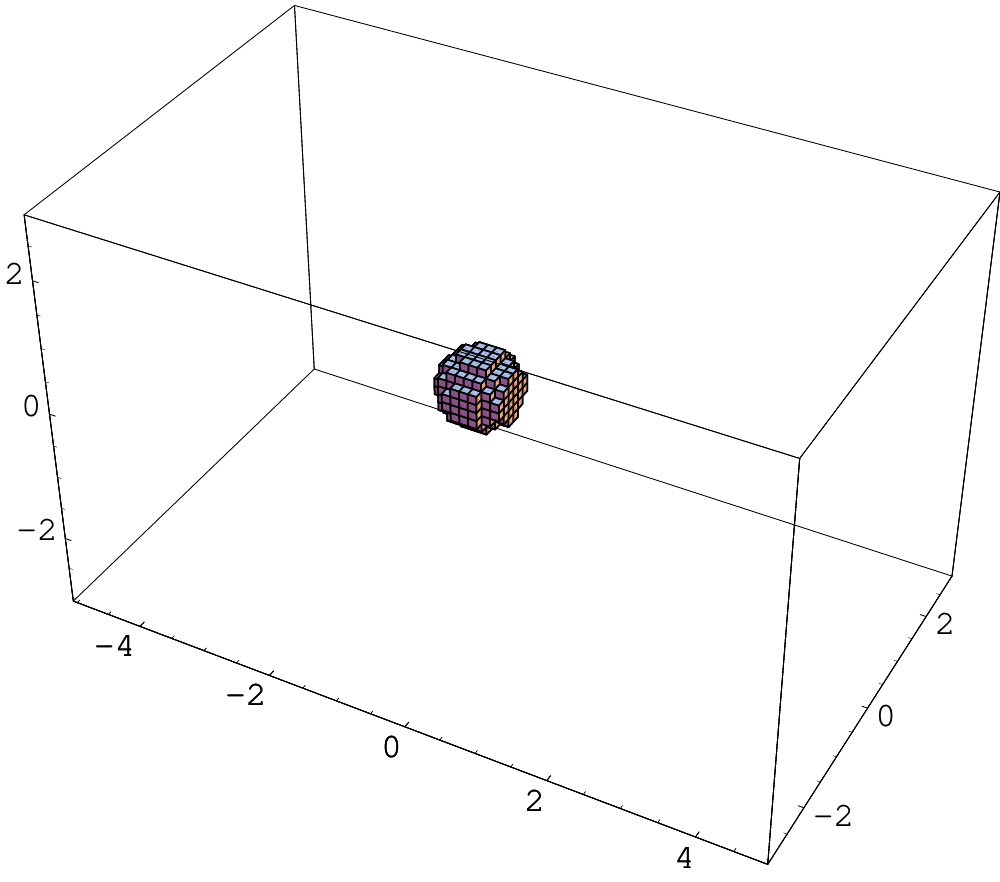}
\includegraphics[width=0.24\linewidth,]{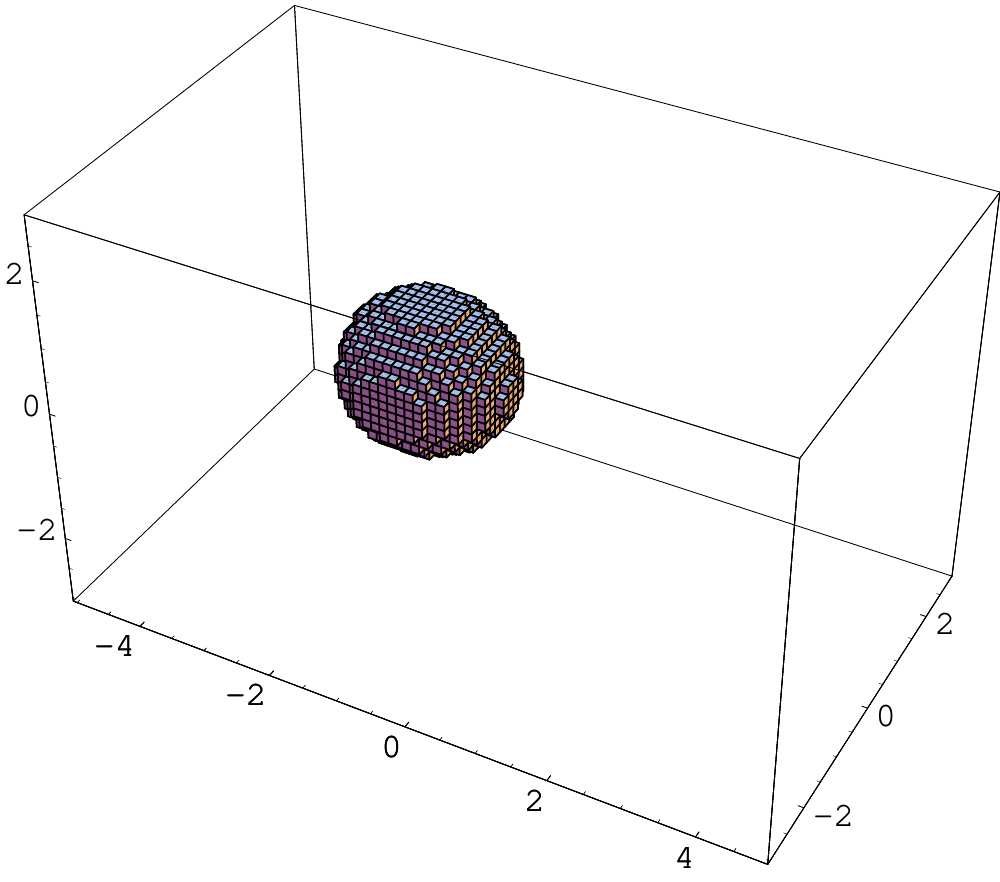}
\includegraphics[width=0.24\linewidth,]{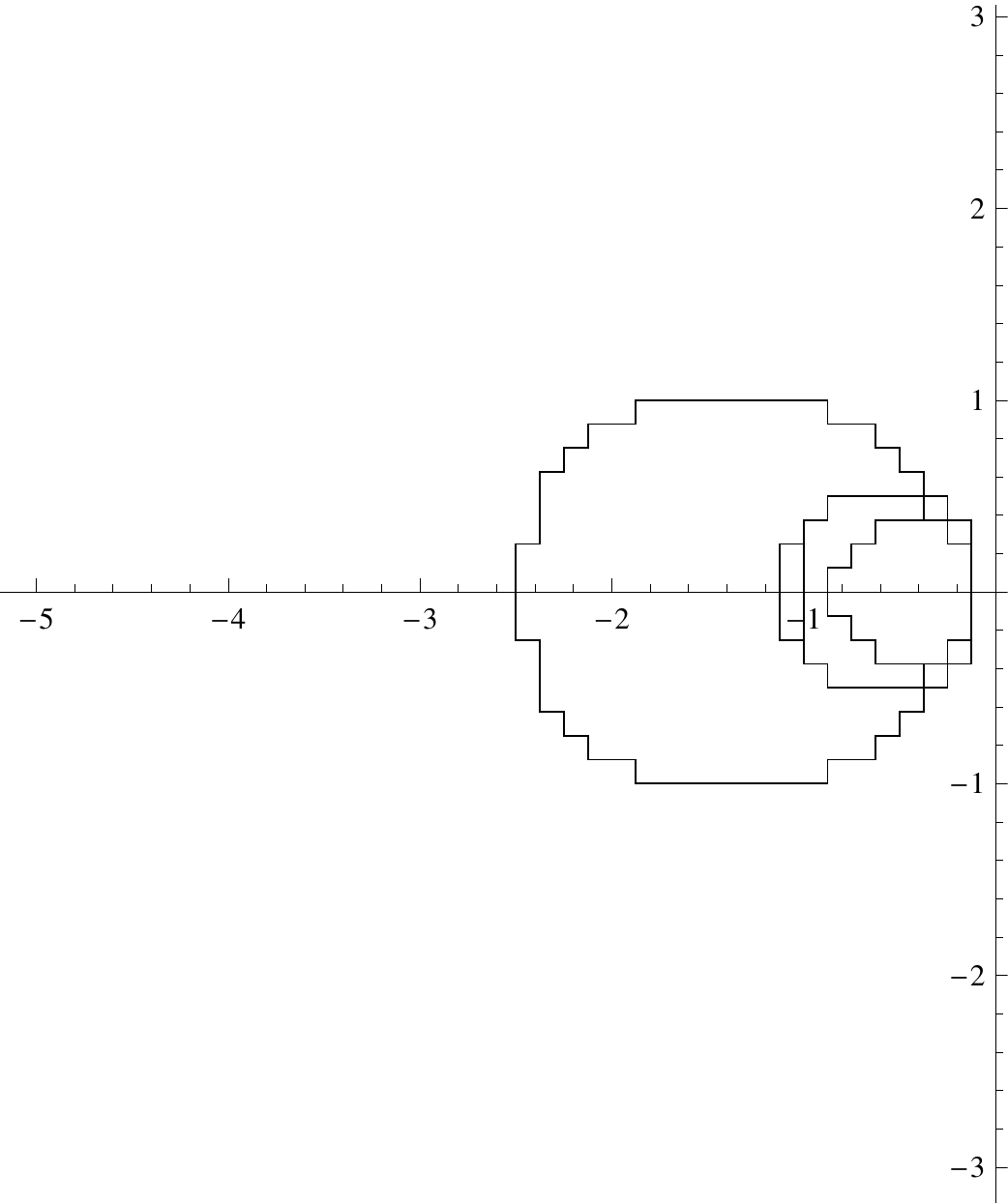}
\caption{Spatial part of the vortex (``bubble'') for calorons of fixed 
intermediate holonomy $\o=0.12$ and sizes from left to 
right: $\rho=0.6\beta,\,0.7\beta,\,0.9\beta$. The panel on the very right shows a 
summary of the bubbles for $\rho=0.6\beta,\,0.7\beta,\,0.8\beta,\,0.9\beta$ at $x_1=0$. 
That the bubble for $\rho=0.9\beta$ is much bigger than that 
for $\rho=0.7\beta,\,0.8\beta$
is probably a finite volume effect. 
For small sizes $\rho$ (and also in the limiting cases of holonomy $\o$ close to the trivial values $0$ and $1/2$) we have met difficulties in resolving 
the corresponding small bubbles in the lattice dicretization.}
\label{fig:Figure12}
\end{figure*}

\subsection{Twist-induced vortex}
\label{subsect:SpatialSpatialVortices}

In this section we will discuss the second part of the LCG vortex surfaces we 
found for individual calorons.
We start again by discussing the singlet case. The twist-induced vortex in the singlet case appears at a fixed time slice and hence is a purely spatial vortex. 
In contrast to the space-time part, this vortex surface does not 
contain the monopole/dyon worldlines.
Hence it is not obvious that this part of the vortex structure is caused by them.

The properties of this spatial vortex depend strongly on the holonomy, 
which will be very important for the 
percolation of vortices in 
caloron ensembles in Sect.~\ref{sect:VorticesCaloronEnsembles}.

In short,
our finding is that the twist-induced part of the vortex is a closed surface around 
the twisting dyon as long as the holonomy parameter 
is 
$\o<1/4$, and becomes a 
closed surface around the static dyon for $\o>1/4$, we will refer to these surfaces as ``bubbles''. 
For maximal nontrivial 
holonomy $\o=1/4$ the vortex is the $x_3=0$ plane, i.e.\ the midplane
perpendicular to the axis connecting the dyons, we will refer to it as ``degenerate bubble''.

The bubble depends on the holonomy $\o$ as shown in \fig{fig:Figure11}. 
For two complementary holonomies $\o=\o_0$ and $\o=\frac{1}{2}-\o_0$ the bubbles are of same 
shape just reflected at the origin, 
thus one of them encloses the static dyon and another encloses the twisting dyon. 
This is to be expected from the symmetry of the 
underlying calorons. 
In the limit of $\o\to 1/4$ the bubbles grow to become a flat plane which enables to turn over to the other dyon.

In our $\o=0.25$ data we find another piece of the vortex near the boundary of the lattice, see Fig.~\ref{fig:Figure11}. 
It is an artefact of the finite periodic volume. 
Likewise, very large bubbles in our results have deformations since they come close to the 
boundary of the lattice. The intermediate bubbles shown in these figures
are generally free from 
discretization artefacts and can easily be extrapolated (at least 
qualitatively) to these limits.

The size of the bubble also depends on the size parameter $\rho$ of 
the caloron, i.e.\ the distance 
between the dyons, as shown in \fig{fig:Figure12}.

The time-coordinate of LCG bubbles in large calorons 
is always
consistent with $x_0=0.5\beta$. 
For small calorons, on the other hand, $x_0=0$ is the exclusive time slice: the action density peaks there and the LAG monopoles are circling around it (cf.\ Fig.~\ref{fig:Figure5}).
However, the bubbles of small calorons are too small to be detected.\\ 

In the case of the first excited mode being a doublet, similar bubbles have been found.
They also enclose one of the dyons and degenerate to the midplane for $\o=1/4$.
Their sizes, however, may be different and they are distributed over several time slices.
Considering the collection of all time slices, these fragments add up to full bubbles.

\begin{figure}[b]
\includegraphics[width=0.7\linewidth]{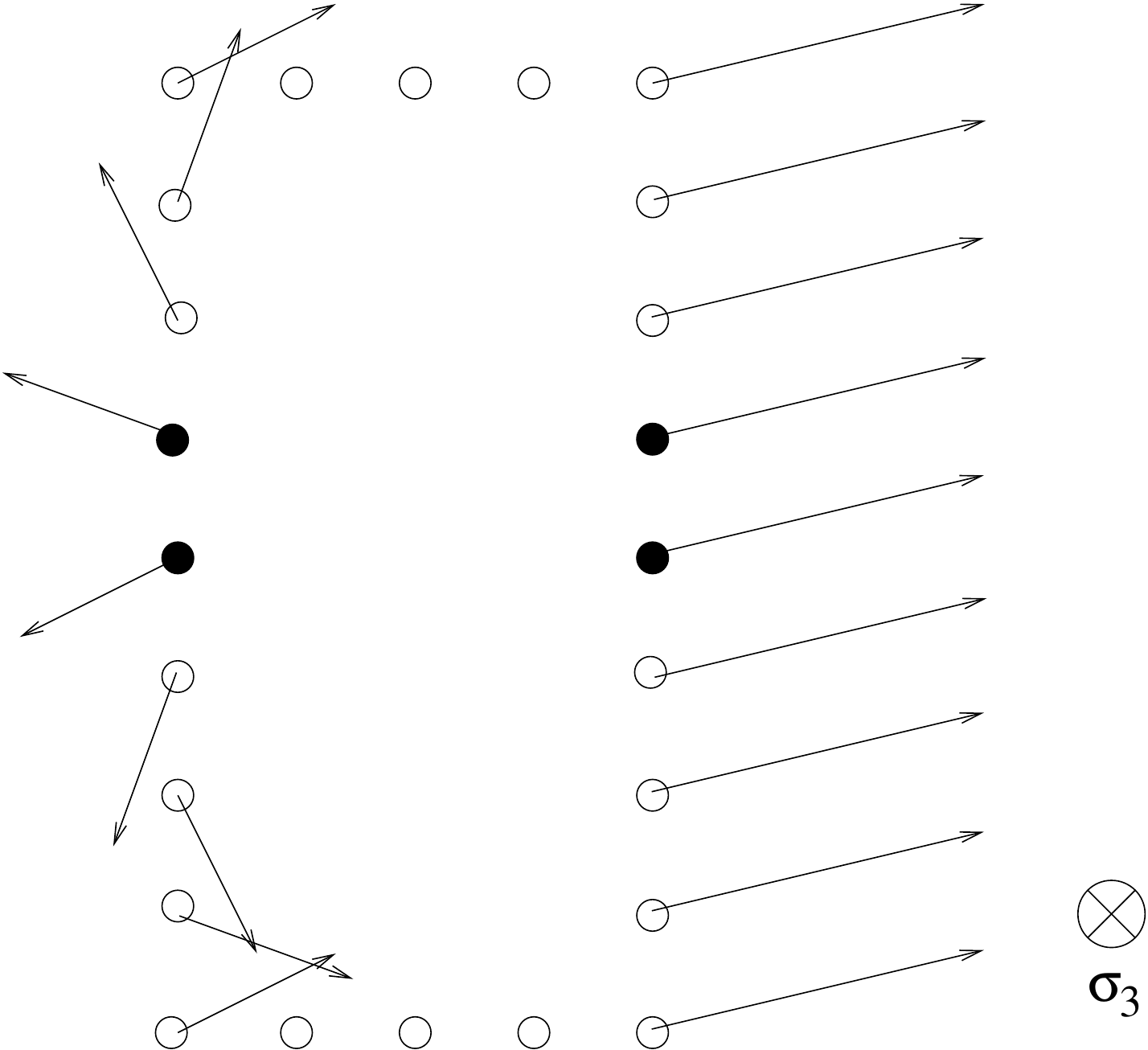}
\caption{Behaviour of the 
nondiagonal elements of $V$ tranformed first excited
mode $^V\!\phi_1$ in the 
twisting region (left) and in the static region (right) with time $x_0$ 
evolving upwards. The lattice sites inbetween are indicated only at $x_0=0$.
On the entire discretised rectangle the field has winding number 1, meaning it contains the twist-induced vortex. More precisely, it is the ``plaquette'' marked with filled circles that contains the winding (in analogy to Fig.~\protect\ref{fig:Figure2}) and thus the vortex (in all other plaquettes the field performs a partial winding but then winds back).}
\label{fig:Figure13}
\end{figure}


\subsubsection{Analytic considerations}

In the following we present two analytic arguments -- relying on the twist -- 
that support the existence of the bubbles (playing the role of spatial vortices) 
and help to estimate their sizes.

The first one is specific for vortices in LCG. As we have demonstrated in 
Sect.~\ref{sect:IndividualCalorons}, 
the lowest mode $\phi_0$ twists near the twisting 
dyon and is static near the static dyon. The same holds for the first excited 
mode $\phi_1$, see Fig.~\ref{fig:Figure6}.

Then a topological argument shows that they have to be (anti)parallel somewhere 
inbetween, cf.~Fig.~\ref{fig:Figure13}.
As $\phi_0$, $\phi_1$ and the diagonalising gauge transformation $V$ are static 
around the static dyon, so is $^V\!\phi_1$ and its projection along the third 
direction (see the right part of Fig.~\ref{fig:Figure13}). We assume that 
this projection is nonzero, otherwise the two states are obviously 
(anti)parallel and the point would belong to the vortex already\footnote{in 
particular to the space-time part since then the two modes are (anti)parallel 
for all $x_0$}.

In the twisting region called $S$, the two lowest modes behave like (suppressing arguments $\vec{x}$)
\begin{eqnarray}
  \phi_{0}(x_0)=\T(x_0)\phi_{0}(x_0=0)\T^\dagger(x_0) \nonumber \\
  \phi_{1}(x_0)=\T(x_0)\phi_{1}(x_0=0)\T^\dagger(x_0)
\end{eqnarray}
with the twisting transformation/rotation from Eq.~(\ref{eq:eqn_twist_T}). 
The time dependence of the diagonalising $V$ can be deduced 
easily\footnote{The first factor is necessary, otherwise $V$ is singular around the north pole and nonperiodic.},
\begin{equation}
  V(x_0)=\T(x_0)V(0)\T^\dagger(x_0)\,,
\end{equation}
such that
\begin{equation}
  ^V\!\phi_1(x_0)=\T(x_0)\,^V\!\phi_1(0)\T^\dagger(x_0)\,.
 \label{eq:eqn_rotation_ground_first}
\end{equation}

Again we assume that the two modes are not (anti)parallel at $x_0=0$. Then 
$^V\!\phi_1(0)$ has a nonvanishing component perpendicular to $\sigma_3$. 
According to Eq.~(\ref{eq:eqn_rotation_ground_first}) this component then
rotates in time $x_0$ around the third direction (left 
part of Fig.~\ref{fig:Figure13}). This immediately implies that there 
is a space-time ``plaquette'' (in the sense of Fig.~\ref{fig:Figure2},
marked in Fig.~\ref{fig:Figure13} with filled circles) 
that contains a point 
where the two modes are collinear. 
Notice the similarity of Figs.~\ref{fig:Figure13} and \ref{fig:Figure2}.

This argument applies to all pairs of points with one point in the twisting 
region $S$ and one point in the static region (its complement) $\bar{S}$: on 
any line connecting the two there exists a point which belongs to the vortex. 
This results in a closed surface at the boundary between $S$ and $\bar{S}$ (see 
below). The time-coordinate of this surface is not determined by these considerations.

Our argument can be extended to vortices beyond LCG. 
For that aim we mimic the caloron gauge field by 
$A_0=0,\,A_i=0$ in the static region $\bar{S}$ and 
$A_0=-\sigma_3 (\pi/\beta),\,A_i=0$ in the twisting region $S$ (cf. Eq.~(\ref{eq:eqn_relation_A_3})) \cite{Zhang:2009et}.
In this simplified gauge field vortices can be located directly by the definition that $-1$ Wilson loops are linked with them.
Obviously rectangular Wilson loops connecting $(0,\vec{x}_1)$, $(\beta,\vec{x}_1)$, $(\beta,\vec{x}_2)$, $(0,\vec{x}_2)$ and $(0,\vec{x}_1)$ are $-1_2$ if and only if $\vec{x}_1$ belongs to $S$ and $\vec{x}_2$ belongs to $\bar{S}$ (or vice versa).
This again predicts spatial vortices at the boundary between the twisting and the static region.

Actually, this argument is exact if one chooses for the points $\vec{x}_{1,2}$ the 
dyon locations $\vec{z}_{1,2}$: the 
path ordered exponentials at fixed $\vec{x}_{1,2}$ are the Polyakov loops $\mp 1_2$ 
and the remaining spatial parts are inverse to each other because of periodicity and cancel.
Hence there should always be a spatial vortex between the two dyons. 

Thus \emph{the twist in 
the gauge field} of the caloron itself \emph{gives rise to a spatial vortex}. This 
vortex extends in the two spatial directions perpendicular to lines connecting 
$S$ and $\bar{S}$, just like a bubble. 

Note that the two arguments above do not work purely within the twisting region or
purely within the static region.\\

\begin{figure}[b]
\includegraphics[width=0.9\linewidth]{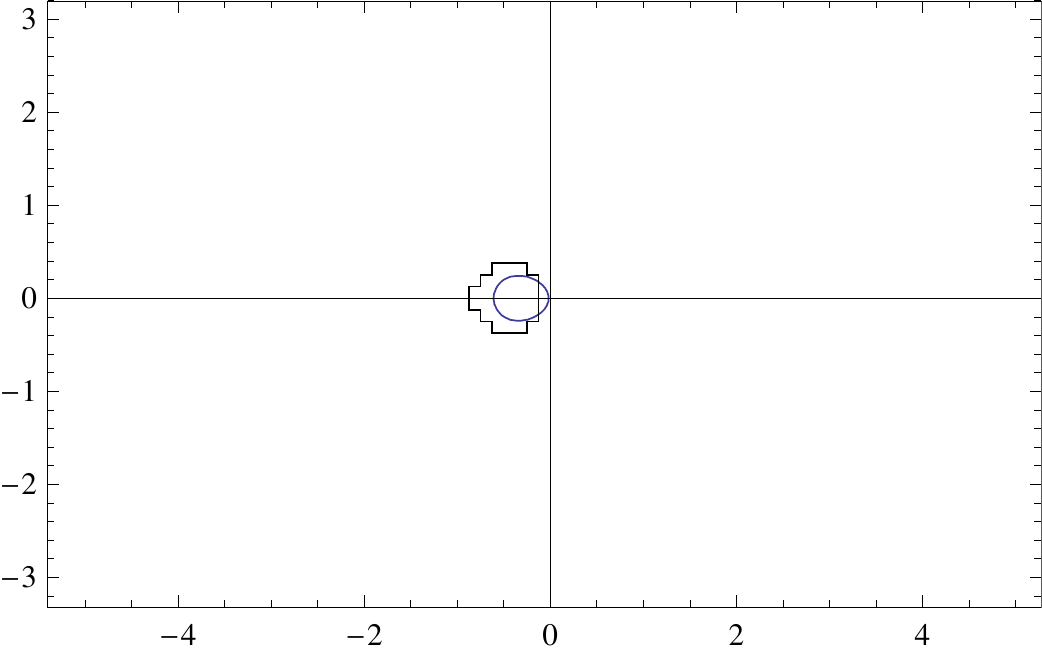}
\includegraphics[width=0.9\linewidth]{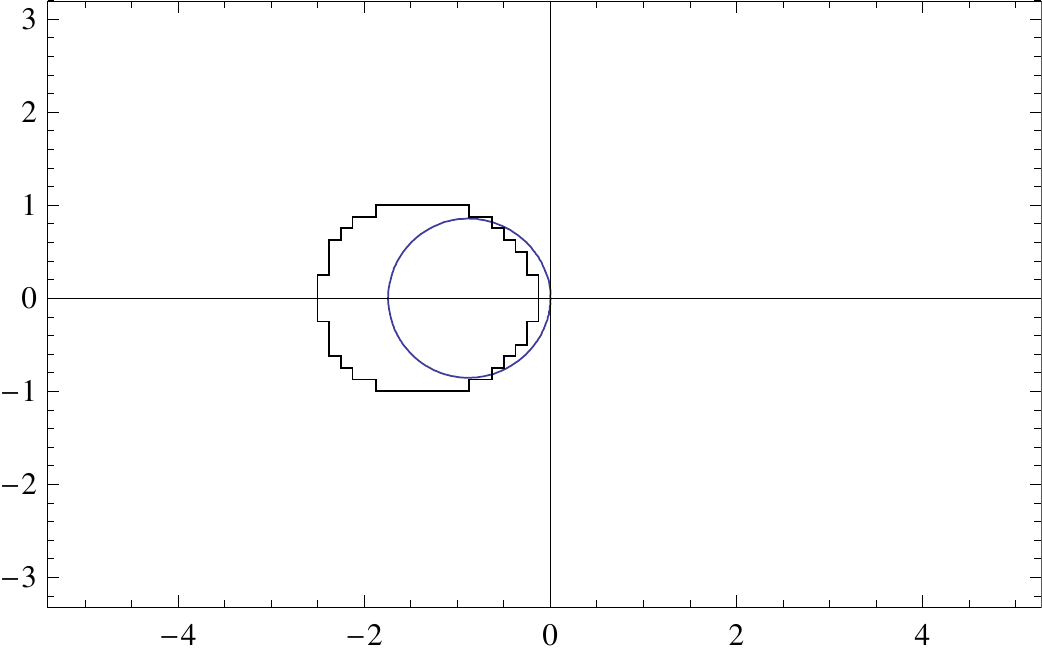}
\caption{The bubbles measured for calorons with holonomy $\o=0.12$, size
$\rho=0.6\beta$ (top) and 
$\rho=0.9\beta$ (bottom) respectively as a function of $x_2$ (vertically) 
and $x_3$ (horizontally) at  $x_1=0$ compared to the boundary of the 
twisting region $S$, the smooth curve computed from the equality in \protect\Eqn{eq:eqn_S_def}.}
\label{fig:Figure14}
\end{figure}

It remains to be specified where the boundary between the twisting region $S$ 
and its complement $\bar{S}$ is.
To that end one should consider the competing terms -- twisting vs.\ static -- in
the relevant function $\tc$, see \Eqn{eqn_chihat_exact}. Actually its derivatives
enter the off-diagonal gauge fields, see \Eqn{eq:eqn_A_caloron}. In the periodic
gauge we have used so far, there is an additional term proportional to $\tc$ itself.
To decide whether the static or the twisting part dominates (at a given point) it is
better to go over to the algebraic gauge, where this term is absent and where $\tc$
must be replaced by $\chi=\exp(4\pi i\o x_0/\beta)\tc$ \cite{Kraan:1998pm}. The two
competing terms become
\begin{eqnarray}
 e^{4\pi i\o x_0/\beta}\,\frac{\sinh(\r)}{\psi r}   & \equiv & f_{\rm static}\\
 e^{-4\pi i\bo x_0/\beta}\,\frac{\sinh(\s)}{\psi s} & \equiv & f_{\rm twist}
\end{eqnarray}
with $\psi$ given in Eqn.~(\ref{eqn_psi_exact}).
Note that the time dependence of these functions still differs by a factor $\exp(2\pi i  x_0/\beta)$.

We finally define the twisting region $S$ 
as where the gradient of $f_{\rm twist}$ dominates
\begin{equation}
 |\partial_\mu f_{\rm twist}|^2\geq |\partial_\mu f_{\rm static}|^2\,.
\label{eq:eqn_S_def}
\end{equation}
and its boundary where the equality holds.

In two particular cases this can be determined analytically. For the case of maximally nontrivial holonomy $\o=\bo=1/4$, the two functions $f_{\rm static,twist}$ only differ by the arguments $r$ vs.\ $s$. Then the  boundary of $S$ is obviously $r=s$, which gives the midplane between the dyons. This indeed amounts to our numerical 
finding, the degenerate bubble for $\o=1/4$.

In the large caloron limit and if we further assume
the solutions of the equality in Eq.~(\ref{eq:eqn_S_def}) to obey $\o r/\beta, \bo s/\beta \gg 1$,
it is enough to compare in $f_{\rm static,twist}$ the exponentially large terms in
sinh and $\psi$. This yields for the boundary of $S$ the equation 
$\bo r = \o s$, which can be worked out to give 
\begin{equation}
 x_1^2+x_2^2+(x_3-\Omega d)^2=(\Omega d)^2\,,\quad \Omega\equiv\frac{2\o\bo}{\o-\bo}
\label{eq:eqn_boundaryS_farfield}
\end{equation}
Thus, the boundary of the twisting region $S$ is a sphere with midpoint
$(0,0,\Omega d)$ and radius  $|\Omega| d$. This sphere always 
touches the origin, is centered at negative and positive $x_3$ for 
$\o<1/4$ and $\o>1/4$, respectively, and again degenerates to the 
midplane of the dyons for maximally nontrivial holonomy $\o=\bo=1/4$. 

In Fig.~\ref{fig:Figure14} we compare the 
boundary of $S$ obtained from the equality in \Eqn{eq:eqn_S_def} to the 
numerically obtained bubbles in LCG for two different values of the caloron 
parameter $\rho$.
The graphs agree qualitatively.

One could also think of characterising the locations $\vec{x}$  of the twist-induced vortex by a fixed value of the traced Polyakov loop, say $\tr\P(\vec{x})=0$. This also encloses one of the dyons and becomes the midplane for $\o=1/4$. In the large separation limit, however, this surface is that of a single dyon of fixed size set by $\beta$ and $\o$ (just like the topological density). 
It does not grow with the separation $d$, which however seems to be the case for the measured vortices as well as for the boundary of $S$ using the far field limit, \Eqn{eq:eqn_boundaryS_farfield}. 
Hence the local Polyakov loop seems not a perfect pointer to the spatial vortex.





\subsection{Intersection and topological charge}
\label{subsect:TopologicalCharge}


To a good approximation the dyon charge induced vortex extends in space and time connecting the dyons twice, whereas the twist induced vortex is purely spatial around one of the dyons. This results in two intersection points generating topological charge as we will describe now.

The notion of topological charge also exists for (singular) vortex sheets. 
In order to illustrate that
let us choose a local coordinate system and denote the two directions perpendicular 
to the vortex sheet, in which a Wilson loop is $-1$, by $\mu$ and $\nu$. The Wilson 
loop can be generated by a circular Abelian gauge field decaying with the  inverse 
distance, which generates a gauge field $F_{\mu\nu}$ (the magnetic field, 
say $B_3\propto F_{12}$ for a static vortex in the $x_3$-direction, 
is tangential to the vortex, respectively). The corresponding flux is via an 
Abelian Stokes' Theorem connected to the Wilson loop and is nothing but the 
winding number used in LCG to detect the vortex.

\begin{figure}[b]
\includegraphics[width=\linewidth]{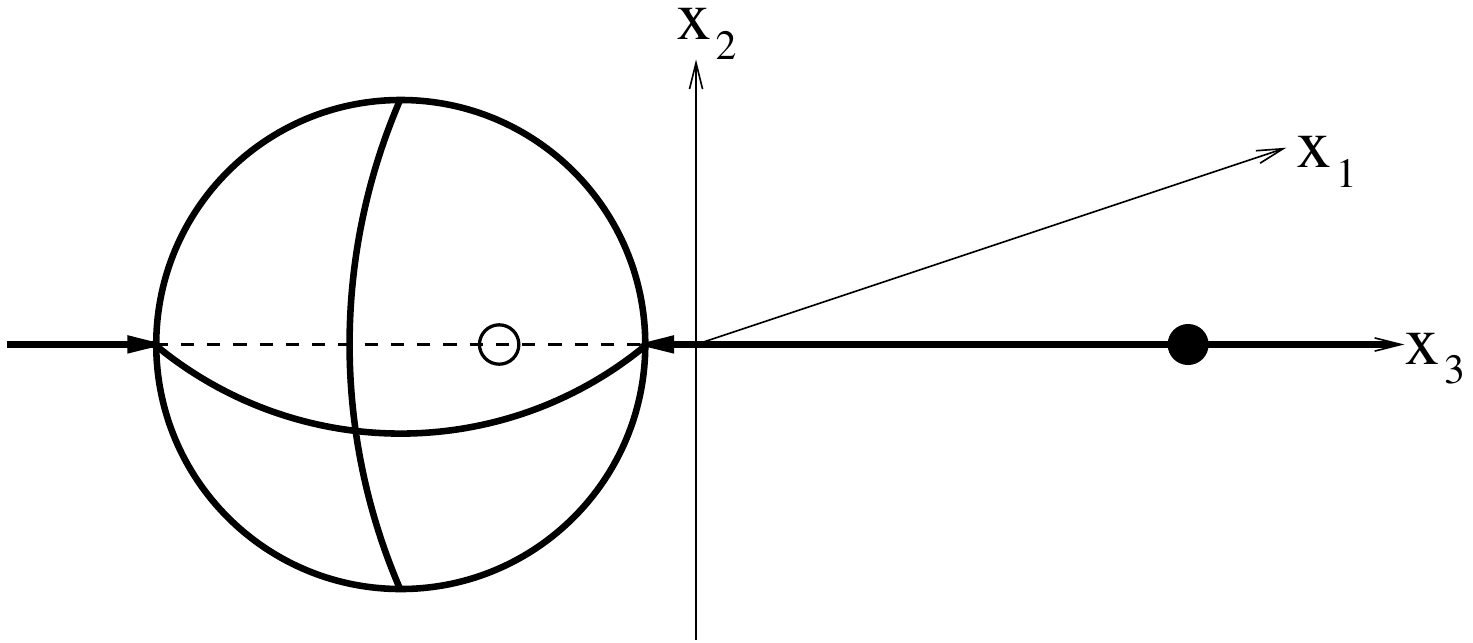}\\
\includegraphics[width=\linewidth]{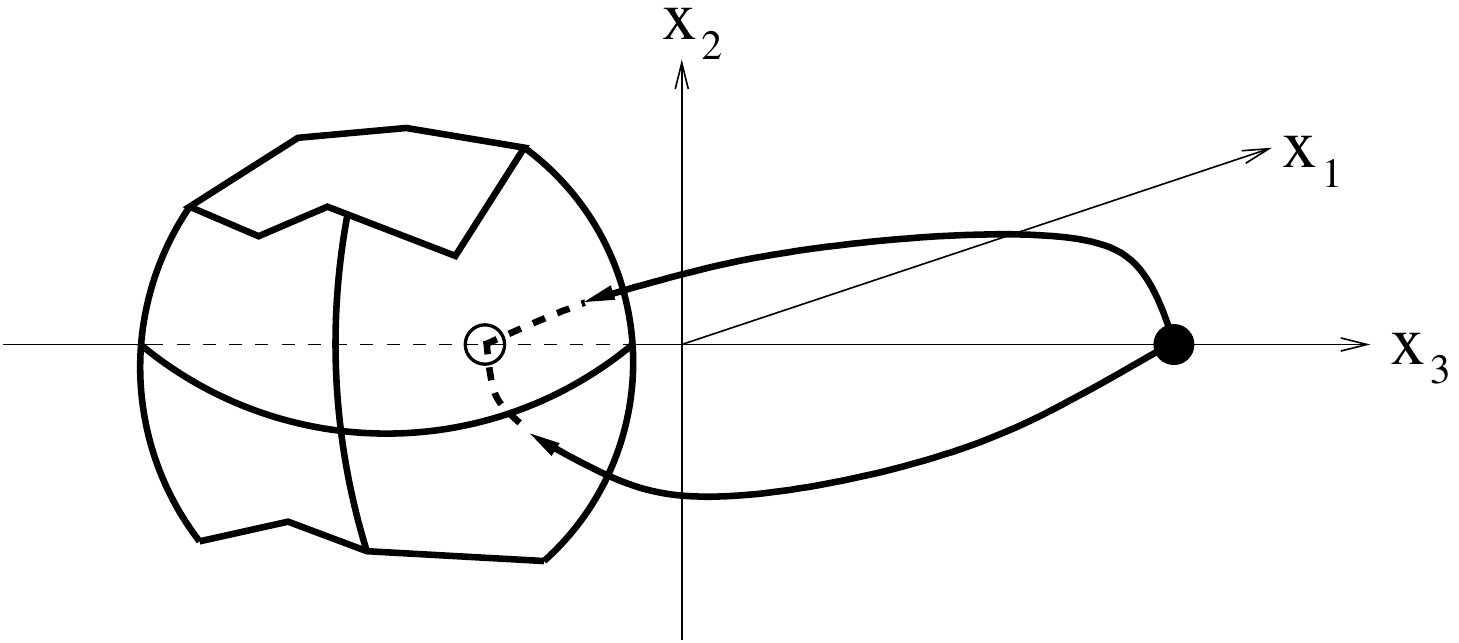}
\caption{The intersection of the spatial bubble (at fixed $x_0=\beta/2$)
for $\o<1/4$ with the space-time part of the vortex from the singlet (top) and 
the doublet (bottom) excited mode (in the doublet case the bubble is distributed over several time slices).}
\label{fig:Figure15}
\end{figure} 

In order to generate topological charge proportional to 
$\epsilon^{\mu\nu\rho\sigma}F_{\mu\nu}F_{\rho\sigma}$, 
the vortex thus needs to ``extend in all directions''. 
This is made more precise by the geometric objects 
called writhe and self-intersection. 
The relation to the topological charge including example configurations has been 
worked out for vortices consisting of hypercubes in \cite{Engelhardt:2000wc} and
for smooth vortices in \cite{Engelhardt:1999xw}. 
The result is that a (self)intersection point -- where two branches of the vortex 
meet such that the combined tangential space is four-dimensional -- contributes 
$\pm 1/2$ to the topological charge. The contribution of the writhe is related 
to gradients of the vortex' tangential and normal space w.r.t.\ the two coordinates 
parametrising the vortex. 
Two trivial examples are important for vortices in a caloron: a two-dimensional 
plane as well as a  two-dimensional sphere embedded in four-dimensional space 
have no writhe. Since the two parts of our vortex are of these topologies, we 
immediately conclude that the topological charge of vortices in calorons comes 
exclusively from intersection points. 

We first discuss the position of the intersection points in the singlet case, cf.\ Fig.~\ref{fig:Figure15} top panel. The twist induced bubble occurs at a fixed time slice and so does any intersection point. The dyon induced vortex consists of two static straight lines from one of the dyon to another and therefore intersects the bubble twice on the $x_3$-axis.

There are two exceptions to this fact:  for small calorons (small $\rho$) the dyon induced vortex exists in some time slices only and the number of intersection points depends on whether the time-coordinate of the bubble is within that time-interval (but the bubble is usually too small to detect when $\rho$ is small).

The case of maximal nontrivial holonomy is particular because for the corresponding degenerate bubble there is only one intersection point at 
the center of mass of the caloron, the other one moved to $x_3\to\pm\infty$ 
as $\o\to 1/4\pm 0$ in infinite volume.

Concerning the sign of the contributions, it is essential that 
the relative sign of the vortex flux is determined:
the magnetic flux of the dyon induced vortex flips at the dyons and hence is of opposite sign at the intersection points on the bubble.
One can depict 
the flux on the bubble by an electric field normal to the bubble (i.e.\ hedgehog-like).
It follows that in LCG 
the contributions of the intersection points to the topological 
charge of the vortex are both $+1/2$. 

The vortex in the caloron is thus an example for a general statement, 
that a non-orientable vortex surface is needed for a nonvanishing total  
topological charge. 
In our case the two branches of the 
dyon charge induced vortex have been glued together at the dyons in a non-orientable 
way: the magnetic fluxes start or end at the dyons as LAG-monopoles 
(this construction is impossible for the bubble as the dyons are not 
located on them).
Thus, vortices without monopoles on them possess trivial total topological 
charge.

In the doublet case with its fragmented bubbles there are still two intersection points (cf.\ Fig.~\ref{fig:Figure15} bottom panel) which again contribute topological charges of $+1/2$ each.

To sum up this section we have demonstrated that the vortex has unit 
topological charge like the caloron background
gauge field. This result is not 
completely trivial as there is to our knowledge no general proof that 
the topological charge from the gauge field persists for its vortex 
``skeleton'' after center projection (P-vortices). 
Moreover, the topological density 
of the caloron is not maximal at the two points where the topological 
density of the vortex is concentrated and the total topological charge 
of the caloron is split into fractions of $2\o$ and $2\bo$ whereas that 
of the vortex always comes in equal fractions $1/2$ from two intersection 
points, close to the static dyon if $\o < 1/4$ and close to the twisting 
dyon if $\o > 1/4$.

\subsection{Results from Maximal Center Gauges}
\label{sect:direct_gauges}

We have performed complementary studies of vortices
both in the Direct and in the Indirect 
Maximal Center Gauges (DMCG \cite{Del Debbio:1998uu} and IMCG \cite{Del Debbio:1996mh},
respectively). The DMCG in $SU(2)$ 
is defined by the maximization of the functional
\begin{equation}
F_{\rm DMCG}[U] = \sum_{\mu, x} \left( \tr {}^gU_\mu(x)\right)^2 \, ,
\label{eq:maxfunc_1}
\end{equation}
with respect to gauge transformations $g(x) \in SU(2)$. 
$U_\mu(x)$ are the lattice
links and ${}^gU_\mu(x)=g^{\dag}(x)U_\mu(x)g(x+\hat{\mu})$ the gauge
transformed ones.
Maximization of (\ref{eq:maxfunc_1}) minimizes the distance to center elements and fixes the gauge up to a $Z(2)$ gauge
transformations. 
The corresponding, projected $Z(2)$ links are
defined as
\begin{equation}
Z_\mu(x) = {\rm sign} \left( \tr {}^gU_\mu(x) \right) \, .
\label{eq:Zdef}
\end{equation}
The Gribov copy problem is known to spoil gauges with maximizations such as DMCG.
In practice we also applied random $SU(2)$ gauge transformations before maximizing $F_{\rm DMCG}$ and selected the configuration with the largest value of that functional.

The IMCG goes an indirect way. At first, one fixes the maximal Abelian 
gauge (MAG \cite{'tHooft:1981ht}) by maximizing the functional
\begin{equation}
F_{\rm MAG}[U] = \sum_{\mu, x} \tr \left( {}^gU_\mu(x)\sigma_3 ({}^gU_\mu(x))^{\dag}
\sigma_3\right) \, ,
\label{eq:maxfunc_3}
\end{equation}
with respect to gauge transformations $g \in SU(2)$. 
The MAG minimizes the 
off-diagonal elements of the links and fixes the
gauge up to $U(1)$. Therefore, the following projection to a $U(1)$ gauge
field through the phase of the diagonal elements of the links,
$\theta_\mu(x) = \arg \left( ({}^gU_\mu(x))^{11} \right)$,
is not unique.
Exploiting the remaining $U(1)$ gauge freedom, which amounts to a shift
$\theta_\mu(x) \to {}^{\alpha}\theta_\mu(x)
= - \alpha(x) + \theta_\mu(x) + \alpha(x+\hat{\mu})$, one 
maximize the 
IMCG functional
\begin{equation}
F_{\rm IMCG}[U] = \sum_{\mu, x} \left( \cos( {}^{\alpha}\theta_\mu(x)) \right)^2 \, ,
\label{eq:maxfunc_4}
\end{equation}
that serves the same purpose as $F_{\rm DMCG}$ in (\ref{eq:maxfunc_1}).
Finally, the projected $Z(2)$ gauge links are defined as
\begin{equation}
Z_\mu(x) = {\rm sign} \left( \cos( {}^{\alpha}\theta_\mu(x) ) \right) \, .
\label{eq:ZdefICG}
\end{equation}

Finally, the $Z(2)$ 
links
are used to form $Z(2)$ plaquettes.
All dual plaquettes of the negative $Z(2)$ plaquettes form closed two dimensional
surfaces 
-- the vortex surfaces.\\

In the background of calorons we tried to confirm both dyon charge induced and 
twist-induced vortices seen in LCG. In DMCG, the dyon charge induced vortices are observed and 
the twist induced part splits into several parts in adjacent time slices.
Choosing the best among random gauge copies, the dyon charge induced part disappears and the twist-induced vortex bubble occurs at fixed time slice.
In both cases, the bubble is much smaller 
than that found in LCG.

In IMCG, the situation for dyon charge induced vortices is 
rather stable: 
we find always a space-time vortex connecting the dyons or better to say the Abelian monopoles representing the dyons in MAG \cite{Brower:1998ep,Ilgenfritz:2004zz}.
Similar to the LCG doublet case two vortex lines pass
near the center
of mass of the caloron.
This structure propagates either statically
or nonstatically in time, depending on the distance between dyons in the caloron.

Straight lines through the center of mass and through the outer space, as found in the LCG singlet case, 
were never observed. One can convince oneself, that it is actually impossible to get such 
vortex structures
from $Z(2)$ link configurations.

The situation with twist-induced vortices is unstable, as a rule they do not
appear in IMCG. The reason for this could be partially understood in IMCG considerations.
Let us consider two gauge equivalent Abelian configurations that generate local 
Polyakov loops $\frac{1}{2}\tr\P(\vec{x})=\cos(a({\vec x}))$ and how their 
temporal links contribute to the Abelian gauge functional $F_{\rm IMCG}$ 
given in Eq.~(\ref{eq:maxfunc_4}).
In the 
quasi-temporal gauge when all 
subsequent temporal links are the same, they give a contribution equal to 
$N_0\cos(a({\vec x})/N_0)^2$ 
to the corresponding part of the functional. When, on the other hand,
all but one temporal links are trivial, the contribution is equal to
$N_0 - 1 +\cos(a({\vec x}))^2$. For $\cos(a({\vec x}))\ne -1$ and for sufficiently large $N_0$ we have
$N_0\cos(a({\vec x})/N_0)^2 \simeq N_0 -a({\vec x})^2/N_0 > N_0 - \sin(a({\vec x}))^2 $.
This means that after maximizing the functional (\ref{eq:maxfunc_4}) and projecting onto
$Z(2)$, we get all temporal links trivial in all points ${\vec x}$ 
where the Polyakov loop is not equal to $-1$. So, 
the twist-induced vortex shrinks to one point where the (untraced) Polyakov loop
is equal to $-1_2$. 

Maximization of the functional (\ref{eq:maxfunc_4}) is equivalent to the 
minimization of the functional
\begin{equation}
F[U] = \sum_{\mu, x} \left( \sin( {}^{\alpha}\theta_\mu(x) ) \right)^2 \, .
\label{eq:maxfunc_5}
\end{equation}
If we would replace it by the functional
\begin{equation}
F^{\prime}[U] = \sum_{\mu, x} \sqrt{\left( \sin( {}^{\alpha}\theta_\mu(x) ) \right)^2} \, .
\label{eq:maxfunc_6}
\end{equation}
the situation with the $Z(2)$ projected Polyakov loop would change because
$N_0\sqrt{\sin(a({\vec x})/N_0)^2} > \sqrt{\sin(a({\vec x}))^2}$ 
on points where $\cos(a({\vec x}))<0$ and now
we would have 
$-1_2$ temporal links in some time slice and trivial
temporal links in all other time slices in the 
spatial region where the 
Polyakov loop
is negative 
as well as trivial temporal $Z(2)$ links in all time slices in the region
where the 
Polyakov loop is positive. Numerical studies support the
appearance of twist-induced vortex on the boundary where
initial Polyakov loop changes the sign from negative to
positive. 

One may conclude from this section that the Gribov copy problems of DMCG and IMCG persist for the smooth caloron backgrounds. The basic features of the vortices can be reproduced, but for clarity we stick to the vortices obtained in the Laplacian Center gauge.

\section{Vortices in caloron ensembles}
\label{sect:VorticesCaloronEnsembles}

In this section we present the vortex content of ensembles of calorons. 
The generation of the latter has been described in 
Sect.~\ref{subsect:CaloronEnsembles}. 
We superposed 6 calorons and 6 anticalorons with an average size of $\bar{\rho}=0.6\beta$ on a $8\times 64^3$ lattice.

The most important feature of these ensembles is their holonomy 
$\Pi=\exp\left(2\pi i\o \sig_3\right)$. Under the conjecture mentioned in the 
introduction we will use $\frac{1}{2}\,\tr \Pi=\cos(2\pi\o)$ as equivalent to 
the order parameter $\langle \frac{1}{2}\,\tr \P\rangle$. In particular, 
caloron ensembles with  maximally nontrivial holonomy $\o=1/4$ mimic the 
confined phase with $\langle \frac{1}{2}\,\tr \P\rangle=0$.

For each of the holonomy parameters $\o=\{0.0625, \,0.0125,\,0.01875,\,0.25\}$ we considered one caloron ensemble with otherwise equal parameters.
Again we computed the lowest adjoint modes in these backgrounds and used the 
routines based on winding numbers to detect the LCG vortex content.

\begin{figure*}[t]
\hspace{1.0cm}
\includegraphics[width=0.23\textwidth,viewport=60 -20 369 290]{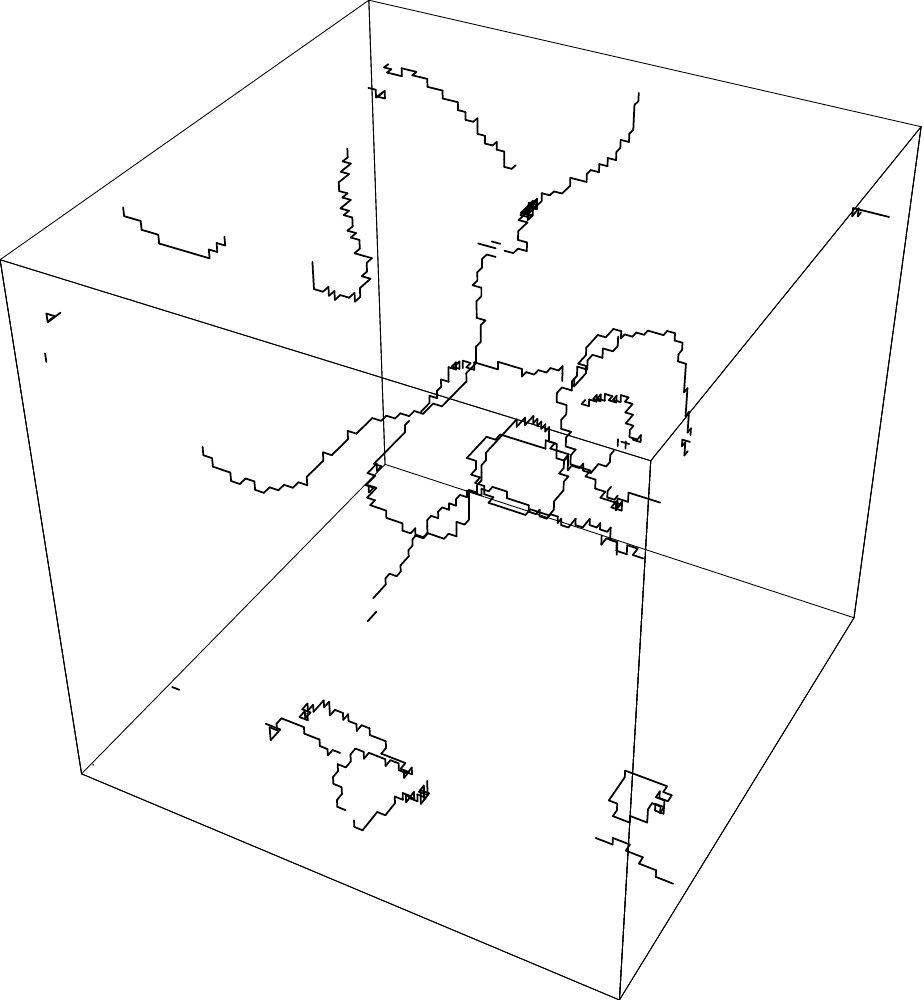}
\includegraphics[width=0.23\textwidth,viewport=60 -20 369 290]{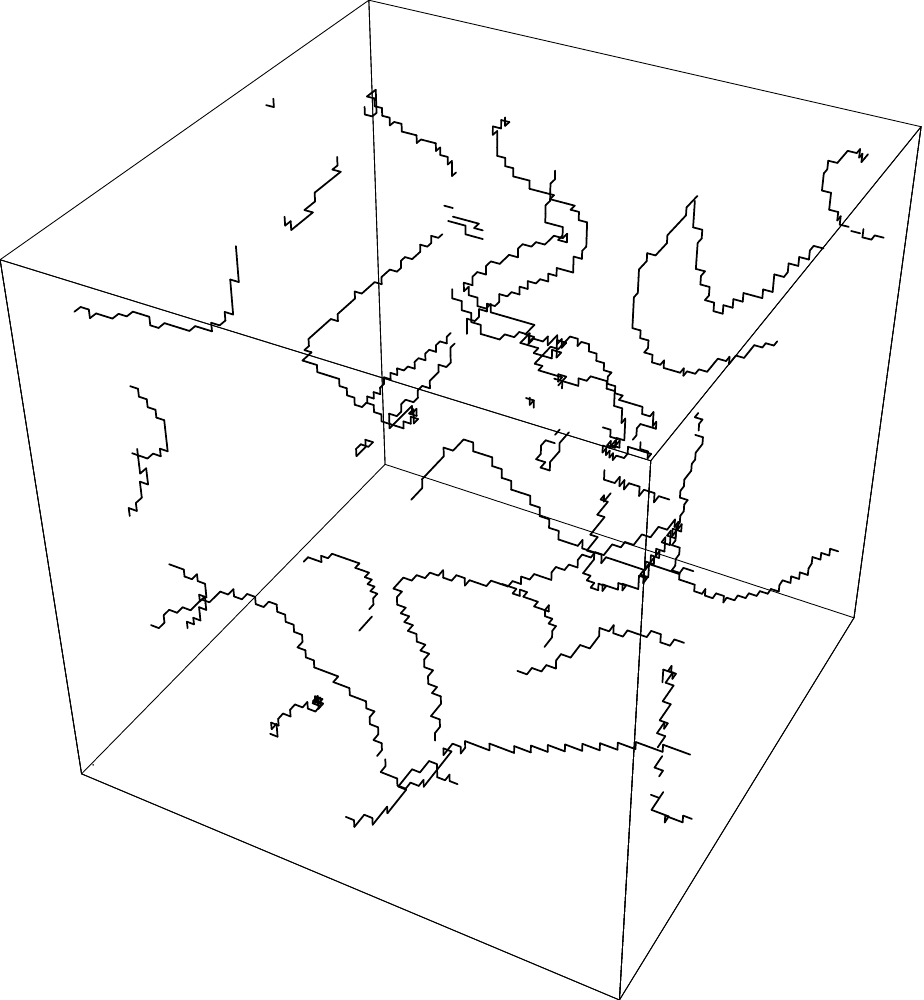}
\includegraphics[width=0.23\textwidth,viewport=60 -20 369 290]{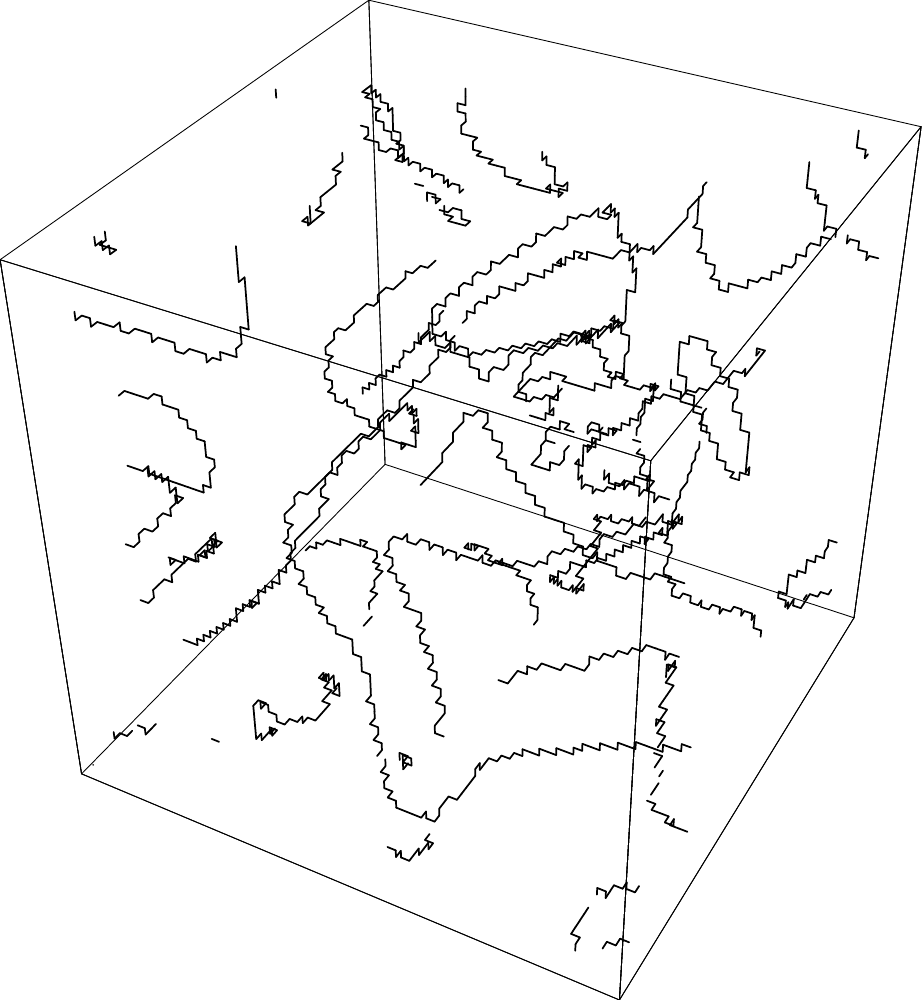}
\includegraphics[width=0.23\textwidth,viewport=60 -20 369 290]{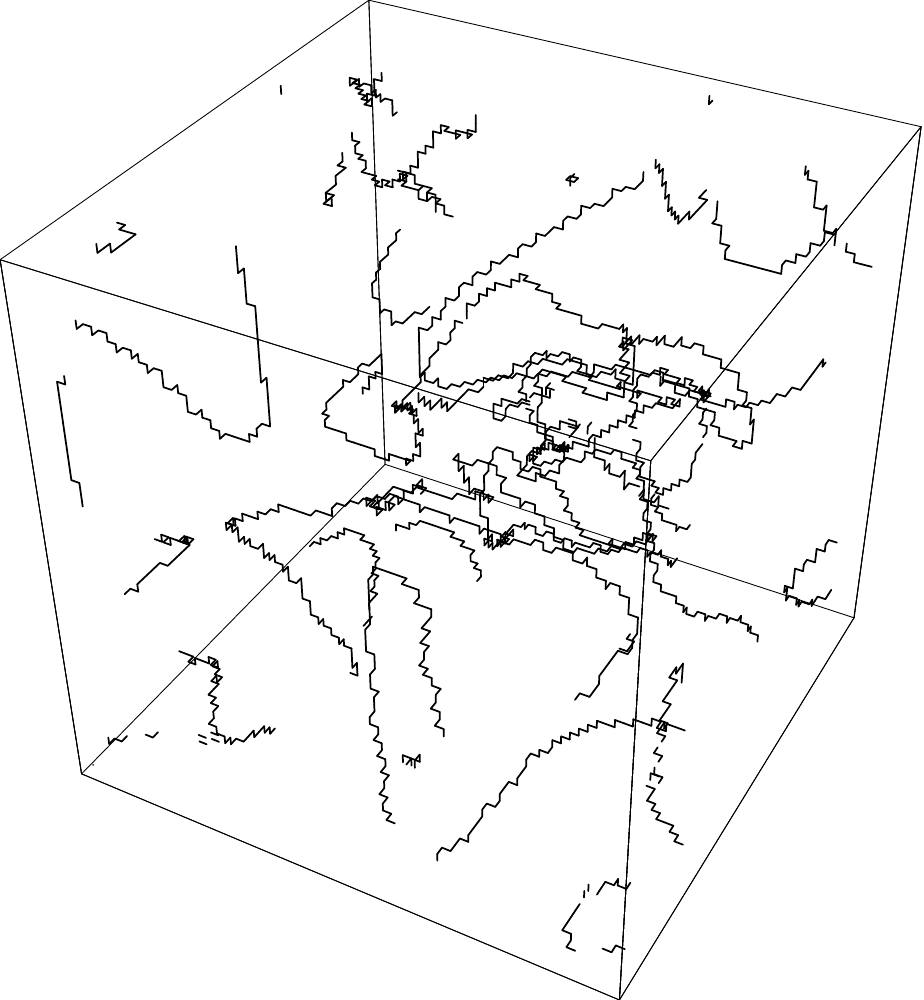}\\
\hspace{1.0cm}
\includegraphics[width=0.23\textwidth,viewport=60 -20 369 290]{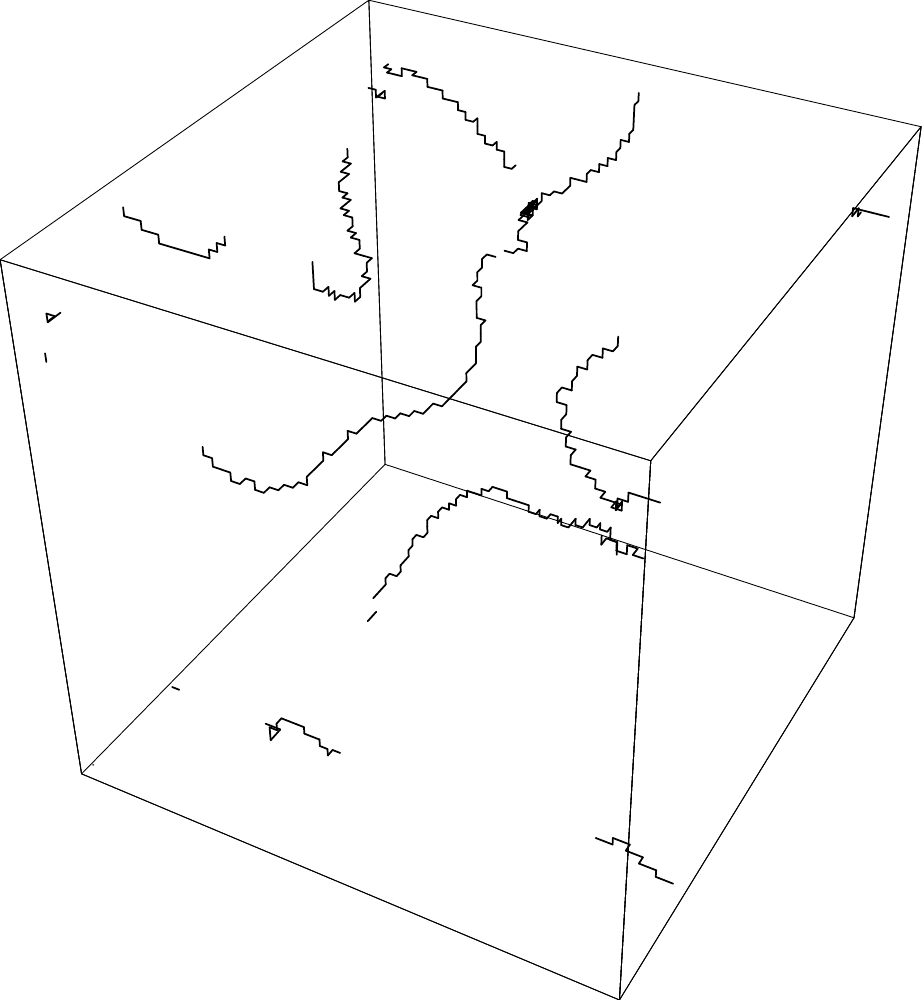}
\includegraphics[width=0.23\textwidth,viewport=60 -20 369 290]{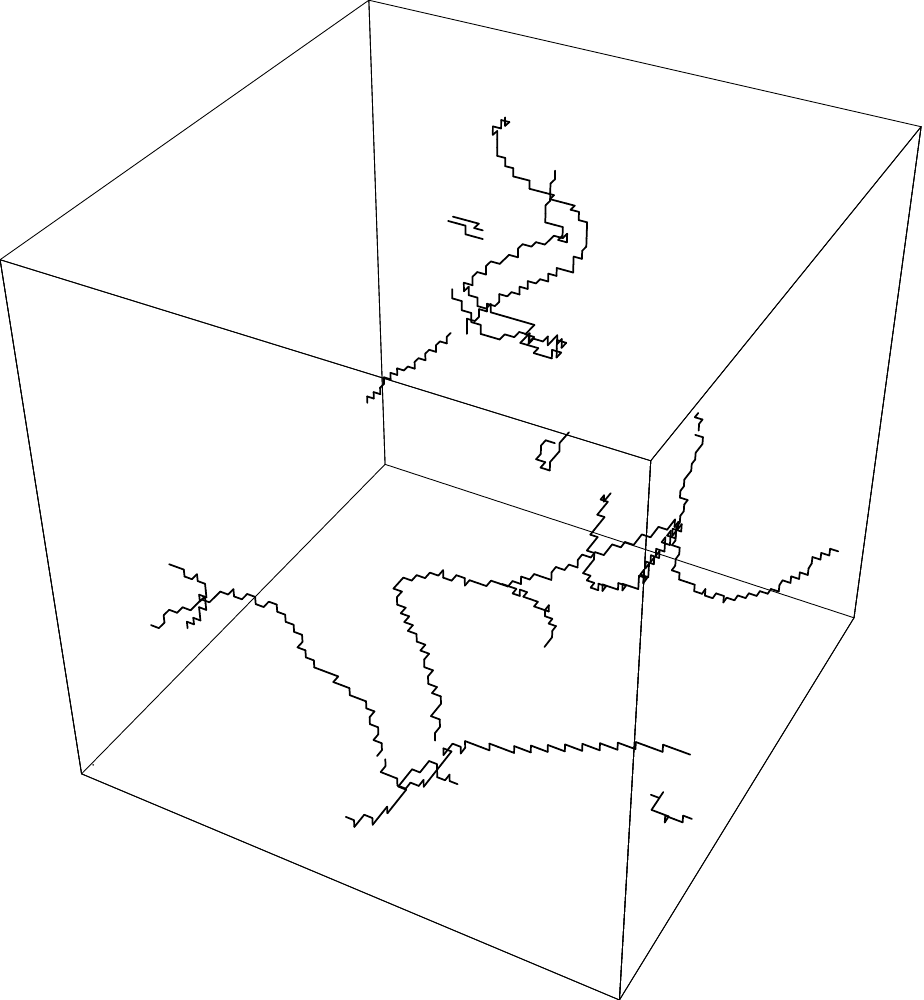}
\includegraphics[width=0.23\textwidth,viewport=60 -20 369 290]{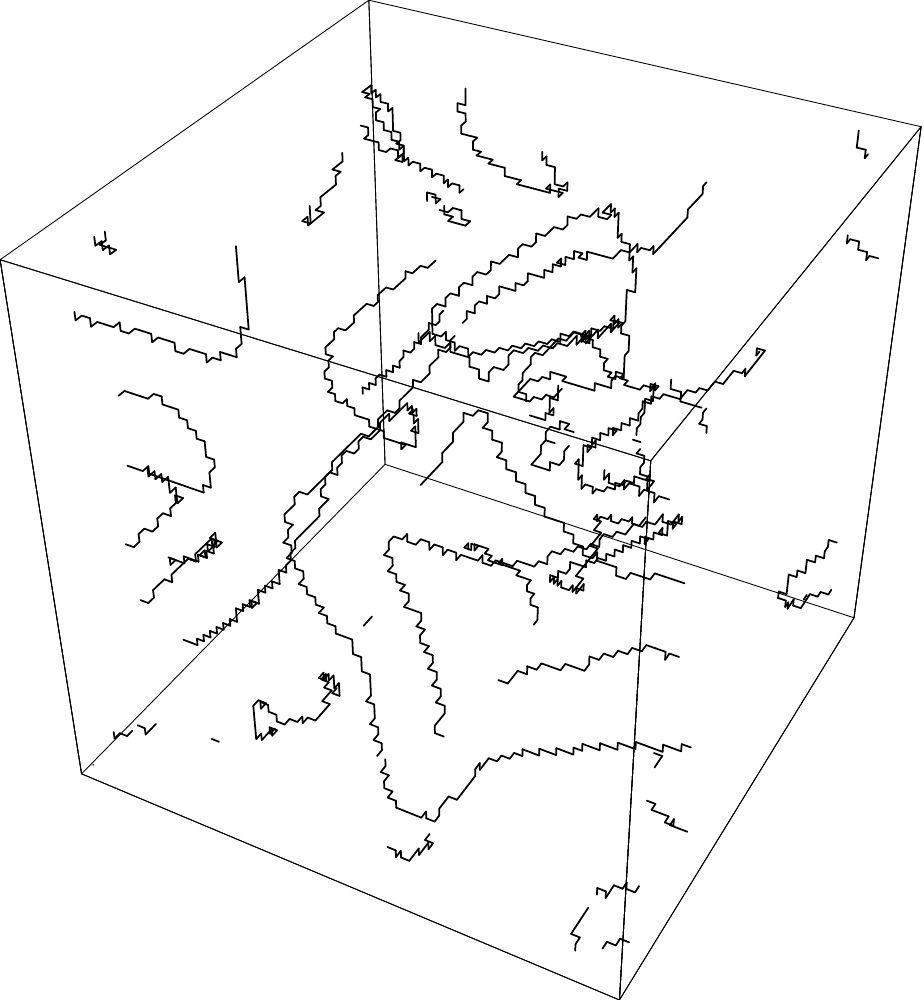}
\includegraphics[width=0.23\textwidth,viewport=60 -20 369 290]{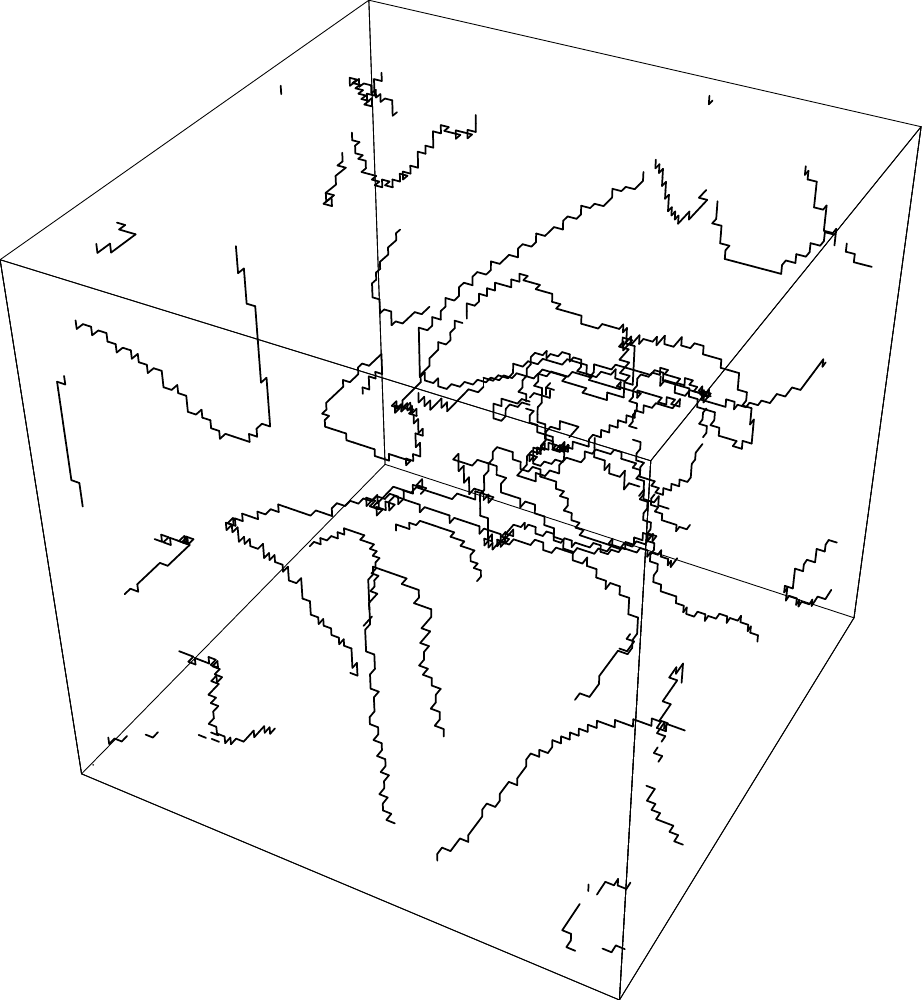}
\caption{The space-time part of vortices in caloron ensembles in a fixed time slice. Only the holonomy varies from left to right: $\o=\{0.0625,\,0.0125,\,0.01875,\,0.25\}$ (from deconfined phase to confined phase). The upper row shows the entire vortex content in each caloron ensemble, the lower row shows the corresponding biggest vortex cluster.}
\label{fig:Figure16}
\end{figure*}

\begin{figure*}
\hspace{1.0cm}
\includegraphics[width=0.23\textwidth,viewport=60 -20 369 290]{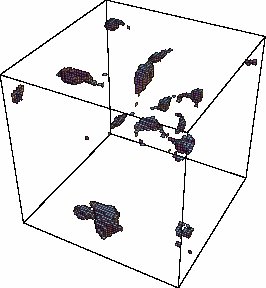}
\includegraphics[width=0.23\textwidth,viewport=60 -20 369 290]{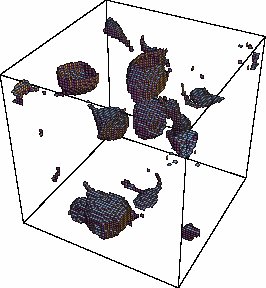}
\includegraphics[width=0.23\textwidth,viewport=60 -20 369 290]{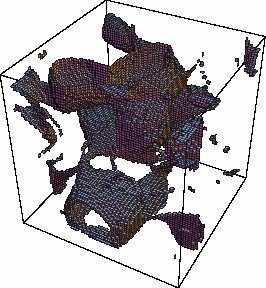}
\includegraphics[width=0.23\textwidth,viewport=60 -20 369 290]{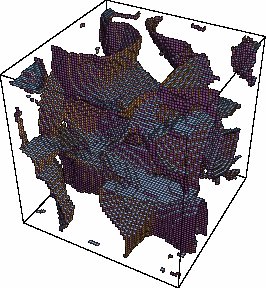}\\
\hspace{1.0cm}
\includegraphics[width=0.23\textwidth,viewport=60 -20 369 290]{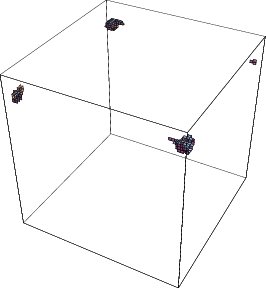}
\includegraphics[width=0.23\textwidth,viewport=60 -20 369 290]{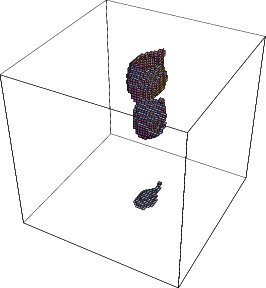}
\includegraphics[width=0.23\textwidth,viewport=60 -20 369 290]{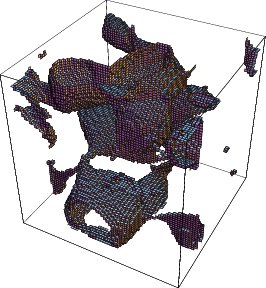}
\includegraphics[width=0.23\textwidth,viewport=60 -20 369 290]{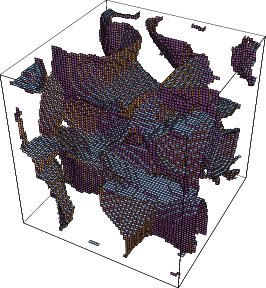}
\caption{The spatial part of vortices in the caloron ensembles of 
Fig.~\protect\ref{fig:Figure16} (with the same values of the holonomy $\omega$) 
summed over all time slices. Again the upper row shows the entire vortex content 
and the lower row the corresponding biggest vortex cluster. }
\label{fig:Figure17}
\end{figure*} 

\begin{figure}
\includegraphics[width=0.7\linewidth]{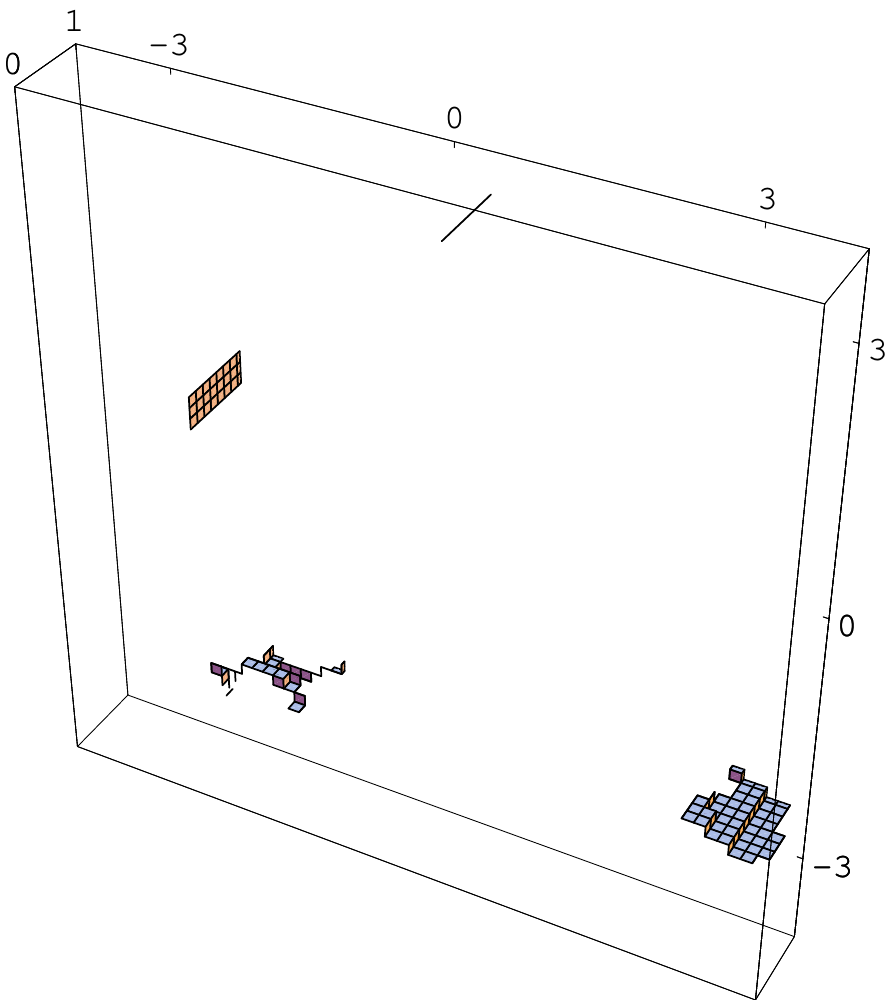}\\
\includegraphics[width=0.7\linewidth]{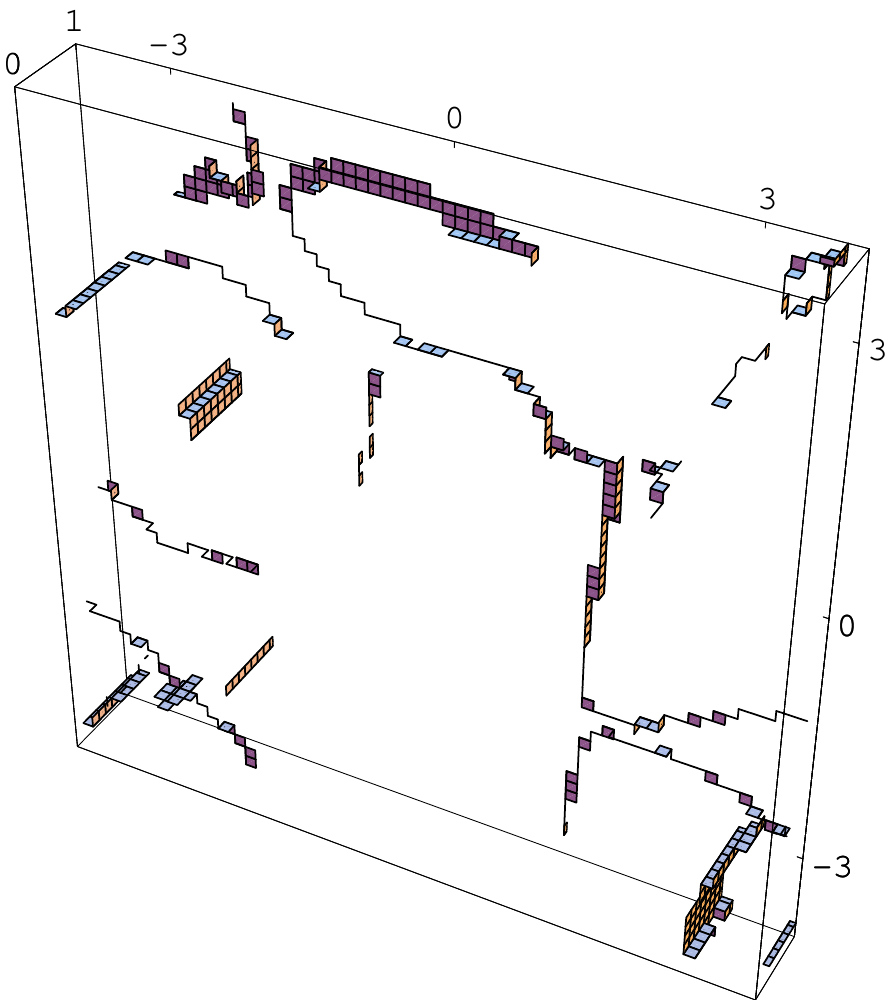}
\caption{The vortex content of a caloron ensemble in the deconfined phase 
(upper panel, mimicked by holonomy $\o=0.0625$ close to trivial) and in the 
confined phase (lower panel, maximal nontrivial holonomy $\o=0.25$) in a 
lattice slice at fixed spatial coordinate.
The short direction is $x_0$ while the other directions are the remaining spatial 
ones, all given in units of $\beta$.}
\label{fig:Figure18}
\end{figure} 

In Figs.\ \ref{fig:Figure16} and \ref{fig:Figure17} we show the space-time part 
respectively the purely spatial part of the corresponding vortices as a function 
of the holonomy $\o$. 
One can clearly see that \emph{with holonomy 
approaching the confining 
value $1/4$ the vortices grow in size, especially the spatial vortices start to 
percolate}, which will be quantified below.


Fig.~\ref{fig:Figure18} shows another view on this property. 
In these plots we have fixed one of the spatial coordinates 
to a particular value, 
such that vortices become line-like or remain surfaces 
(and may 
appear
to be non-closed, when they actually close through other slices 
than the fixed one).
These plots should be compared to Fig.~7 of 
Ref.~\cite{Engelhardt:1999fd}, which however does not show a particular 
vortex configuration, but the authors' interpretation of measurements 
(in addition, the authors of \cite{Engelhardt:1999fd} seem to have overlooked 
that vortices cut at fixed spatial coordinate still have surface-like parts, 
i.e.\ dual plaquettes).

We find that vortices in the deconfined phase tend to align in the time-like 
direction, while in the confined phase vortices percolate in the spatial 
directions. We remind the reader that we distinguish the different phases 
by the values of the holonomy, $\o\simeq 0,\,\o\simeq 1/2$ vs. $\o=1/4$, and 
not by different temperatures, which would lead to different caloron density and
size distribution.

\begin{figure*}[t]
\includegraphics[width=0.24\linewidth]{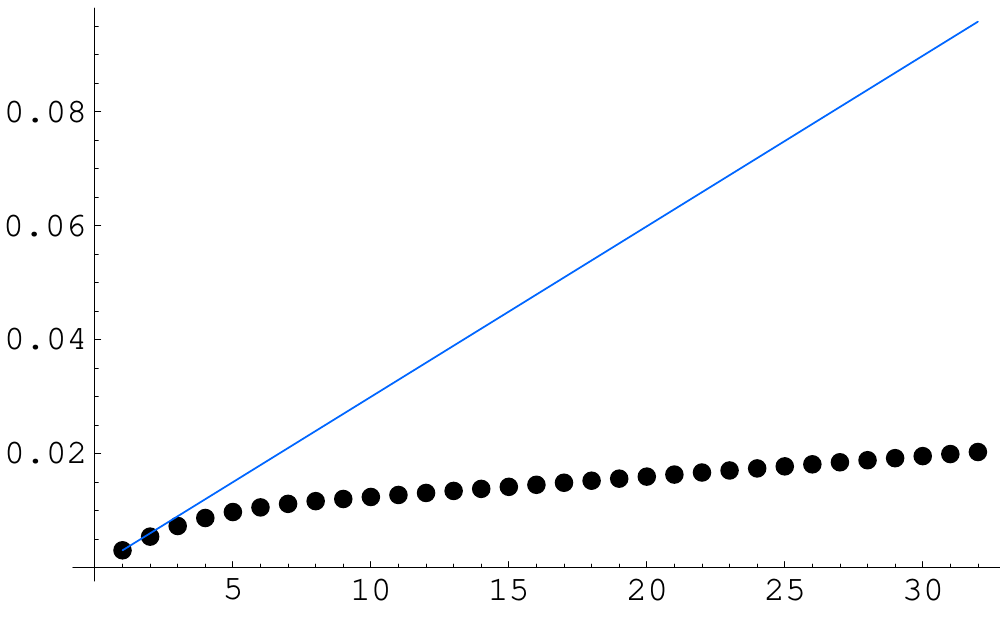}
\includegraphics[width=0.24\linewidth]{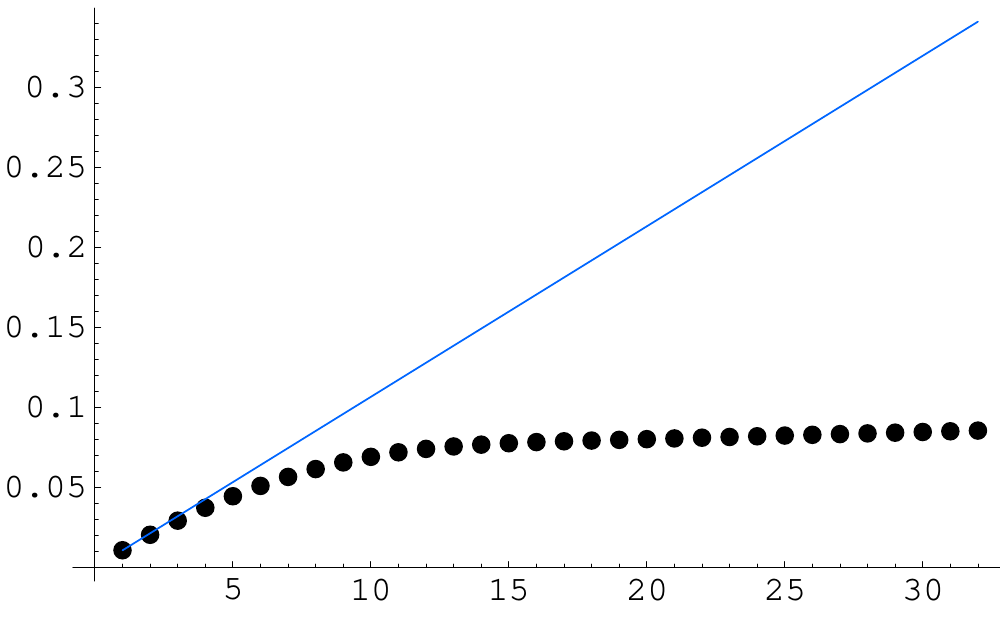}
\includegraphics[width=0.24\linewidth]{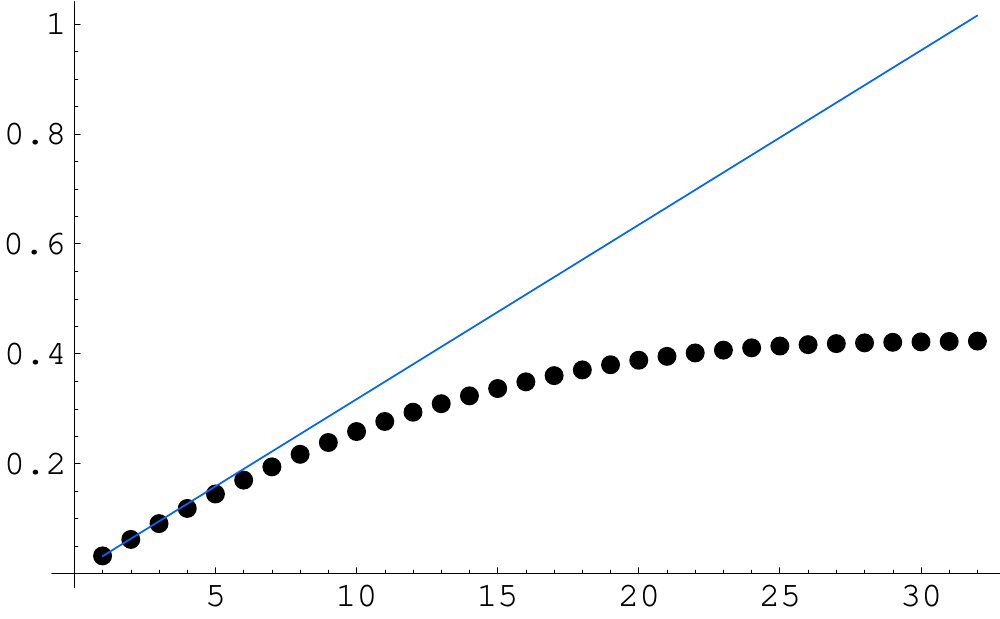}
\includegraphics[width=0.24\linewidth]{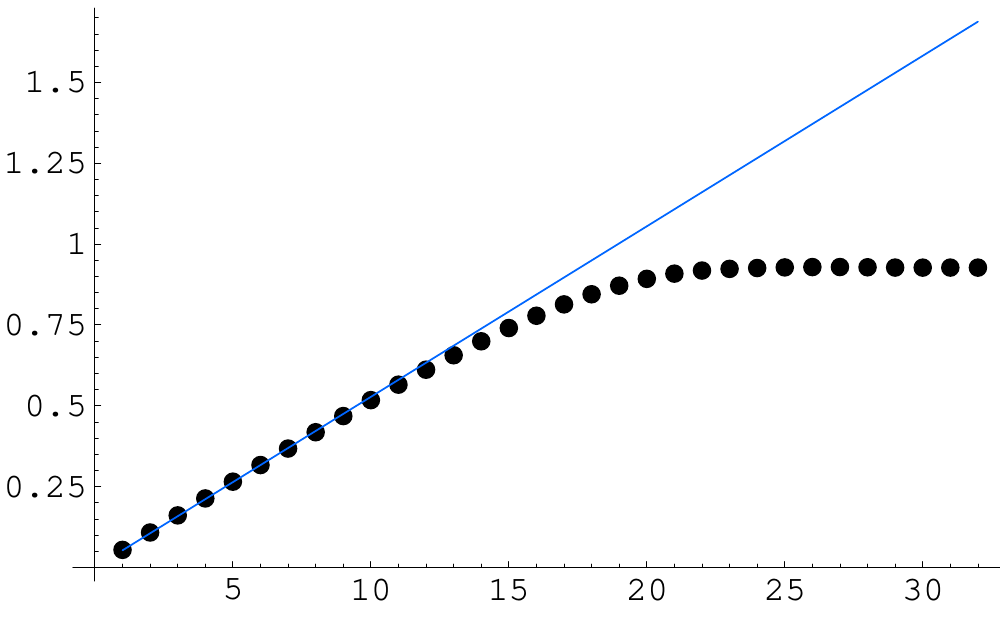}
\caption{The observable 
$-\log\langle W(L,\beta)\rangle$ 
as a function of $L$ in units of lattice spacing $a=\beta/8$
from vortices in caloron ensembles 
with holonomies as in Figs.~\protect{\ref{fig:Figure16}} and \protect{\ref{fig:Figure17}} (from left to right: $\o=\{0.0625,\,0.0125,\,0.01875,\,0.25\}$). Note the very different scales.
Percolating vortices, i.e.\ random penetrations, would give a linear behavior.
The line shown here is a linear extrapolation of the first data point.
The caloron vortices follow this line for higher and higher distances $L$ when the holonomy approaches the confinement value $1/4$, thus inducing confinement in the Polyakov loop correlator (see text).}
\label{fig:Figure19}
\end{figure*}

We start our interpretation of these results by the fact that the caloron 
background is dilute in the sense that the topological density 
is well approximated by the sum over the constituent dyons of individual calorons 
(of course, the long-range $A_\mu^{a=3}$ components still ``interact'' with the 
short-range $A_\mu^{a=1,2}$ components inside other dyon cores).
Therefore it is permissible and helpful to interpret the vortices in caloron 
ensembles as approximate \emph{
recombination of vortices from individual calorons} presented in the previous sections.

Indeed, the space-time vortices resemble the dyon-induced vortices which are 
mostly space-time like and therefore line-like at fixed $x_0$. 
The spatial vortices, on the other hand, resemble the twist-induced bubbles
in individual calorons. 

Following the 
recombination interpretation, the bubbles should become larger and larger when the holonomy approaches the
confining holonomy $\o=1/4$, where they degenerate to flat planes. This is indeed the case in caloron 
ensembles: towards $\o=1/4$ the individual vortices merge to from one big vortex, see also the schematic
plot Fig.~7 in \cite{Zhang:2009et}.
In other words, \emph{the maximal nontrivial holonomy has the effect of forcing 
the spatiall vortices to percolate}. Consequently the vortices yield 
exponentially decaying Polyakov loop correlators, the equivalent of the Wilson
loop area law at finite temperature (see below).

Note also that 
there is no similar scenario for the space-time vortices.
The dyon-induced vortices in each caloron are either always as large as the lattice
(in the singlet case)
or are always confined 
to the interior of the caloron
(in the doublet case).
This is consistent with the physical picture, that spatial Wilson loops 
do not change much across the phase transition.\\

In order to quantify the percolation we measured 
two observables. 
The first one concerns the spatial and space-time extensions of the largest vortices.
We have included plots of them in the second row of Figs.~\ref{fig:Figure16}
and \ref{fig:Figure17}, respectively.

For the spatial extension we superpose the purely spatial vortex plaquettes of all time slices in one 3d lattice, 
whereas for the space-time extension we remove all purely spatial vortex plaquettes.
Then we pick the largest connected cluster in the remaining vortex structure.

Table \ref{tab:Table1} shows the 
extension of largest spatial and space-time vortex clusters for different holonomy 
parameters of otherwise identical caloron ensembles.
The spatial-spatial vortex cluster extension changes drastically with the 
holonomy parameter $\o$ whereas the space-time one almost keeps to percolate.

The second 
row in Table \ref{tab:Table1}
is related to confinement generated by 
vortices. 
If center vortices penetrate a Wilson loop with extensions $T$ and $L$ 
\emph{randomly}, 
the probability to find $n$ vortices penetrating the area $A=TL$ is given by the 
Poisson distribution
\begin{equation}
 P_r(n;T,L)=\frac{(pA)^n}{n!}\,e^{-p A}
\end{equation}
where $p$ is the density of (spatial) vortices. 
The index $r$ characterizes the perfect randomness of this distribution,
in order to distinguish it from an arbitrary empirical distribution.
The average Wilson 
loop in such a
center configuration is given by an alternating sum
\begin{equation}
 \langle W(T,L)\rangle=\sum_n (-1)^n P(n;T,L)
\end{equation}
Obviously, for the Poisson distribution $P_r$ 
one obtains an area law, $\log \langle W\rangle \varpropto TL$,
with a string tension $\sigma=2 p$.

\begin{table}[t]
\centering
\begin{tabular}{r | c c c c }
\hline\hline
 holonomy parameter   & $0.0625$ & $0.125$ & $0.1875$ & $0.25$ \\ \hline
 space-time extension & $47$      & $56$    & $56$     & $56$ \\ 
 spatial extension    & $20$      & $35$    & $56$     & $56$ \\ \hline\hline 
\end{tabular}
\caption{Extensions of the largest vortex cluster (see text) in the caloron ensembles of Figs.\protect{\ref{fig:Figure16}} and \protect{\ref{fig:Figure17}}.
Note that the largest extension on a $8\cdot 64^3$ lattice is $\sqrt{(8/2)^2+3\cdot(32/2)^2}=55.6$.}
\label{tab:Table1}
\end{table}

We explore space-time Wilson loops $\langle W(L,\beta)\rangle$, which amount to  Polyakov loop 
correlators at distance $L$ and probe confinement,
as well as purely spatial Wislon loops as a function of their area, $\langle W(L,L')\rangle\equiv \langle W(A=LL')\rangle$.
As \fig{fig:Figure19} clearly shows, the values of 
$\log \langle W(L,\beta)\rangle$ show a confining linear behaviour 
(like from random vortices) reaching
larger and larger
distances $L$ when the holonomy approaches the confinement value $\o=1/4$.
At the same time the corresponding string tensions also grow by an order of magnitude.


One can also explain the deviation from the confining 
behaviour in holonomies far from $1/4$ . The probability 
$P(2;L,\beta)$ for two vortices penetrating the Wilson loop is found much larger 
than in the case of Poisson distribution [not shown]. 
This comes from small bubbles, which very likely penetrate a given 
rectangular twice.

\begin{figure}[t]
\includegraphics[width=0.48\linewidth]{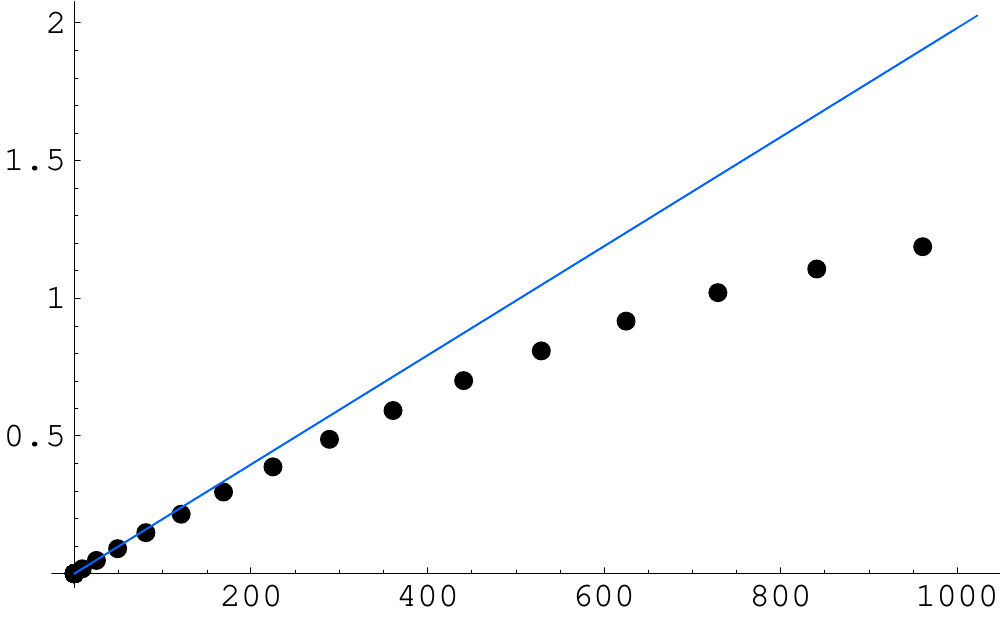}
\includegraphics[width=0.48\linewidth]{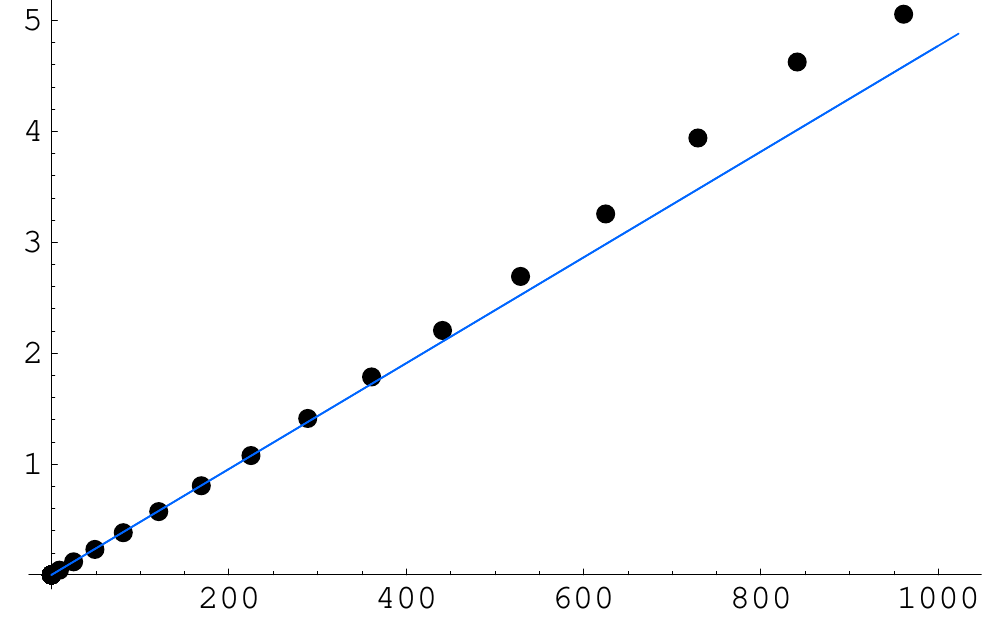}
\caption{ The observable $-\log\langle W(A)\rangle$ as a function of the area $A$ in lattice units $a^2=\beta^2/64$ from vortices in caloron ensembles
with holonomy parameters $\omega=0.0625$ (left) and $\omega=0.25$ (right).
Percolating vortices, i.e.\ random penetrations, would give a linear behavior.
The line shown here is a linear extrapolation of the first data point.} 
\label{fig:Figure20}
\end{figure}

The behaviour of $\log \langle W(A)\rangle$ on the other hand changes only slightly for different holonomies, see Fig.~\ref{fig:Figure20}.
The corresponding slopes ("string tensions") vary by a factor of approximately 2. For holonomy $\o=0.25$ we find an exponential decay stronger than proportional to the area. A quantitative analysis of this effect needs to include suitable caloron densities and size distributions around the critical temperature.

\section{Summary and Outlook}
\label{sect:SummaryOutlook}

In this paper we have extracted the vortex content of $SU(2)$ calorons and 
ensembles made of them, mainly with 
the help of the Laplacian Center Gauge, and studied the 
properties of the emerging vortices. Our main results are\\

(1) The constituent dyons of calorons induce zeroes of the lowest adjoint 
mode and therefore 
appear as monopoles in the Laplacian Abelian Gauge. 
The corresponding worldlines are either two static lines or one closed loop
for large and small calorons, respectively.\\ 

(2)  One part of the caloron's vortex surface contains the dyon/monopole
worldlines. The vortex changes its flux there (hence the surface should 
be viewed as non-\-orien\-table). These are general properties of LCG 
vortices. The specific shapes of these dyon-induced vortices 
depend on the caloron size as well as on the lattice extensions. 
These vortices are predominantly space-time like.\\

(3) Another part of the vortex surface consists of a ``bubble'' around 
one of the dyons, depending on the holonomy. The bubble degenerates into 
the midplane of the dyon ``molecule'' 
in the case of maximal nontrivial holonomy $\o=1/4$.
This part 
is 
predominantly spatial. We have argued that it is induced by the relative twist between different dyons in the caloron.\\

(4) Both parts of the vortex together reproduce the unit topological 
charge of the caloron by 2 intersection points with contributions $1/2$.\\

(5) In dilute caloron ensembles -- that differ only in the holonomy mimicking 
confined and deconfined phase -- the vortices can be described to a 
good approximation by 
recombination 
of vortices from individual calorons.
The spatial vortices in ensembles with deconfining holonomies $\o\simeq 0$ 
form small bubbles.
With the holonomy approaching
the confinement value $\o=1/4$, the spatial bubbles grow 
and merge with each other, i.e.\ they percolate in spatial directions.
We have quantified this by 
the extension of the largest 
cluster on one hand and by the quark-antiquark
potential revealed by the Polyakov loop correlator
on the other.\\

In particular the last finding is in agreement with 
the (de)confinement mechanism 
based on the percolation of center vortices.

We postpone a more detailed analysis of this mechanism as well as working out
the picture for the more physical case of gauge group $SU(3)$ 
to a later paper.

\section*{Acknowledgements}

We are grateful to Michael M\"uller-Preussker, Stefan Olejnik and Pierre van Baal 
for useful comments and to Philippe de Forcrand for discussions in an early stage 
of the project. We also thank Philipp Gerhold for making his code for the generation
of caloron ensembles available.
FB and BZ have been supported by DFG (BR 2872/4-1) and BM by RFBR grant 09-02-00338-a.

\renewcommand{\thesection}{}
\renewcommand{\thesubsection}{\Alph{subsection}}
\renewcommand{\theequation}{\Alph{subsection}.\arabic{equation}}
\section*{Appendix}

\setcounter{equation}{0}

\subsection{Caloron gauge fields}
\label{app_functions}

Here we give the functions necessary for the gauge fields of 
calorons \cite{Kraan:1998pm} including $\beta$ and their form 
in the limits described in Sect.~\ref{sect:Calorons}. 
The first set of two auxiliary dimensionless functions are
\begin{widetext}
\begin{eqnarray}
\psi&=&-\cos(2\pi x_0/\beta)+\cosh(\r)\cosh(\s)+\frac{r^2+s^2+d^2}{2rs}\sinh(\r)\sinh(\s)\nonumber\\
&&+ d\left(\frac{\sinh(\r)}{r}\,\cosh(\s)+\cosh(\r)\,\frac{\sinh(\s)}{s}\right)\label{eqn_psi_exact}\\
\hat{\psi}&=&-\cos(2\pi x_0/\beta)+\cosh(\r)\cosh(\s)+\frac{r^2+s^2-d^2}{2rs}\sinh(\r)\sinh(\s) \, .
\end{eqnarray}
We remind the reader that $r=|\vec{x}-\z|$ and $s=|\vec{x}-\zz|$ are the distances to the dyon locations and $d=|\z-\zz|=\pi\rho^2/\beta$ is the distance 
between the dyon locations, the ``size of the caloron''.
The next set of two auxiliary functions entering Eqn. (3) are 
\begin{eqnarray}
 \phi=\frac{\psi}{\hat{\psi}}\,,&\qquad&
 \tc=\frac{1}{\psi}\,d\,\left(\frac{\sinh(\r)}{r}+e^{-2\pi i x_0/\beta}\frac{\sinh(\s)}{s}\right)
\label{eqn_chihat_exact}
\end{eqnarray}
\end{widetext}

For the twist we have analyzed the limit of large size $d\gg \beta$ in 
Sect.~\ref{sect:Calorons}. For points $\vec{x}=\z+\vec{\delta}$ near the 
location of the first dyon, 
$r=|{\vec \delta}|$ 
is small and $s=d-\delta_3+O(\delta^2/d)$ is large. 
Hence the argument $\s$ is much larger 
than 1 (unless trivial holonomy $\o=0$) and the hyperbolic functions can be 
replaced by exponential functions with exponentially small corrections. 
On the other hand, no manipulations are made in all functions with 
argument $\r$, such that we get the exact expressions in terms of the 
distance $r$,
\begin{eqnarray}
 \psi&=&e^{\s}\frac{d}{r}\sinh(\r)\\
 \hat{\psi}&=&\frac{1}{2}e^{\s}\left(\cosh(\r)-\frac{\delta_3}{r}\sinh(\r)\right)\nonumber
\end{eqnarray}
The exponentially large prefactors cancel in the functions $\phi$ and $\tilde{\chi}$:
\begin{eqnarray}
 \phi(\vec{x}=\z+\vd)&=&\frac{2d}{\delta\coth(4\pi\bo \delta/\beta)-\delta_3}\\
 \tc(\vec{x}=\z+\vd)&=&e^{-2\pi i x_0/\beta}\frac{1}{2d}\frac{\delta}{\sinh(4\pi\bo \delta/\beta)}
\end{eqnarray}
where we have replaced $r$ by 
$|{\vec \delta}|$. 

For points $\vec{x}=\zz+\vec{\delta}$ near the location of the second dyon, 
$s=|{\vec \delta}|$ 
is small and $r= d+\delta_3+O(\delta^2/d)$ is large leading to 
\begin{eqnarray}
 \psi&=&e^{\r}\,d\,\frac{\sinh(\s)}{s}\\
\hat{\psi}&=&\frac{1}{2}e^{\r}\left(\cosh(\s)+\frac{\delta_3}{s}\sinh(\s)\right)\nonumber
\end{eqnarray}
and
\begin{eqnarray}
\phi(\vec{x}=\zz+\vd)&=&\frac{2d}{\delta\coth(4\pi\o\delta/\beta)+\delta_3}\\
 \tc(\vec{x}=\zz+\vd)&=&\frac{1}{2d}\frac{\delta}{\sinh(4\pi\o \delta/\beta)}
\end{eqnarray}

\end{document}